\documentclass[12pt]{article}

\usepackage{epsfig,latexsym,amsfonts,amsmath,amsthm,amssymb,amsbsy,multirow,slashed,wasysym,textcomp,dsfont,comment,mathtools,cancel,cite,diagbox,datetime,appendix,BOONDOX-calo}
\usepackage{tocloft}
\usepackage{bbold}
\usepackage{sidecap}
\usepackage{adjustbox}
\usepackage{pgfplots}
\usepackage{pgfkeys}
\usepackage{oplotsymbl}
\usepackage{tikzpagenodes}
\usepackage{adforn}
\tikzstyle{bag} = [align=center]
\usetikzlibrary{shapes,backgrounds,arrows.meta,patterns,decorations.markings,bending,positioning, decorations.pathmorphing,cd} 
\tikzset{snake it/.style={decorate, decoration=snake}}
\usepackage{adjustbox}
\usepackage[normalem]{ulem}
\usepackage[mathscr]{eucal}
\usepackage[version=3]{mhchem}
\usepackage{physics}

\usepackage[hidelinks]{hyperref}
\usepackage{subcaption}

 \newcommand{\badat}{\begin{alignedat}}
 \newcommand{\eadat}{\end{alignedat}}
 
 \def\be{\begin{equation}}
\def\ee{\end{equation}}

\usepackage{color}

\newcommand{\pink}[1]{\textcolor{\pink}{#1}}

\definecolor{dblue}{rgb}{0.2,0.50,0.80}

\usepackage[utf8]{inputenc}

\tikzset{snake it/.style={decorate, decoration=snake}}

\usepackage{tensor}
\usepackage{amsmath}
\usepackage{graphicx}

\usepackage{amssymb}
\usepackage{mathrsfs}
\usepackage[normalem]{ulem}
\usepackage{bbm} 
\usepackage{graphicx}
\usepackage{stmaryrd}
\usepackage{xcolor}
\usepackage{mathtools}
\usepackage{xfrac}

\theoremstyle{definition}

\newtheorem{thm}{Theorem}

\numberwithin{equation}{section} 
\pgfplotsset{compat=1.17} 
\begin{document}

\begin{titlepage}

\unitlength = 1mm
\ \\
\vskip 2cm
\begin{center}
{\LARGE{\textsc{Fortuitous ABJM Operators as Black Holes}}}

\vspace{0.8cm}
Connor Behan$^{1,2}$ and Leonardo Pipolo de Gioia$^2$
\vspace{0.7cm}\\
\small{${}^{1}$\textit{Perimeter Institute for Theoretical Physics, 31 Caroline St. N, N2L 2Y5,  Waterloo, Canada}
\vspace{5pt}\\
${}^2$\textit{ICTP South American Institute for Fundamental Research\\IFT-UNESP, S\~{a}o Paulo, SP Brazil 01440-070}} 
\vspace{20pt}

\begin{abstract}
Given a superconformal gauge theory at a fixed loop order, BPS states can be studied in two ways --- by enumerating cohomology classes of a nilpotent supercharge or by diagonalizing the anomalous dimension matrix.
The second one allows for a deeper look at how holographic theories distinguish between operators
based on whether they are dual to horizonless geometries or black holes.
To this end, we carry out the diagonalization in $\mathrm{U}(N)_k \times \mathrm{U}(N)_{-k}$ ABJM theory with $N$ as a formal parameter. This makes it possible to define trajectories of operators and focus on the fortuitous ones which are characterized by being BPS at only finitely many values of $N$. In contrast to $\mathcal{N} = 4$ Super Yang-Mills, where it has been numerically prohibitive to study more than one of these, ABJM gives us access to 18 such trajectories along with their descendants. We show that they take an especially simple form at large $N$ where they agree with the predictions of integrability. We also show that many of them, as with black holes, can only be dressed by certain types of gravitons and comment on how this exercise in cohomology also has implications for properties of the full operators. Finally, we show that the gap to the first non-BPS state in ABJM theory qualitatively differs from that of $\mathcal{N} = 4$ Super Yang-Mills in the expected way.
\end{abstract}
\vspace{0.5cm}
\end{center}
\vspace{0.8cm}

\end{titlepage}
\tableofcontents

\section{Introduction}
\label{sec:intro}

Black holes are among the most important and mysterious objects in the universe, with some of their unique properties making them ideal systems for obtaining insights on quantum gravity. Notably, it was understood that they behave as thermal systems with an entropy scaling with the area instead of the volume \cite{b73}. This hinted upon the notion of holography \cite{gr-qc/9310026,hep-th/9409089}, which lies today at the cornerstone of our understanding of quantum gravity in anti-de Sitter (AdS) space. Another important property of black holes is that classically they obey the no-hair theorem: a black hole solution in classical general relativity is completely characterized by a small number of parameters, such as its mass, charge and angular momentum \cite{Hawking:1971vc}. Moreover, in flat spacetimes, black holes are expected to evaporate as a result of the particle emission known as Hawking radiation \cite{h75}, which leads to the information problem \cite{Hawking:1976ra}.

To better understand these phenomena at a more fundamental level,
and aspects of quantum gravity that follow from them,
it is extremely desirable to have at our disposal concrete examples of quantum black hole microstates. The holographic correspondence \cite{hep-th/9711200} between AdS quantum gravity and a conformal field theory (CFT) provides a promising approach since it gives the quantum gravity Hilbert space a very concrete realization through the operator formalism of CFT. In fact, the duality between type IIB superstring theory on AdS$_5\times S^5$ with $N$ units of five-form flux and ${\cal N}=4$ Super Yang-Mills (SYM) with gauge group ${\rm SU}(N)$ on $\mathbb{R}^{1,3}$ was the first example of top-down holography, and it was in this example that a proposal to identify minimally supersymmetric (BPS) quantum black hole states first emerged.

The need for two qualitatively different types of BPS states in $\mathcal{N} = 4$ SYM can be argued based on the superconformal index \cite{hep-th/0510251}, a signed counting of BPS states which is independent of the coupling. In the case of imperfect cancellation, which is the generic expectation, the unrefined index
\begin{equation}
I_N(t) = \text{Tr} \Big [ (-1)^F t^{2(E + J_L)} \Big ] = \sum_n d_N(n) t^n
\end{equation}
should exhibit $O(N^2)$ entropy which is the result of applying the Bekenstein-Hawking formula to known extremal black holes \cite{hep-th/0401042,hep-th/0401129}. Due to a series of papers culminating in \cite{1810.11442,1810.12067}, the $d_N(n)$ are now known to behave in this way if one takes $N,n \to \infty$ while keeping $n/N^2$ fixed. On the other hand, taking the $N \to \infty$ limit first only produces $\log |d_\infty(n)| = O(\sqrt{n})$, which is closer to the entropy for an ideal gas of gravitons. The simplest explanation for this behaviour, which had already been predicted in \cite{1305.6314}, is that it reflects a competition between two types of contributions: \textit{monotone states} defined to be BPS for all $N$, and \textit{fortuitous states} defined to be BPS for $N$ sufficiently small. Since monotone operators can be matched to coherent states of gravitons in a precise sense, black holes and all other exotic geometries must be fortuitous.

The first fortuitous operator was found in the computerized search of \cite{2209.06728}, which showed that the sensitive dependence on $N$ is a manifestation of trace relations. This search used the fact that BPS states, annihilated by a nilpotent supercharge $Q$ and its BPZ adjoint $Q^\dagger$, are in one-to-one correspondence with $Q$-cohomology classes.
Further developments in this approach were made shortly afterwards, first
at the cohomology level in \cite{2209.12696,2304.10155,2312.16443} and later at the operator level in \cite{2306.04673,2306.04693}, which can be said to contain the first concrete examples of black hole microstates. The monotone-fortuitous dichotomy has since been vastly generalized \cite{2402.10129} and applied to a variety of 4d \cite{2512.12764,2605.16254}, 3d \cite{2511.03105,2512.04146,2512.23603}, 2d \cite{2501.05448,2505.14888,2511.23294,2604.20663} and 1d \cite{2412.06902,2504.14181,2511.00790,2608.12160,2604.27164} theories. In particular, \cite{2512.04146,2512.23603} have studied the fortuitous cohomologies in 3d ${\cal N}=6$ ABJM theory \cite{0806.1218} and unveiled an unexpected feature: they appear at a much lower level and appear to be much simpler than in ${\cal N}=4$ SYM. ABJM theory is therefore a natural arena for studying quantum black holes through the low-lying fortuitous Hilbert space.
This paper will investigate structures in this Hilbert space which are relevant for two long-term goals: (1) seeing how the unique classical properties of black holes become manifest at the quantum level, and (2) interpreting black holes as bound states arising from more fundamental degrees of freedom.

An appealing starting point for the first goal is the classical no-hair theorem.
Quantum mechanically, a notion of hair can be defined by looking for the possible ways of dressing a black hole state.
A recent example, motivated by the asymptotic symmetries of flat space, is the suggestion that black holes could admit soft hair \cite{1601.00921}:
dressing by soft bosons such as photons and gravitons. Of course, it is hard to make this proposal precise in flat space, since we do not know a rigorous way to formulate a holographic principle, let alone determine which states on the associated Hilbert space are to be understood as black holes. We can, however, study this kind of phenomenon with the current AdS/CFT machinery.

This direction was pursued in \cite{2304.10155} which found many examples of no-hair behaviour in ${\cal N}=4$ SYM: multiplying a fortuitous cohomology by single gravitons most often yields a trivial cohomology. Put differently, multiplying representatives yields a $Q$-exact operator, which is a null state. The authors called this a partial no-hair theorem. We will find that they are common in ABJM theory as well. In terms of the lowest fortuitous operator $\mathcal{O}^a_b$, to be discussed at length in the later sections, one of our simplest results is that the graviton dressings
\begin{equation}
\epsilon_{ac} \epsilon^{bd} \mathcal{O}^a_b \text{Tr}(\bar{\phi}^c \phi_d), \quad \mathcal{O}^{\{a}_{\{b} \text{Tr}(\bar{\phi}^{c\}} \phi_{d\}})
\end{equation}
are allowed while
\begin{equation}
\epsilon_{ac} \mathcal{O}^a_{\{b} \text{Tr}(\bar{\phi}^c \phi_{d\}}), \quad \epsilon^{bd} \mathcal{O}^{\{a}_b \text{Tr}(\bar{\phi}^{c\}} \phi_d)
\end{equation}
are not. Although we have obtained these results in a brute-force level-by-level way, partial no-hair theorems can also be seen using the superconformal index. Indeed, what \cite{2304.10155} found was that the index in a certain truncation of $\mathcal{N} = 4$ SYM with $N = 2$ could be fully explained by a sequence of ``core primaries'' (which are known in cohomology) and dressings by a very restricted set of gravitons. This truncation of (one-loop) $\mathcal{N} = 4$ SYM is known as the BMN sector due to its relation with the BMN matrix quantum mechanics \cite{hep-th/0202021}. The fortuity perspective has influenced several works on this model in recent years \cite{2404.18442,2602.22163,2605.25560,2606.05388}. These studies, along with a set of composite operators we identified in \cite{2512.23603}, indicate that BMN quantum mechanics might have something to say about ABJM theory as well.

\begin{figure}[h]
\centering
\includegraphics[scale=0.075]{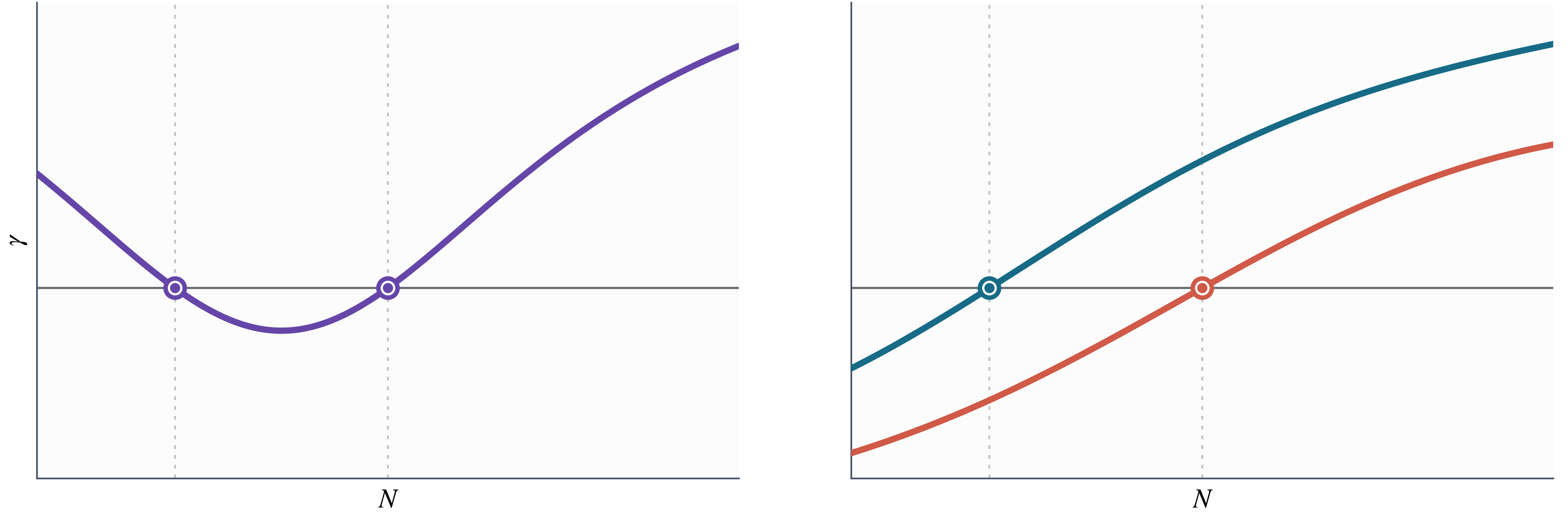}
\caption{Two schematic plots of the anomalous dimension $\gamma$ versus the matrix size $N$. Both of them show two fortuitous operators with the same spin and R-symmetry charges. On the left, analytic continuation can be used to move from one to the other. On the right, the two are instead part of distinct trajectories. We will only find examples of the later in the charge sectors we will study in this paper.}
\label{fig:branch-scenarios}
\end{figure}

To make progress on the second goal, we go beyond cohomology and construct the actual fortuitous BPS operators in ABJM theory. To do so, we embed these operators in continuous families at non-integer $N$ as was done in \cite{2306.04693} for the lowest example in $\mathcal{N}=4$ SYM. This once more showcases how ABJM theory makes it easier to study fortuity:
we manage to construct the analytic continuation in closed form for the lowest fortuitous operator in ABJM with $E = \frac{5}{2}$ previously identified in \cite{2512.23603,2512.04146}. More than that, we numerically compute the trajectories in $N$ for all of the so-called centralizer primaries that were enumerated in \cite{2512.23603}, thus giving a total of 18.
Various features of gauge theories happen to be manifest in these trajectories --- single-traces and multi-traces become independent at large $N$ and states below the unitarity bound decouple at integer $N$. The results also display features which would have been hard to predict \textit{a priori}. Up to the level that we investigated, when a charge sector contains fortuitous operators that are BPS for different values of $N$, they are part of different analytically continued families, \textit{i.e.} the second scenario in Figure \ref{fig:branch-scenarios}. In all such cases, the lightest member of this set is the one that becomes single-trace at large $N$. We are also able to survey a large set of eigenvalues, as done in \cite{2306.04673} for $\mathcal{N} = 4$ SYM, and see that the gap between the unitarity bound and the first non-BPS operator dominates over other gaps in the spectrum.

This paper is organized as follows. Section \ref{sec:conventions} reviews some essential features of ABJM theory including the field content, supercharge actions and BPS counting. Section \ref{sec:dressing} then explores the idea of dressing the BPS black hole cohomologies found in \cite{2512.23603} with graviton hair. We explain the fact that a descendant graviton gives new information if and only if the corresponding primary is subject to a no-hair theorem. Detecting them algorithmically leads to compact expressions for many new fortuitous cohomology representatives. Section \ref{sec:hamiltonian}, which is the heart of the paper, focuses on operators which lift above the BPS bound. Their anomalous dimensions, depending on $N$, are obtained as eigenvalues of the two-loop Hamiltonian in radial quantization which we construct using the superconformal algebra. Additionally, the eigenvectors allow us to single out the canonical representative for each operator that is fortuitous at $N = 2$ or $N = 3$. As in \cite{2306.04693}, we follow observables from these small values of $N$ to the planar limit where ABJM becomes integrable. Section \ref{sec:integrability} then exploits this connection with integrability to carry out several non-trivial checks. Section \ref{sec:conc} finally summarizes our results and speculates about possible interpretations.

\textbf{Note added:} Shortly after the first version of this paper appeared, \cite{2609.19297} showed that the monotone-fortuitous dichotomoy is too coarse in many theories with bifundamental matter, including ABJM. The authors introduced a new fortuity criterion to help decide whether operators are dual to black holes or branes wrapping internal cycles \cite{1103.1163}. Some of the operators we have looked at are fortuitous in this stronger sense and we will elaborate on this in the conclusion.

\section{BPS Operators in ABJM Theory}
\label{sec:conventions}
The BPS sector of a supersymmetric gauge theory has a rich mathematical structure. This is captured by the holomorphic twist in 4d \cite{1111.4234} and the holomorphic-topological twist in 3d \cite{2005.00083}. These constructions are based on a nilpotent supercharge $Q$ which exists in any theory with at least four supercharges. The choice of $Q$ induces a natural splitting of the spacetime symmetry algebra which is the 3d $\mathcal{N} = 6$ superconformal algebra in the case of ABJM theory. In this section, we will discuss these algebraic considerations and review some results of \cite{2512.04146,2512.23603} on the monotone and fortuitous operators that are known so far.

We will work at one loop in what follows. When powers of the Chern-Simons level $k$ are restored, this means truncating the action of the supercharge as
\begin{equation}
Q = Q^{(0)} + k^{-1} Q^{(1)} + O(k^{-2}). \label{q-expansion}
\end{equation}
Although the superconformal index is protected from perturbative corrections by design, more refined probes of the BPS sector are not. The operators themselves will depend on both the coupling and the renormalization scheme when they are expressed in terms of fundamental fields. It is also expected that the \textit{number} of BPS operators, as encoded in a partition function for instance, will receive corrections at non-trivial loop orders. This has been verified by explicit calculations in 4d theories \cite{2306.01039,2506.13887,2510.24008,2511.09519,2512.07771,2606.27955} and it is likely that they can be repeated in 3d.

\subsection{Supercharge actions}
The 3d superconformal algebra $\mathfrak{osp}(\mathcal{N} | 4)$ may be presented using the anti-commutators
\begin{align}
\begin{split} \label{anticomms}
& \{ Q_{\alpha r}, Q_{\beta s} \} = 2 \delta_{rs} P_{\alpha \beta}, \quad \{ S^\alpha_r, S^\beta_s \} = -2 \delta_{rs} K^{\alpha \beta}, \\
& \;\;\;\; \{ Q_{\alpha r}, S^\beta_s \} = 2i [\delta_{rs} (M_\alpha^{\;\;\beta} + \delta_\alpha^\beta D) - i \delta_\alpha^\beta R_{rs}]
\end{split}
\end{align}
along with commutators involving the bosonic generators (written for instance in \cite{2011.05728}). As long as $\mathcal{N} \geq 2$, it is possible to construct the supercharges
\begin{equation}
Q \equiv Q_{-1} + iQ_{-2}, \quad Q^\dagger = S^-_1 - iS^-_2
\end{equation}
which are nilpotent as implied by \eqref{anticomms}. The case of relevance for ABJM theory is $\mathcal{N} = 6$.

\begin{figure}[h]
\centering
\begin{tikzpicture}[
    gauge/.style={circle, draw, thick, minimum size=1.3cm, align=center},
    matter/.style={->, thick, >=stealth},
    midarrow/.style={
        thick,
        postaction={
            decorate,
            decoration={
                markings,
                mark=at position 0.5 with {\arrow{stealth}}
            }
        }
    },
    label/.style={font=\small},
    rhs/.style={font=\Large}
]

\node[gauge] (U1) at (0,0) {${\rm U}(N)_{+k}$};
\node[gauge] (U2) at (3,0) {${\rm U}(N)_{-k}$};

\draw[midarrow]
  (U1.north east) to[bend left=18]
  node[label, above] {$\Phi_A, \Psi^A$}
  (U2.north west);

\draw[midarrow]
  (U2.south west) to[bend left=18]
  node[label, below] {$\bar{\Phi}^A, \bar{\Psi}_A$}
  (U1.south east);

\node[rhs, align=center] at (8.2,0) {
  $\mathcal{W}
  =
  \epsilon^{ab}\epsilon_{cd}
  \operatorname{Tr}(\phi_a \bar{\phi}^c \phi_b \bar{\phi}^d)$
};

\end{tikzpicture}
\caption{The defining data of ABJM as a 3d $\mathcal{N} = 2$ superconformal Chern-Simons theory. The fundamental scalars are $\Phi_A$ and $\bar{\Phi}^A$. The BPS letters $\phi_a$ and $\bar{\phi}^a$ are particular rescaled components defined below. The direction of the arrows indicates that barred and unbarred fields must alternate inside a trace to produce a gauge invariant operator.}
\label{fig:quiver-w}
\end{figure}
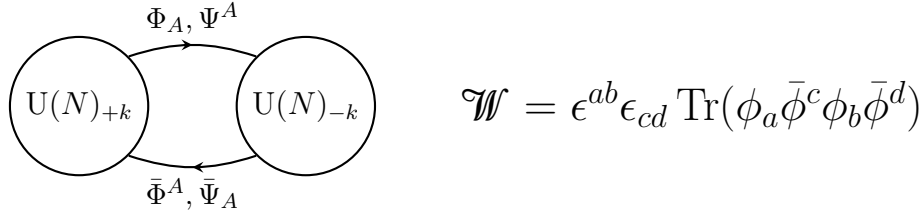
The Lagrangians for ABJM and several related theories were written explicitly in \cite{0806.4977}. It is also possible to specify ABJM theory with the quiver and superpotential shown in Figure \ref{fig:quiver-w}. The matter multiplets arrange themselves into
\begin{equation}
\Phi_A, \quad \bar{\Phi}^A, \quad \Psi^A_\alpha, \quad \bar{\Psi}_{A\alpha}, \quad A \in \{ 1,2,3,4 \} \label{abjm-fields}
\end{equation}
which transform as either spinors or conjugate spinors of ${\rm SO}(6)$ as indicated by the position of the index. Components of these fields may be labelled by their energy $E$, spin $J$ and ${\rm SO}(6)$ Cartans $(H_1, H_2, H_3)$. Special attention should be paid to the components annihilated by
\begin{equation}
\{ Q, Q^\dagger \} = 4(E - J - H_1). \label{bps-cond}
\end{equation}
This is known as the \textit{BPS condition}. Non-negativity of \eqref{bps-cond}, combined with discreteness of $J$ and $H_1$, has a simple consequence --- all fields (also referred to as ``letters'') in a BPS operator must be BPS individually. If even one of them were non-BPS, \eqref{bps-cond} would differ from zero by a finite amount and quantum corrections would not be able to change that.\footnote{An important structural fact about the expansion \eqref{q-expansion} is that $Q^{(n)}$ must be suitably restricted before it becomes a nilpotent operator. In particular, its cohomology can only be defined as a subspace of the $Q^{(n - 1)}$-cohomology \cite{2306.01039,2512.07771}. Starting all enumerations from the set of BPS letters is just the $n = 1$ case of this result.}

The BPS components of \eqref{abjm-fields} may be identified by considering the $4 \times 4$ Weyl matrices $\Gamma_{AB}$ and $\bar{\Gamma}^{AB}$. Specifically, the combinations $i(\Gamma_1 \bar{\Gamma}_2)_A^{\;\;B}$ and $i(\bar{\Gamma}_1 \Gamma_2)^A_{\;\;B}$ have eigenvalues of $\pm 1$. Writing $A=(L,a)\oplus(R,a)$ with $a\in\{1,2\}$, the $+1$ projections are $\Phi_{La}$, $\bar\Phi_R^a$, $\Psi_{R+}^a$, and $\bar\Psi_{La+}$. We follow the normalization convention of \cite{2512.23603} and define the BPS letters by
\begin{equation}
\phi_a \equiv 2\Phi_{La},\qquad
\bar{\phi}^a \equiv 2\bar{\Phi}_R^a,\qquad
\psi^a \equiv -2i\Psi_{R+}^a,\qquad
\bar{\psi}_a \equiv -2i\bar{\Psi}_{La+}.
\label{abjm-letters}
\end{equation}
They may also  be freely acted upon by the covariant derivative $D \equiv iD_{++}$.
In the rest of this paper, we will only write quantities made from lowercase letters which all have $H_1 = \frac{1}{2}$. For the purposes of Appendix \ref{app:gram}, however, we should note that the BPS letters only have two-point functions with different components of \eqref{abjm-fields} having $H_1 = -\frac{1}{2}$.
With the normalization \eqref{abjm-letters}, one can use the SUSY transformation laws in \cite{0807.0880} to determine that $Q$ acts as
\begin{align}
& [Q, \phi_a] = 0, \quad \quad \quad \quad \quad [Q, \bar{\phi}^a] = 0 \label{q-action} \\
& \{ Q, \psi^a \} = \epsilon^{bc} \phi_b \bar{\phi}^a \phi_c, \quad \{ Q, \bar{\psi}_a \} = \epsilon_{bc} \bar{\phi}^b \phi_a \bar{\phi}^c \nonumber \\
& [Q, D \chi \} = (\epsilon^{ab} \phi_a \bar{\psi}_b - \epsilon_{ab} \psi^a \bar{\phi}^b) \chi - (-1)^{|\chi|} \chi (\epsilon_{ab} \bar{\phi}^a \psi^b - \epsilon^{ab} \bar{\psi}_a \phi_b) + D[Q, \chi \} \nonumber \\
& [Q, D \bar{\chi} \} = (\epsilon_{ab} \bar{\phi}^a \psi^b - \epsilon^{ab} \bar{\psi}_a \phi_b) \bar{\chi} - (-1)^{|\chi|} \bar{\chi} (\epsilon^{ab} \phi_a \bar{\psi}_b - \epsilon_{ab} \psi^a \bar{\phi}^b) + D [Q, \bar{\chi} \}. \nonumber
\end{align}
In what follows, it will also be important to understand the action of four additional supercharges.

For this last calculation, we would like to consider the generators which have vanishing graded commutators with $Q$ and $Q^\dagger$. These form the \textit{centralizer algebra} which is a symmetry of the BPS sector. For the case of ABJM theory, the centralizer was found to be
\begin{equation}
\mathfrak{osp}(4|2) \oplus \mathfrak{u}(1|1) \subset \mathfrak{osp}(6|4) \label{cen-3d}
\end{equation}
in \cite{2512.23603} which includes the fermionic generators $Q_{+r}$ and $S^+_r$ for $r \geq 3$. Defining $\sigma_r \equiv (i, \vec{\sigma})$ and $\bar{\sigma}_r \equiv (-i, \vec{\sigma})$, the matter fields transform according to
\begin{align}
\begin{split}
& [Q_{+r}, \phi_a] = i (\sigma_r)_{ab} \psi^b, \quad \{ Q_{r+}, \psi^a \} = -i (\bar{\sigma}_r)^{ab} D \phi_b \\
& [Q_{+r}, \bar{\phi}^a] = i (\bar{\sigma}_r)^{ab} \bar{\psi}_b, \quad \{ Q_{r+}, \bar{\psi}_a \} = -i (\sigma_r)_{ab} D \bar{\phi}^b
\end{split}
\end{align}
while covariant derivatives transform trivially. Deriving this is straightforward after slightly changing the convention in \cite{0807.0880} for Weyl matrices. For our purposes, it will be more convenient to use a different basis of supercharges which act as
\begin{align}
\begin{split} \label{cen-actions}
& [Q^a_b, \phi_c] = \epsilon_{cb} \psi^a, \quad \{ Q^a_b, \psi^c \} = -\epsilon^{ca} D \phi_b \\
& [Q^a_b, \bar{\phi}^c] = \epsilon^{ca} \bar{\psi}_b, \quad \{ Q^a_b, \bar{\psi}_c \} = -\epsilon_{cb} D \bar{\phi}^a.
\end{split}
\end{align}
Although it is manifestly true that \eqref{cen-actions} acts within the space of BPS letters, the superconformal algebra only closes up to field-dependent gauge transformations.

\subsection{Monotone and fortuitous cohomologies}
When searching for BPS operators with large quantum numbers, it is very helpful to use the following theorem.
\begin{thm}
If $\mathcal{H}$ is a Hilbert space and $Q \in \text{End}(\mathcal{H})$ is a nilpotent operator, there is an isomorphism between $\text{ker} \{ Q, Q^\dagger \}$ and $\text{ker} Q / \text{im} Q$. In particular, there is exactly one BPS operator in each cohomology class.
\end{thm}
This means that the $Q$-action \eqref{q-action} is enough to count BPS operators and obtain expressions for them up to a $Q$-exact ambiguity. One only needs to invoke $Q^\dagger$ to resolve this ambiguity in a second pass.
Before carrying out this two-step process for ABJM theory (as done in \cite{2306.04673,2306.04693} for $\mathcal{N} = 4$ SYM), there are three key features of the cohomology problem which should be reviewed.

First, for a fixed value of $N$, a BPS operator may be $Q$-closed in one of two ways. These are written as
\begin{equation}
[Q, \mathcal{O} \} = 0, \quad \text{or} \quad [Q, \mathcal{O} \} \in I_N
\end{equation}
where $I_N$ is the ideal of trace relations. This leads to the distinction emphasized in the introduction. When a cohomology class or its BPS operator is of the former type, it is referred to as \textit{monotone} because it is part of a family which exists for any value of $N$. All other cohomology classes containing BPS operators are only $Q$-closed for certain values of $N$ and have therefore been termed \textit{fortuitous}. The work of \cite{2209.06728,2209.12696} led to a surge of interest in fortuitous operators since finite-$N$ effects are needed to describe chaotic systems like black holes. The superconformal index and BPS partition function can both be split into their monotone and fortuitous contributions.
\begin{equation}
I_N = I_N^{\text{mon}} + I_N^{\text{fort}}, \quad Z_N = Z_N^{\text{mon}} + Z_N^{\text{fort}}
\end{equation}

A second well known insight is that the monotone sector is composed entirely of multi-graviton operators.\footnote{Strictly speaking, this is only true when all nodes in the quiver have a rank scaling with $N$. In the situation explored by \cite{2511.03105}, which involves $1 \times N$ and $N \times 1$ matrices, gravitons and mesons lead to distinct types of monotone operators.} One can see this by going to $N = \infty$ so that all trace relations disappear except (super)cyclicity. In this case, multi-trace cohomology classes are given by polynomials in single-trace or \textit{cyclic cohomology} classes \cite{2306.01039}. When $N$ is made finite, the number of multi-graviton operators in a given charge sector decreases due to trace relations while the new BPS operators that appear are fortuitous by definition. This pattern of losing monotone operators and gaining fortuitous operators leads to the question of which effect will dominate. The answer is that, upon lowering $N$, the number of BPS states counted by the index experiences a net decrease for small global charges and a net increase for large global charges \cite{2005.10843}. This is in line with the statement that the inverse Laplace transform of the index exhibits Bekenstein-Hawking growth in the double scaling limit $N \to \infty$ with $E/N^2$ fixed \cite{1810.11442}.

Finally, let us point out that although the partition function
\begin{align}
Z(x, \textbf{y}) = \text{Tr} \Big [ x^{E + J} y^{H_1} y_+^{H_2 + H_3} y_-^{H_2 - H_3} \Big ]
\end{align}
is not as well understood as the index
\begin{align}
I(x, \textbf{y}) = \text{Tr} \Big [ (-1)^F x^{E + J} y_+^{H_2 + H_3} y_-^{H_2 - H_3} \Big ],
\end{align}
its form is strongly constrained by representation theory. This is because the multiplicities in each charge sector are not only required to be non-negative. It must also be possible to arrange them into a non-negative number of centralizer multiplets through the use of characters. These manipulations are slightly subtle because fortuitous operators are long, single-graviton operators are short and multi-graviton operators can be either long or short. Nevertheless, the results in \cite{2512.23603} lead to a consistent picture.
Computing one more power of $x$ in this work, the fortuitous partition function counting $\mathfrak{osp}(4|2)$ primaries is
\begin{align}
Z_2^{\text{fort}} \chi^{-1} &= x^3 y^2 \chi^+_1 \chi^-_1 + x^4 y^3 (1 + \chi^+_2 \chi^-_2) + x^5 \Big [ y^2 \chi^+_1 \chi^-_1 + y^3 (2 \chi^+_1 \chi^-_1 + \chi^+_1 \chi^-_3 + \chi^+_3 \chi^-_1) + y^4 \chi^+_3 \chi^-_3 \Big ] \nonumber \\
&+ x^6 \Big [ y^2 (1 + 4y) (1 + \chi^+_2)(1 + \chi^-_2) + y^4 (\chi^+_2 \chi^-_4 + \chi^+_4 \chi^-_2 + 2 \chi^+_2 \chi^-_2) + y^5 \chi^+_4 \chi^-_4 \Big ] + O(x^7) \label{zfort-n2}
\end{align}
for $N = 2$ and
\begin{align}
Z_3^{\text{fort}} \chi^{-1} &= x^4 y^3 (1 + \chi^+_2 \chi^-_2) + x^5 y^4 (2 \chi^+_1 \chi^-_1 + \chi^+_1 \chi^-_3 + \chi^+_3 \chi^-_1 + \chi^+_3 \chi^-_3) \label{zfort-n3} \\
&+ x^6 \Big [ y^3 \chi^+_2 \chi^-_2 + y^4 (2 + 3 \chi^+_2 + 3 \chi^-_2 + 3 \chi^+_2 \chi^-_2 + \chi^+_4 + \chi^-_4 + \chi^+_2 \chi^-_4 + \chi^+_4 \chi^-_2) \nonumber \\
&+ y^5 (2 + \chi^+_4 + \chi^-_4 + 2 \chi^+_4 \chi^-_4 + \chi^+_2 \chi^-_4 + \chi^+_4 \chi^-_2 + 3 \chi^+_2 \chi^-_2) \Big ] + O(x^7) \nonumber
\end{align}
for $N = 3$.
These expressions use
\begin{align}
\chi(x, y_+, y_-) = \frac{(1 + x y_+ y_-)(1 + x y_+ y_-^{-1})(1 + x y_+^{-1} y_-)(1 + x y_+^{-1} y_-^{-1})}{1 - x^2}
\end{align}
to account for $\mathfrak{osp}(4|2)$ descendants and
\begin{align}
\chi_{2j}(y_\pm) = \sum_{m = -j}^j y_\pm^{2m}
\end{align}
to account for $\mathfrak{su}(2) \oplus \mathfrak{su}(2)$ descendants. Knowing this representation content, \cite{2512.23603} was able to write down explicit representatives for the $O(x^3)$ primary in \eqref{zfort-n2} and the $O(x^4)$ primaries in \eqref{zfort-n3} essentially by trial and error. Among the higher-order terms we have shown, it turns out that only a few of them need to have their representatives constructed from scratch in a similar way. The others allow us to exploit a recursive structure which is more efficient and also physically interesting.

\section{Enumeration Through Dressing}\label{sec:dressing}
In a sense, \cite{2512.04146,2512.23603} already provide infinitely many examples of fortuitous cohomology classes. Given a representative of an $\mathfrak{osp}(4|2)$ primary, it is straightforward to apply supercharges from \eqref{cen-actions}, referred to as $Q'$ here, to build representatives for descendants. This action clearly preserves fortuity --- if $Q \mathcal{O} \left | 0 \right >$ is a trace relation then $Q Q' \mathcal{O} \left | 0 \right >$ simply produces $Q'$ acting on that same trace relation. A more interesting question is how one can generate new fortuitous primaries.

When $Q$ satisfies the Leibniz rule, as is the case at $O(k^{-1})$, a promising strategy comes from multiplying primaries. This is simply a matter of iterating
\begin{align}
(\mathcal{O}_1 + Q \chi_1) (\mathcal{O}_2 + Q \chi_2) \left | 0 \right > = \mathcal{O}_1 \mathcal{O}_2 \left | 0 \right > + Q (\chi_1 \mathcal{O}_2 + \mathcal{O}_1 \chi_2 + \chi_1 Q \chi_2) \left | 0 \right >. \label{dress-schematic1}
\end{align}
When $\mathcal{O}_1$ say is fortuitous, we can be sure that \eqref{dress-schematic1} is not monotone but this is not enough to make it a fortuitous primary. Indeed, one has to worry about whether $\mathcal{O}_1 \mathcal{O}_2$ is $Q$-exact --- in such cases, \eqref{dress-schematic1} is $Q$-exact as well. More generally $\mathcal{O}_1 \mathcal{O}_2$ could be given by a sum of exact and monotone operators. A remarkable result of \cite{2304.10155} is that this scenario is ubiquitous in $\mathcal{N} = 4$ SYM with gauge group ${\rm SU}(2)$ --- starting with any of the known fortuitous primaries in this theory, it is never possible to make a new one by combining it with a product of single-graviton primaries.
Instead, one must consider single-graviton descendants of a particular form.

This section will classify all of the terms shown in \eqref{zfort-n2} and \eqref{zfort-n3} according to whether or not they can be built up by dessing the lower-order terms with graviton operators. This is the property which defines the \textit{core primaries} in the language of \cite{2304.10155}. To identify them, we will start at the top level of a graviton multiplet and look for dressings which lie in the span of monotone and $Q$-exact operators. The results then indicate which dressings at the next level will need to be checked. Indeed, consider a fortuitous primary $\mathcal{O}_1$, a graviton primary $\mathcal{O}_2$ and a conformal supercharge $S' \in \mathfrak{osp}(4|2)$. The relation
\begin{align}
S' \mathcal{O}_1 Q' \mathcal{O}_2 \left | 0 \right > = \# \mathcal{O}_1 \mathcal{O}_2 \left | 0 \right > + Q \chi \left | 0 \right > \label{dress-schematic2}
\end{align}
does not necessarily indicate that $\mathcal{O}_1 Q' \mathcal{O}_2 \left | 0 \right >$ is a representative for a centralizer descendant. If the exactness check passes on the right-hand side then $S'$ kills the representative on the left-hand side and a new exactness check needs to be done.

\subsection{Results in $\mathcal{N} = 4$ SYM}
\begin{figure}[h]
\centering
\begin{tikzpicture}[
  box/.style={
    draw,
    rounded corners,
    align=center,
    minimum width=2.6cm,
    minimum height=0.9cm
  },
  arr/.style={->, thick},
  x=2.2cm,
  y=1.4cm
]

\node[box] (A) at (0,0) {$[0]_n^{(n, 0)}$};

\node[box] (B1) at (-0.7,-1) {$[\tfrac{1}{2}]_{n + 1/2}^{(n - 1, 0)}$};
\node[box] (B2) at ( 0.7,-1) {$[0]_{n + 1/2}^{(n - 1, 1)}$};

\node[box] (C1) at (-1.4,-2) {$[0]_{n + 1}^{(n - 2, 0)}$};
\node[box] (C2) at ( 0.0,-2) {$[\frac{1}{2}]_{n + 1}^{(n - 2, 1)}$};
\node[box] (C3) at ( 1.4,-2) {$[0]_{n + 1}^{(n - 1, 0)}$};

\node[box] (D1) at (-0.7,-3) {$[0]_{n + 3/2}^{(n - 3, 1)}$};
\node[box] (D2) at ( 0.7,-3) {$[\tfrac{1}{2}]_{n + 3/2}^{(n - 2, 0)}$};

\node[box] (E) at (0,-4) {$[0]_{n + 2}^{(n - 3, 0)}$};

\draw[arr] (A) -- (B1);
\draw[arr] (A) -- (B2);

\draw[arr] (B1) -- (C1);
\draw[arr] (B1) -- (C2);
\draw[arr] (B2) -- (C2);
\draw[arr] (B2) -- (C3);

\draw[arr] (C1) -- (D1);
\draw[arr] (C2) -- (D1);
\draw[arr] (C2) -- (D2);
\draw[arr] (C3) -- (D2);

\draw[arr] (D1) -- (E);
\draw[arr] (D2) -- (E);

\end{tikzpicture}
\caption{The conformal primaries in the $B_1 \bar{B}_1 [0]_n^{(n, 0)}$ multiplet generated by a trace of $n$ scalars. The null states are $[\tfrac{1}{2}]_{n + 1/2}^{(n + 1, 0)}$ and $[0]_{n + 1/2}^{(n, 1)}$ while the supercharges transform as $[\tfrac{1}{2}]_{1/2}^{(0, 1)}$ and $[0]_{1/2}^{(1, 0)}$.}
\label{fig:4d-multiplet}
\end{figure}
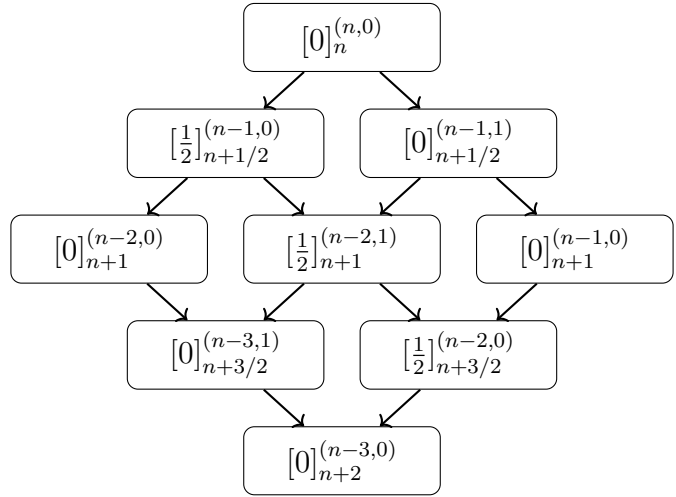

Before discussing ABJM theory further, it will be helpful to review the search for core primaries of $\mathcal{N} = 4$ SYM in a less schematic way. As has been known for a long time, the BPS letters of this theory are
\begin{equation}
\lambda_{\dot{\alpha}}, \quad \phi^i, \quad \psi_i, \quad f, \quad i \in \{ 1, 2, 3 \}
\end{equation}
and their $Q$-actions are
\begin{align}
\{ Q, \lambda_{\dot{\alpha}} \} &= 0 \quad [Q, \phi^i] = 0, \quad \{ Q, \psi_i \} = -i \epsilon_{ijk} [\phi^j, \phi^k], \quad [Q, f] = i[\phi^i, \psi_i] \nonumber \\
[ Q, D_{\dot{\alpha}} \zeta \} &= -i[ \lambda_{\dot{\alpha}}, \zeta \} + D_{\dot{\alpha}} [ Q, \zeta \}. \label{n4-qactions}
\end{align}
The centralizer is
\begin{equation}
\mathfrak{psu}(1, 2 | 3) \oplus \mathfrak{u}(1 | 1) \subset \mathfrak{psu}(2, 2 | 4) \label{cen-4d}
\end{equation}
and the actions of the other supercharges in it are written in \cite{0803.4183}. Modulo $Q$-exact terms, they preserve the number of fields in a trace (which is also known as the bonus charge). This suggests that any single-graviton operator lives in a centralizer multiplet whose primary only has scalars. Due to \eqref{n4-qactions}, a trace of scalars is always $Q$-closed and the order of them is immaterial in cohomology. To construct the BPS operator, it is not difficult to guess that we should take the symmetrized combination
\begin{align}
\text{Tr} \left ( \phi^{\{ i_1} \dots \phi^{i_n \}} \right ). \label{sg-primary-4d}
\end{align}
These operators generate short multiplets of \eqref{cen-4d} and their representation content was worked out in \cite{hep-th/0510251}. Figure \ref{fig:4d-multiplet} summarizes this in the notation of \cite{1612.00809}.

The partition function for the cyclic cohomology is found by including one such multiplet for each $n \in [1, \infty)$ and using the $\mathfrak{su}(3)$ characters
\begin{align}
\chi_{(R_1, R_2)}(y_1, y_2) = \text{det} \left (
\begin{bmatrix}
(1/y_2) & y_1 & (y_2/y_1) \\
(1/y_2)^{-1} & y_1^{-1} & (y_2/y_1)^{-1} \\
1 & 1 & 1
\end{bmatrix}^{-1}
\begin{bmatrix}
(1/y_2)^{R_1 + 1} & y_1^{R_1 + 1} & (y_2/y_1)^{R_1 + 1} \\
(1/y_2)^{-R_2 - 1} & y_1^{-R_2 - 1} & (y_2/y_1)^{-R_2 - 1} \\
1 & 1 & 1
\end{bmatrix}
\right ).
\end{align}
Due to the BPS condition, $E$ and $J_L$ can be taken to have the same fugacity without loss of generality. This leads to the result
\begin{align}
Z_\infty^{\text{cyc}} &= \text{Tr} \Big [ x_1^{2(E + J_L)} x_2^{2J_R} y^Y y_1^{H_1 - H_2} y_2^{H_2 - H_3} y_3^{H_2 + H_3} \Big ] \label{n4-cyc-z} \\
&= \frac{x_1^2 y}{(x_2 - x_1^3) (1 - x_1^3 x_2) (1 - x_1^2 y y_1) (y_2 - x_1^2 y y_3) (y_1 - x_1^2 y y_2 y_3)} \Big [ x_1^{10} x_2 y^2 y_1 y_2 y_3^2-x_1^7 x_2^2 y^2 y_1 y_2 y_3^2 \nonumber \\
&+x_1^7 x_2^2 y y_1 y_2 y_3-x_1^7 y^2 y_1 y_2 y_3^2+x_1^7 y y_1 y_2 y_3+x_1^6 x_2 y y_1^2 y_2 y_3+x_1^6 x_2 y y_1 y_3^2+x_1^6 x_2 y y_2^2 y_3^2+x_1^5 x_2^2 y y_1^2 y_3 \nonumber \\
&+x_1^5 x_2^2 y y_1 y_2^2 y_3+x_1^5 x_2^2 y y_2 y_3^2+x_1^5 y y_1^2 y_3+x_1^5 y y_1 y_2^2 y_3+x_1^5 y y_2 y_3^2+x_1^4 x_2 y^2 y_1 y_2 y_3^2+x_1^4 x_2 y y_1 y_2 y_3^2 \nonumber \\
&-x_1^4 x_2 y y_1 y_2 y_3-x_1^4 x_2 y_1 y_2 y_3+x_1^4 x_2 y_1 y_2-x_1^2 x_2 y y_1^2 y_3-x_1^2 x_2 y y_1 y_2^2 y_3-x_1^2 x_2 y y_2 y_3^2+x_1^2 x_2 y_1^2 \nonumber \\
&+x_1^2 x_2 y_1 y_2^2+x_1^2 x_2 y_2 y_3+x_1 x_2^2 y_1 y_2 y_3+x_1 y_1 y_2 y_3+x_2 y_1^2 y_2+x_2 y_1 y_3+x_2 y_2^2 y_3 \Big ]. \nonumber
\end{align}
The original calculation in \cite{hep-th/0510251} set $y = 1$ and used a different convention for which Cartan to eliminate. Nevertheless, their result is consistent with \eqref{n4-cyc-z} as can be seen by setting
\begin{equation}
x = x_1 y_3^{1/3}, \quad z = x_1 y_3^{-2/3}, \quad y = x_2, \quad v = y_1 y_3^{-2/3}, \quad w = y_2 y_3^{-1/3}.
\end{equation}
One can also compare \eqref{n4-cyc-z} to the partition function in \cite{1305.6314} which again used different fugacities --- those which are adapted to derivatives in an on-shell superspace formalism instead of Cartans. After the change of variables
\begin{equation}
x = y y_3, \quad a = x_1^3 x_2, \quad b = x_1^3 x_2^{-1}, \quad u = x_1^2 y_1 y_3^{-1}, \quad v = x_2^2 y_2 y_1^{-1}, \quad w = x_1^2 y_2^{-1},
\end{equation}
the results are seen to match once more. Finally, \eqref{n4-cyc-z} agrees with a bulk calculation since the \eqref{sg-primary-4d} operators (in addition to being centralizer primaries) transform as primaries in $B_1 \bar{B}_1 [0; 0]_n^{(0, n, 0)}$ representations of $\mathfrak{psu}(2, 2 | 4)$.
These multiplets are the only ones consistent with the absence of higher-spin particles and it has been known since \cite{krv85} that each of them appears in the AdS$_5 \times S^5$ Kaluza-Klein spectrum exactly once.

At finite $N$, the simplest consequence of trace relations is that single-gravitons can have at most $N$ fields. For the case of $N = 2$, there is only one graviton multiplet and the conformal primaries in it can be associated with the representatives
\begin{align}
& \text{Tr}(\phi^i \phi^j), \quad \text{Tr}(\phi^i \lambda_{\dot{\alpha}}), \quad \text{Tr} \left ( \epsilon^{\dot{\alpha} \dot{\beta}} \lambda_{\dot{\alpha}} \lambda_{\dot{\beta}} \right ), \nonumber \\
& \text{Tr} \left ( \phi^i \psi_j - \frac{1}{3} \delta^i_j \phi^k \psi_k \right ), \quad \text{Tr} \left ( \lambda_{\dot{\alpha}} \psi_i - \epsilon_{ijk} \phi^j D_{\dot{\alpha}} \phi^k \right ), \label{sym-su2-traces} \\
& \text{Tr} \left ( \phi^i f - \frac{1}{4} \epsilon^{ijk} \psi_j \psi_k \right ), \quad \text{Tr} \left ( \lambda_{\dot{\alpha}} f - \frac{2}{3} \psi_i D_{\dot{\alpha}} \phi^i + \frac{1}{3} \phi^i D_{\dot{\alpha}} \psi_i \right ) \nonumber
\end{align}
worked out in \cite{2304.10155}. These should be used to dress fortuitous operators, of which the first example is
\begin{equation}
\mathcal{O} = \epsilon^{ijk} \text{Tr}(\phi^{l} \psi_{i}) \text{Tr}(\phi^{m} \psi_{j}) \text{Tr}(\psi_{k} [\psi_{l}, \psi_{m}]). \label{sym-fort}
\end{equation}
While it is possible to check that various products of the traces in \eqref{sym-su2-traces} become $Q$-exact when multiplied by $\mathcal{O}$, the authors of \cite{2304.10155} were able to show something stronger by considering the so called BMN sector which omits covariant derivatives $D_{\dot{\alpha}}$ and gaugini $\lambda_{\dot{\alpha}}$ from the set of BPS letters.\footnote{Another construction is to take s-wave fields in $\mathcal{N} = 4$ SYM which survive compactification on $S^3 \times \mathbb{R}$ \cite{hep-th/0306054}. Since we are not considering loop corrections, the result can be equivalently described by the BMN matrix model \cite{hep-th/0202021}.} Specifically, they showed that for any fortuitous primary $\mathcal{O}'$, the dressings
\begin{equation}
\mathcal{O}' \text{Tr} \left ( \phi^{i_1} f - \frac{1}{4} \epsilon^{i_1 j_1 k_1} \psi_{j_1} \psi_{k_1} \right ) \dots \text{Tr} \left ( \phi^{i_n} f - \frac{1}{4} \epsilon^{i_n j_n k_n} \psi_{j_n} \psi_{k_n} \right ) \label{n4-allowed-dressing}
\end{equation}
are the only ones seen by the fortuitous BMN index --- an index which is calculable for gauge group ${\rm SU}(2)$.\footnote{Generalizing to other gauge groups is an interesting open problem. For ${\rm SU}(3)$, a fortuitous index capturing R-symmetry singlets in the BMN sector was found in \cite{gklm25}.} Replacing any trace in \eqref{n4-allowed-dressing} with copies of $\text{Tr}(\phi^i \phi^j)$ or $\text{Tr} \left ( \phi^i \psi_j - \frac{1}{3} \delta^i_j \phi^k \psi_k \right )$ either produces a $Q$-exact term up to monotone corrections or leads to cancellations in the index. This infinite sequence of partial no-hair theorems allowed \cite{2304.10155} to identify the charge sectors for all core primaries which can play the role of $\mathcal{O}'$ and construct explicit representatives analogous to \eqref{sym-fort} for each one.

\subsection{Cyclic cohomology in ABJM}
\begin{figure}[h]
\centering
\begin{tikzpicture}[
  box/.style={
    draw,
    rounded corners,
    minimum width=2.2cm,
    minimum height=0.9cm,
    align=center
  },
  arr/.style={->, thick},
  x=2.4cm,
  y=1.5cm
]

\node[box] (A) at (0,0) {$[n, n]_n$};

\node[box] (B1) at (-2,-1) {$[n + 1, n - 1]_{n + 1/2}$};
\node[box] (B2) at ( 0,-1) {$[n - 1, n - 1]_{n + 1/2}$};
\node[box] (B3) at ( 2,-1) {$[n - 1, n + 1]_{n + 1/2}$};

\node[box] (C1) at (-2,-2) {$[n, n - 2]_{n + 1}$};
\node[box] (C2) at ( 0,-2) {$[n, n]_{n + 1}$};
\node[box] (C3) at ( 2,-2) {$[n - 2, n]_{n + 1}$};

\node[box] (D) at (0,-3) {$[n - 1, n - 1]_{n + 3/2}$};

\draw[arr] (A) -- (B1);
\draw[arr] (A) -- (B2);
\draw[arr] (A) -- (B3);

\draw[arr] (B1) -- (C1);
\draw[arr] (B3) -- (C3);

\draw[arr] (B1) -- (C2);
\draw[arr] (B2) -- (C1);
\draw[arr] (B2) -- (C3);
\draw[arr] (B3) -- (C2);

\draw[arr] (C1) -- (D);
\draw[arr] (C2) -- (D);
\draw[arr] (C3) -- (D);

\end{tikzpicture}
\caption{The conformal primaries in the $A_1[n, n]_n$ multiplet generated by a trace of $n$ barred and $n$ unbarred scalars. The null states are $[n + 1, n + 1]_{n + 1/2}$ while the supercharges transform as $[1, 1]_{1/2}$.}
\label{fig:3d-multiplet}
\end{figure}
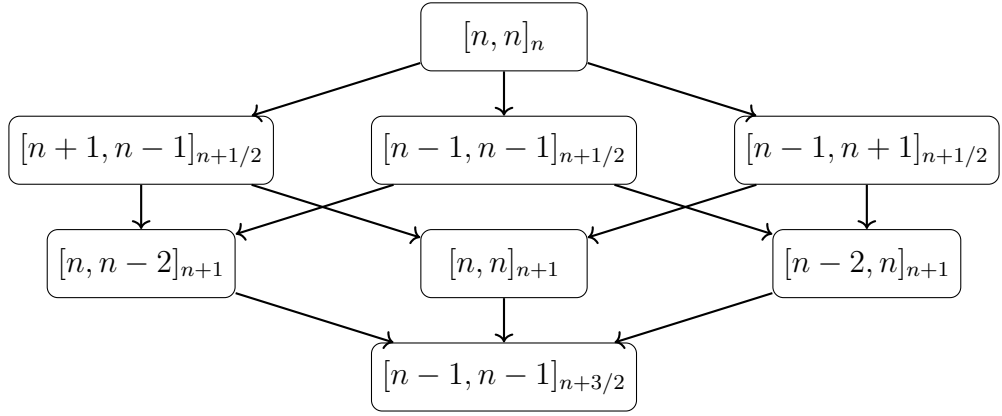

The single-graviton multiplets in ABJM theory turn out to be highly analogous to those in $\mathcal{N} = 4$ SYM. First, \eqref{q-action} makes it clear that whenever a trace consists of only scalars, the barred and unbarred strings may be separately symmetrized without loss of generality. All other orderings differ from this one by a $Q$-exact term. Starting with
\begin{equation}
\text{Tr} \Big ( \bar{\phi}^{\{a_1} \phi_{\{b_1} \dots \bar{\phi}^{a_n\}} \phi_{b_n\}} \Big ), \label{sg-primary-3d}
\end{equation}
it is easy to act with centralizer supercharges and observe that there are null states. The short $\mathfrak{osp}(4|2)$ multiplet, of which \eqref{sg-primary-3d} is the primary, may be denoted by $A_1[n, n]_n$.\footnote{Since $\mathfrak{osp}(4|2)$ does not have Lorentz spin, the notation of \cite{1612.00809} would lead to empty brackets. This is awkward so we have switched to the notation of \cite{1911.10391}.} The character for this superconformal multiplet was obtained from recombination rules in \cite{1911.10391}. Although this leads to an expression taking the form of an infinite sum, it may be simplified to a finite sum consisting of the conformal characters shown in Figure \ref{fig:3d-multiplet}. A reasonable assumption is that there are no other single-trace primaries contributing to the cyclic cohomology. Assuming this leads to the partition function
\begin{align}
Z_\infty^{\text{cyc}} &= \left [ \frac{x y y_+ y_-}{1 - x y y_+ y_-} \frac{(x + y_+ y_-) (x y_+ + y_-) (x y_- + y_+)}{(1 - y_+^2) (1 - y_-^2)} + (y_+ \leftrightarrow y_+^{-1}) \right ] + (y_- \leftrightarrow y_-^{-1}). \label{abjm-cyc-z}
\end{align}
Explicitly enumerating the $Q$-cohomology on single traces for increasing values of $n$ leads to strong evidence that this is correct.

There are $N$ of these multiplets in the ${\rm U}(N)_k \times {\rm U}(N)_{-k}$ theory. To build cohomology classes for their operators, it is enough to treat the $Q^a_b$ supercharges as anti-commuting and act with them in a fixed order since total derivatives are $Q$-exact. Nevertheless, it is also straightforward to construct conformal primaries as done below for $n = 1$.\footnote{To impose the conformal primary condition, it is important to remember that the unique bosonic Cartan of $\mathfrak{osp}(4|2)$ is a combination of dilation and R-symmetry generators from $\mathfrak{osp}(6|4)$ \cite{2512.23603}.}
\begin{align}
& \text{Tr}(\bar{\phi}^a \phi_b), \quad \text{Tr}(\bar{\phi}^{\{a} \psi^{b\}}), \quad \text{Tr}(\bar{\psi}_{\{a} \phi_{b\}}), \quad \epsilon_{ab} \text{Tr}(\bar{\phi}^a \psi^b) - \epsilon^{ab} \text{Tr}(\bar{\psi}_a \phi_b), \label{abjm-graviton-n1} \\
& 2 \text{Tr}(\bar{\psi}_a \psi^b) + \text{Tr}(\phi_a D \bar{\phi}^b) - \text{Tr}(D \phi_a \bar{\phi}^b), \quad \epsilon_{ab} \text{Tr}(\bar{\phi}^a D \psi^b - 3 D \bar{\phi}^a \psi^b) + \epsilon^{ab} \text{Tr}(D \bar{\psi}_a \phi_b - 3 \bar{\psi}_a D \phi_b) \nonumber
\end{align}
For $n = 2$, all parts of Figure \ref{fig:3d-multiplet} contribute and we have
\begin{align}
& \text{Tr}(\bar{\phi}^{\{a} \phi_{\{b} \bar{\phi}^{c\}} \phi_{d\}}), \quad \text{Tr}(\bar{\phi}^a \phi_{\{b} \bar{\psi}_c \phi_{d\}}), \quad \text{Tr}(\phi_a \bar{\phi}^{\{b} \psi^c \bar{\phi}^{d\}}), \label{abjm-graviton-n2} \\
& \epsilon_{cd} \text{Tr}(\bar{\phi}^a \phi_b \bar{\phi}^c \psi^d) + \epsilon_{cd} \text{Tr}(\phi_b \bar{\phi}^a \psi^d \bar{\phi}^c) + \epsilon^{cd} \text{Tr}(\phi_b \bar{\phi}^a \phi_c \bar{\psi}_d) + \epsilon^{cd} \text{Tr}(\bar{\phi}^a \phi_b \bar{\psi}_d \phi_c), \nonumber \\
& \epsilon_{cd} \text{Tr}(\psi^c \bar{\phi}^d \phi_{\{a} \bar{\psi}_{b\}}) + \epsilon_{cd} \text{Tr}(\bar{\psi}_{\{a} \phi_{b\}} \bar{\phi}^c \psi^d) + \epsilon^{cd} \text{Tr}(\bar{\psi}_c \phi_{\{a} \bar{\psi}_{|d|} \phi_{b\}}) - \epsilon_{cd} \text{Tr}(D\bar{\phi}^c \phi_{\{a} \bar{\phi}^d \phi_{b\}}), \nonumber \\
& \epsilon^{cd} \text{Tr}(\bar{\psi}_c \phi_d \bar{\phi}^{\{a} \psi_{b\}}) + \epsilon^{cd} \text{Tr}(\psi^{\{a} \bar{\phi}^{b\}} \phi_c \bar{\psi}_d) + \epsilon_{cd} \text{Tr}(\psi^c \bar{\phi}^{\{a} \psi^{|d|} \bar{\phi}^{b\}}) - \epsilon^{cd} \text{Tr}(D\phi_c \bar{\phi}^{\{a} \phi_d \bar{\phi}^{b\}}), \nonumber \\
& 2 \text{Tr}(\psi^{\{a} \bar{\psi}_{\{b} \phi_{c\}} \bar{\phi}^{d\}}) - 2 \text{Tr}(\bar{\psi}_{\{b} \psi^{\{a} \bar{\phi}^{d\}} \phi_{c\}}) + \text{Tr}(\bar{\phi}^{\{a} \phi_{\{b} \bar{\phi}^{d\}} D\phi_{c\}}) - \text{Tr}(\phi_{\{b} \bar{\phi}^{\{a} \phi_{c\}} D\bar{\phi}^{d\}}), \nonumber \\
& \epsilon_{cd} \text{Tr}(D\psi^c \bar{\phi}^d \phi_a \bar{\phi}^b) + \bar{\phi}^b \phi_a \bar{\phi}^d D\psi^c) - \epsilon^{cd} \text{Tr}(D\bar{\psi}_c \phi_d \bar{\phi}^b \phi_a + \phi_a \bar{\phi}^b \phi_d D\bar{\psi}_c) \nonumber \\
&- \epsilon_{cd} \text{Tr}(D\phi_a \bar{\phi}^c \psi^d \bar{\phi}^b - \bar{\phi}^b \psi^c \bar{\phi}^d D\phi_a) + \epsilon^{cd} \text{Tr}(D\bar{\phi}^b \phi_c \bar{\psi}_d \phi_a - \phi_a \bar{\psi}_c \phi_d D\bar{\phi}^a) \nonumber \\
&- \epsilon^{cd} \text{Tr}(D\phi_c \bar{\psi}_a \phi_d \bar{\phi}^b + 4 D\phi_c \bar{\psi}_d \phi_a \bar{\phi}^b - \bar{\phi}^b \phi_d \bar{\psi}_a D\phi_c - 4 \bar{\phi}^b \phi_a \bar{\psi}_d D\phi_c) \nonumber \\
&+ \epsilon_{cd} \text{Tr}(D\bar{\phi}^c \psi^b \bar{\phi}^d \phi_a + 4 D\bar{\phi}^c \psi^d \bar{\phi}^b \phi_a - \phi_a \bar{\phi}^d \psi^b D \bar{\phi}^c - 4 \phi_a \bar{\phi}^b \psi^d D \bar{\phi}^c) \nonumber \\
&+ 6\epsilon^{cd} \text{Tr}(\phi_a \bar{\psi}_c \psi^b \bar{\psi}_d) - 6\epsilon_{cd} \text{Tr}(\bar{\phi}^b \psi^c \bar{\psi}_a \psi^d). \nonumber
\end{align}

\subsection{${\rm U}(2)_k \times {\rm U}(2)_{-k}$ gauge group}
Starting from a fortuitous primary, multiplying it by a trace of scalars either produces a new fortuitous primary or an example of a partial no-hair theorem. As explained at the start of this section, the latter scenario leaves open the possibility that one can instead build a fortuitous primary operator using a more complicated single-graviton trace --- expressions for these were given in \eqref{abjm-graviton-n1} and \eqref{abjm-graviton-n2}. We will use them to answer the question of how the leading representative
\begin{align}
\mathcal{O}^a_b &= \epsilon^{cd} \left [ 4 \text{Tr}(\bar{\psi}_c \phi_d \bar{\phi}^a \phi_b) - 3 \text{Tr}(\bar{\psi}_c \phi_d) \text{Tr}(\bar{\phi}^a \phi_b) \right ] - \epsilon_{cd} \left [ 4 \text{Tr}(\psi^c \bar{\phi}^d \phi_b \bar{\phi}^a) - 3 \text{Tr}(\psi^c \bar{\phi}^d) \text{Tr}(\phi_b \bar{\phi}^a) \right ] \label{abjm-fort1}
\end{align}
may be dressed by single-graviton operators. We will also discuss a few examples of allowed and disallowed dressings by multiple gravitons. Compared to $\mathcal{N} = 4$ SYM with $N = 2$, ABJM theory with $N = 2$ leads to additional richness because \eqref{abjm-fort1} is not an R-symmetry singlet.

Starting at the first level of the $n = 1$ multiplet, the partition function \eqref{zfort-n2} indicates constraints on how
\begin{equation}
\mathcal{O}^a_b, \quad \text{Tr}(\bar{\phi}^c \phi_d) \label{dressing-ingredients1}
\end{equation}
may be combined. Specifically, it includes the term $x^4 y^3 (1 + \chi_2^+ \chi_2^-)$ but not $x^4 y^3 (\chi_2^+ + \chi_2^-)$. This predicts that mixed contractions of \eqref{dressing-ingredients1} must be expressible as a linear combination of monotone and $Q$-exact operators. We have confirmed this with an explicit membership test. The same test fails for the bi-singlet and bi-triplet combinations, indicating that they are fortuitous. Color-coding these findings, we may write
\begin{align}
\textcolor{teal}{\epsilon_{ac} \epsilon^{bd} \mathcal{O}^a_b \text{Tr}(\bar{\phi}^c \phi_d)}, \quad \textcolor{red}{\epsilon_{ac} \mathcal{O}^a_{\{b} \text{Tr}(\bar{\phi}^c \phi_{d\}})}, \quad \textcolor{red}{\epsilon^{bd} \mathcal{O}^{\{a}_b \text{Tr}(\bar{\phi}^{c\}} \phi_d)}, \quad \textcolor{teal}{\mathcal{O}^{\{a}_{\{b} \text{Tr}(\bar{\phi}^{c\}} \phi_{d\}})}. \label{dressing1}
\end{align}
The existence of combinations in the exact and monotone span (red entries in \eqref{dressing1}) indicates that we need to check the second level. Here, there is a bi-doublet which is sent to a linear combination of these red entries by $\mathfrak{osp}(4|2)$ conformal supercharges which act as
\begin{align}
\begin{split} \label{cen-actions2}
& [S^a_b, \phi_c] = 0, \quad \{ S^a_b, \psi^c \} = \epsilon^{ca} \phi_b \\
& [S^a_b, \bar{\phi}^c] = 0, \quad \{ S^a_b, \bar{\psi}_c \} = \epsilon_{cb} \bar{\phi}^a.
\end{split}
\end{align}
We will write this bi-doublet in green since our code has found that it cannot be made from $Q$-exact and monotone operators. In other words,
\begin{align}
\textcolor{teal}{
\epsilon_{ac} \mathcal{O}^a_b [ \text{Tr}(\bar{\phi}^c \psi^d) + \text{Tr}(\bar{\phi}^d \psi^c) ] - \epsilon^{ac} \mathcal{O}^d_a [ \text{Tr}(\bar{\psi}_b \phi_c) + \text{Tr}(\bar{\psi}_c \phi_b) ]
}
\end{align}
is a representative for one of the two  $x^5 y^3 \chi_1^+ \chi_1^-$ terms in the partition function.

Having reached a level with no disallowed dressings, the $n = 1$ algorithm terminates allowing us to move onto $n = 2$. Starting from
\begin{align}
\mathcal{O}^a_b, \quad \text{Tr}(\bar{\phi}^{\{c} \phi_{\{d} \bar{\phi}^{e\}} \phi_{f\}}), \label{dressing-ingredients2}
\end{align}
the computations are more intense than before but still manageable. Four irreducible representations can be made from \eqref{dressing-ingredients2} and at most one (the fully symmetrized one) is allowed to be fortuitous based on \eqref{zfort-n2}. Checking explicitly, it is indeed fortuitous, thus leaving us with
\begin{align}
\begin{split}\label{dressing2}
& \textcolor{red}{\epsilon_{ac} \epsilon^{bd} \mathcal{O}^a_b \text{Tr}(\bar{\phi}^{\{c} \phi_{\{d} \bar{\phi}^{e\}} \phi_{f\}})}, \quad
\textcolor{red}{\epsilon_{ac} \mathcal{O}^a_{\{b} \text{Tr}(\bar{\phi}^{\{c} \phi_d \bar{\phi}^{e\}} \phi_{f\}})}, \quad \\
& \textcolor{red}{\epsilon^{bd} \mathcal{O}^{\{a}_b \text{Tr}(\bar{\phi}^c \phi_{\{d} \bar{\phi}^{e\}} \phi_{f\}})}, \quad \quad \;
\textcolor{teal}{\mathcal{O}^{\{a}_{\{b} \text{Tr}(\bar{\phi}^c \phi_d \bar{\phi}^{e\}} \phi_{f\}})}.
\end{split}
\end{align}
The next level tells us to consider the 64-dimensional space of operators where $\mathcal{O}^a_b$ multiplies a trace with three bosons and one fermion. To be a candidate for representing a primary, such an operator must become $Q$-exact up to monotone corrections when acted upon by the conformal supercharges \eqref{cen-actions2}. In other words, it must be mapped to red elements of \eqref{dressing2}. This requirement gets us down to a 41-dimensional space which is the direct sum of three degenerate and three non-degenerate irreps. The non-degenerate ones may be labelled by $1 + \chi_4^+ + \chi_4^-$ in terms of $\mathfrak{su}(2) \oplus \mathfrak{su}(2)$ characters. Checking these first, they all lie in the span of monotone and $Q$-exact operators.
\begin{align}
\begin{split}
& \quad \quad \quad \quad \quad \textcolor{red}{\epsilon^{bf} \mathcal{O}^{\{a}_b \text{Tr}(\bar{\phi}^c \psi^d \bar{\phi}^{e\}} \phi_f)}, \quad \textcolor{red}{\epsilon_{af} \mathcal{O}^a_{\{b} \text{Tr}(\phi_c \bar{\psi}_d \phi_{e\}} \bar{\phi}^d)}, \\
& \textcolor{red}{\epsilon^{bf} (\epsilon_{ae} \epsilon_{cd} - \epsilon_{ac} \epsilon_{de}) \mathcal{O}^a_b \text{Tr}(\bar{\phi}^c \psi^d \bar{\psi}^e \phi_f) + \epsilon_{af} (\epsilon^{be} \epsilon^{cd} - \epsilon^{bc} \epsilon^{de}) \mathcal{O}^a_b \text{Tr}(\phi_c \bar{\psi}_d \phi_e \bar{\phi}^f)}
\end{split}
\end{align}
This is also the case for two of the three degenerate irreps, namely $2(\chi_2^+ + \chi_2^-)$.
\begin{align}
\begin{split}
& \textcolor{red}{(\epsilon_{ac} \epsilon_{de} - \epsilon_{ae} \epsilon_{cd}) \mathcal{O}^a_{\{b} \text{Tr}(\bar{\phi}^c \psi^d \bar{\phi}^e \phi_{f\}}) + \epsilon_{ac} \epsilon^{de} \mathcal{O}^a_{\{b} [ \text{Tr}(\phi_{f\}} \bar{\psi}_d \phi_e \bar{\phi}^c) + \text{Tr}(\phi_{|e} \bar{\psi}_{d|} \phi_{f\}} \bar{\phi}^c) ]} \\
& \hspace{4.8cm} \textcolor{red}{+ \alpha \epsilon_{ac} \epsilon^{ed} \mathcal{O}^a_e \text{Tr}(\bar{\phi}^c \phi_{\{d} \bar{\psi}_b \phi_{f\}})}, \\
& \textcolor{red}{(\epsilon^{bc} \epsilon^{de} - \epsilon^{be} \epsilon^{cd}) \mathcal{O}^{\{a}_b \text{Tr}(\phi_c \bar{\psi}_d \phi_e \bar{\phi}^{f\}}) + \epsilon^{bc} \epsilon_{de} \mathcal{O}^{\{a}_b [ \text{Tr}(\bar{\phi}^{f\}} \psi^d \bar{\phi}^e \phi_c) + \text{Tr}(\bar{\phi}^{|e} \psi^{d|} \bar{\phi}^{f\}} \phi_c) ]} \\
& \hspace{4.8cm} \textcolor{red}{+ \alpha \epsilon^{bc} \epsilon_{ed} \mathcal{O}^e_b \text{Tr}(\phi_c \bar{\phi}^{\{d} \psi^a \bar{\phi}^{f\}})}
\end{split}
\end{align}
For $2 \chi_2^+ \chi_2^-$, it turns out that only one linear combination has the $Q$-exact plus monotone form.
\begin{align}
\begin{split}
& \textcolor{red}{\epsilon_{ac} \mathcal{O}^a_{\{b} [ \text{Tr}(\bar{\phi}^c \psi^{\{d} \bar{\phi}^{e\}} \phi_{f\}}) + \text{Tr}(\bar{\phi}^{\{e} \psi^{d\}} \bar{\phi}^c \phi_{f\}}) ] + \epsilon^{ac} \mathcal{O}^{\{d}_a [ \text{Tr}(\phi_c \bar{\psi}_{\{b} \phi_{f\}} \bar{\phi}^{e\}}) + \text{Tr}(\phi_{\{f} \bar{\psi}_{b\}} \phi_c \bar{\phi}^{e\}}) ]}, \\
& \textcolor{teal}{\epsilon_{de} \mathcal{O}^{\{a}_{\{b} [ \text{Tr}(\bar{\phi}^{c\}} \psi^d \bar{\phi}^e \phi_{f\}}) + \text{Tr}(\bar{\phi}^{|e} \psi^{d|} \bar{\phi}^{c\}} \phi_{f\}}) ] - \epsilon^{de} \mathcal{O}^{\{a}_{\{b} [ \text{Tr}(\phi_{f\}} \bar{\psi}_d \phi_e \bar{\phi}^{c\}}) + \text{Tr}(\phi_{|e} \bar{\psi}_{d|} \phi_{f\}} \bar{\phi}^{c\}}) ]} \\
& \hspace{1.8cm} \textcolor{teal}{+ \epsilon^{de} \mathcal{O}^{\{a}_d [ 2 \text{Tr}(\phi_{\{b} \bar{\psi}_{|e|} \phi_{f\}} \bar{\phi}^{c\}}) + 5 \text{Tr}(\phi_e \bar{\psi}_{\{b} \phi_{f\}} \bar{\phi}^{c\}}) + 5 \text{Tr}(\phi_{\{b} \bar{\psi}_{f\}} \phi_e \bar{\phi}^{c\}}) ]} \\
& \hspace{1.8cm} \textcolor{teal}{- \epsilon_{de} \mathcal{O}^d_{\{b} [ 2 \text{Tr}(\bar{\phi}^{\{a} \psi^{|e|} \bar{\phi}^{c\}} \phi_{f\}}) + 5 \text{Tr}(\bar{\phi}^e \psi^{\{a} \bar{\phi}^{c\}} \phi_{f\}}) + 5 \text{Tr}(\bar{\phi}^{\{a} \psi^{c\}} \bar{\phi}^e \phi_{f\}}) ]}
\end{split}
\end{align}
Despite all of the disallowed dressings we have seen so far, every possibility at the third level has at least one $S^a_b$ or $S^a_b S^c_d$ action which is not in their span. The algorithm applied to \eqref{dressing-ingredients2} has thus terminated as well.

The next step in complexity is reached by including two single-graviton operators instead of one. Looking at
\begin{align}
\mathcal{O}^a_b, \quad \text{Tr}(\bar{\phi}^c \phi_d), \quad \text{Tr}(\bar{\phi}^e \phi_f) \label{dressing-ingredients3}
\end{align}
at the first level, there are five possible ways of symmetrizing and anti-symmetrizing the indices. Two of them are just single-graviton dressings of disallowed entries in \eqref{dressing1} and must therefore be disallowed as well. Checking the other three explicitly, the result is
\begin{align}
\begin{split}\label{dressing3}
& \textcolor{red}{\epsilon_{ac} \epsilon^{bd} \mathcal{O}^a_b \text{Tr}(\bar{\phi}^{\{c} \phi_d) \text{Tr}(\bar{\phi}^{e\}} \phi_f) + \alpha \epsilon_{ac} \epsilon^{bd} \mathcal{O}^e_f \text{Tr}(\bar{\phi}^a \phi_b) \text{Tr}(\bar{\phi}^c \phi_d)}, \\
& \hspace{3cm} \textcolor{teal}{\mathcal{O}^{\{a}_{\{b} \text{Tr}(\bar{\phi}^c \phi_d) \text{Tr}(\bar{\phi}^{e\}} \phi_{f\}})}.
\end{split}
\end{align}
Although the fully symmetrized operators in \eqref{dressing2} and \eqref{dressing3} both appear to be allowed on their own, we have checked that it is possible to get a ``red operator'' by taking a linear combination.
This is what ensures consistency with the fortuitous partition function which only has $x^5 y^4 \chi^+_3 \chi^-_3$ appearing once. The second level is 96-dimensional and the requirement of the $S^a_b$ actions being disallowed gets us down to a 73-dimensional space. Decomposing this into irreps, the two that are non-degenerate are disallowed because they come from multiplying a red entry in \eqref{dressing1} by a trace.
There is also a pair of four-fold degenerate irreps which are all disallowed. Three index contractions need to have this checked explicitly.
\begin{align}
\begin{split}
& \textcolor{red}{(\epsilon^{be} \epsilon^{df} + \epsilon^{bf} \epsilon^{de}) \mathcal{O}^{\{a}_b \text{Tr}(\bar{\phi}^{c\}} \phi_d) \text{Tr}(\bar{\psi}_e \phi_f) + \alpha \epsilon^{bd} \epsilon_{ef} \mathcal{O}^{\{a}_b \text{Tr}(\bar{\phi}^{|e} \phi_d) [\text{Tr}(\bar{\phi}^{f|} \psi^{c\}}) + \text{Tr}(\psi^{f|} \bar{\phi}^{c\}})]} \\
& \;\;\; \textcolor{red}{+ \beta \epsilon^{bd} \epsilon_{ef} \mathcal{O}^e_b [ \text{Tr}(\bar{\phi}^{\{a} \phi_d) \text{Tr}(\bar{\phi}^{c\}} \psi^f) + \text{Tr}(\bar{\phi}^{\{a} \phi_d) \text{Tr}(\psi^{c\}} \bar{\phi}^f) + \text{Tr}(\bar{\phi}^f \phi_d) \text{Tr}(\bar{\phi}^{\{a} \psi^{c\}}) ]}, \\
& \textcolor{red}{(\epsilon_{ae} \epsilon_{cf} + \epsilon_{af} \epsilon_{ce}) \mathcal{O}^a_{\{b} \text{Tr}(\bar{\phi}^c \phi_{d\}}) \text{Tr}(\bar{\phi}^e \psi^f) + \alpha \epsilon_{ac} \epsilon^{ef} \mathcal{O}^a_{\{b} \text{Tr}(\bar{\phi}^c \phi_{|e}) [\text{Tr}(\phi_{f|} \bar{\psi}_{d\}}) + \text{Tr}(\bar{\psi}_{f|} \phi_{d\}})]} \\
& \;\;\; \textcolor{red}{+ \beta \epsilon_{ac} \epsilon^{ef} \mathcal{O}^a_e [ \text{Tr}(\bar{\phi}^c \phi_{\{b}) \text{Tr}(\bar{\psi}_{d\}} \phi_f) + \text{Tr}(\bar{\phi}^c \phi_{\{b}) \text{Tr}(\phi_{d\}} \bar{\psi}_f) + \text{Tr}(\bar{\phi}^c \phi_f) \text{Tr}(\bar{\psi}_{\{d} \phi_{f\}}) ]}
\end{split}
\end{align}
The bi-singlet is three-fold degenerate and also fully disallowed. This time, two of the three come from adding one more graviton to \eqref{dressing1} and only
\begin{align}
\textcolor{red}{\epsilon_{ac} \epsilon^{bd} \mathcal{O}^a_b \text{Tr}(\bar{\phi}^c \phi_d) [\text{Tr}(\bar{\phi}^e \psi^f) \epsilon_{ef} + \text{Tr}(\bar{\psi}_e \phi_f) \epsilon^{ef}]}
\end{align}
needs to be checked. The only fortuitous representative at this level is one linear combination of the bi-triplet $4 \chi^+_2 \chi^-_2$.
\begin{align}
\begin{split}
& \;\;\; \textcolor{red}{\epsilon^{bf} \mathcal{O}^{\{a}_b \text{Tr}(\bar{\phi}^{c\}} \phi_{\{d}) \text{Tr}(\bar{\psi}_{e\}} \phi_f) + \epsilon_{bf} \mathcal{O}^b_{\{d} \text{Tr}(\bar{\phi}^{\{a} \phi_{e\}}) \text{Tr}(\psi^{c\}} \bar{\phi}^f)} \\
& \textcolor{red}{ + \alpha \epsilon^{bd} \mathcal{O}^{\{a}_b \text{Tr}(\bar{\phi}^{c\}} \phi_d) \text{Tr}(\bar{\psi}_{\{e} \phi_{f\}}) + \beta \epsilon_{bd} \mathcal{O}^b_{\{e} \text{Tr}(\bar{\phi}^d \phi_{f\}}) \text{Tr}(\psi^{\{a} \bar{\phi}^{c\}})}, \\
& \hspace{1.6cm} \textcolor{teal}{\mathcal{O}^{\{a}_{\{b} \text{Tr}(\bar{\phi}^{c\}} \phi_{d\}}) [\text{Tr}(\bar{\psi}_e \phi_f) \epsilon^{ef} - \text{Tr}(\bar{\phi}^e \psi^f) \epsilon_{ef}]}
\end{split}
\end{align}
We have not gone to the third level using the \eqref{dressing-ingredients3} seed.

We can try adding a single-graviton operator yet again and look at
\begin{align}
\mathcal{O}^a_b, \quad \text{Tr}(\bar{\phi}^c \phi_d), \quad \text{Tr}(\bar{\phi}^e \phi_f), \quad \text{Tr}(\bar{\phi}^g \phi_h).
\end{align}
This time, we will only check dressings at the first level and only those whose status cannot be inferred from \eqref{dressing1} and \eqref{dressing3}. The result is
\begin{align}
\begin{split}\label{dressing4}
& \textcolor{red}{\epsilon_{eg} \epsilon^{fh} [ \mathcal{O}^{\{a}_{\{b} \text{Tr}(\bar{\phi}^{c\}} \phi_{d\}}) \text{Tr}(\bar{\phi}^e \phi_f) \text{Tr}(\bar{\phi}^g \phi_h) + \alpha \mathcal{O}^e_f \text{Tr}(\bar{\phi}^g \phi_{\{b}) \text{Tr}(\bar{\phi}^{\{a} \phi_{|h|}) \text{Tr}(\bar{\phi}^{c\}} \phi_{d\}}) ]}, \\
& \hspace{4cm} \textcolor{teal}{\mathcal{O}^{\{a}_{\{b} \text{Tr}(\bar{\phi}^c \phi_d) \text{Tr}(\bar{\phi}^e \phi_f) \text{Tr}(\bar{\phi}^{g\}} \phi_{h\}})}.
\end{split}
\end{align}

Finally, a similar starting point to the above is
\begin{align}
\mathcal{O}^a_b, \quad \text{Tr}(\bar{\phi}^{\{c} \phi_{\{d} \bar{\phi}^{e\}} \phi_{f\}}), \quad \text{Tr}(\bar{\phi}^g \phi_h).
\end{align}
At the first level, our non-trivial checks have given
\begin{align}
& \textcolor{red}{\epsilon_{ce} \epsilon^{df} \mathcal{O}^{\{a}_{\{b} \text{Tr}(\bar{\phi}^{|c} \phi_{|d}) [ \text{Tr}(\bar{\phi}^{e|} \phi_{f|} \bar{\phi}^{g\}} \phi_{h\}}) + \text{Tr}(\bar{\phi}^{e} \phi_{|h\}} \bar{\phi}^{|g\}} \phi_f) ] + \alpha \epsilon_{ce} \epsilon^{df} \mathcal{O}^{\{e}_{\{f} \text{Tr}(\bar{\phi}^c \phi_d) \text{Tr}(\bar{\phi}^a \phi_b \bar{\phi}^{g\}} \phi_{h\}})} \nonumber \\
& \hspace{1.8cm} \textcolor{red}{ + \epsilon_{ce} \epsilon^{df} [\beta \mathcal{O}^{\{a}_f \text{Tr}(\bar{\phi}^{|c} \phi_{\{d}) \text{Tr}(\bar{\phi}^{e|} \phi_b \bar{\phi}^{g\}} \phi_{h\}}) + \gamma \mathcal{O}^e_{\{b} \text{Tr}(\bar{\phi}^{\{c} \phi_{|d}) \text{Tr}(\bar{\phi}^a \phi_{f|} \bar{\phi}^{g\}} \phi_{h\}})]} \nonumber \\
& \hspace{2.6cm} \textcolor{red}{\epsilon_{ce} \mathcal{O}^{\{a}_{\{b} \text{Tr}(\bar{\phi}^{|c} \phi_d) \text{Tr}(\bar{\phi}^{e|} \phi_f \bar{\phi}^{g\}} \phi_{h\}})}, \quad \textcolor{red}{\epsilon^{df} \mathcal{O}^{\{a}_{\{b} \text{Tr}(\bar{\phi}^c \phi_{|d}) \text{Tr}(\bar{\phi}^e \phi_{f|} \bar{\phi}^{g\}} \phi_{h\}})}, \nonumber \\
& \hspace{6cm} \textcolor{teal}{\mathcal{O}^{\{a}_{\{b} \text{Tr}(\bar{\phi}^c \phi_d \bar{\phi}^e \phi_f) \text{Tr}(\bar{\phi}^{g\}} \phi_{h\}})}. \label{dressing5}
\end{align}
As before, the last lines of \eqref{dressing4} and \eqref{dressing5} must be linearly dependent up to $Q$-exact and monotone operators because $x^6 y^4 \chi^+_4 \chi^-_4$ only appears in the fortuitous partition function once. No new check is needed for this because the previous linear relation with one less trace respects multiplication.

\subsection{${\rm U}(3)_k \times {\rm U}(3)_{-k}$ gauge group}
It is possible to apply much of the same machinery to $N = 3$ ABJM theory as well. The partition function which restricts the possible no-hair behaviour is \eqref{zfort-n3} and this time its leading term contains two fortuitous primaries. Reproducing the expressions from \cite{2512.23603},
\begin{align}
\mathcal{O} &= \epsilon^{ab} \epsilon_{cd} \epsilon^{ef} \left [ 12 \text{Tr}(\bar{\psi}_a \phi_b \bar{\phi}^c \phi_e \bar{\phi}^d \phi_f) - 8 \text{Tr}(\bar{\psi}_a \phi_b \bar{\phi}^c \phi_e) \text{Tr}(\bar{\phi}^d \phi_f) + 3 \text{Tr}(\bar{\psi}_a \phi_b) \text{Tr}(\bar{\phi}^c \phi_e) \text{Tr}(\bar{\phi}^d \phi_f) \right ] \nonumber \\
&- \epsilon_{ab} \epsilon^{cd} \epsilon_{ef} \left [ 12 \text{Tr}(\psi^a \bar{\phi}^b \phi_c \bar{\phi}^e \phi_d \bar{\phi}^f) - 8 \text{Tr}(\psi^a \bar{\phi}^b \phi_c \bar{\phi}^e) \text{Tr}(\phi_d \bar{\phi}^f) + 3 \text{Tr}(\psi^a \bar{\phi}^b) \text{Tr}(\phi_c \bar{\phi}^e) \text{Tr}(\phi_d \bar{\phi}^f) \right ] \nonumber \\
\mathcal{O}^{ab}_{cd} &= \epsilon^{ef} \Big [ 3 \text{Tr}(\bar{\psi}_e \phi_{\{c} \bar{\phi}^{\{a} \phi_{|f|} \bar{\phi}^{b\}} \phi_{d\}}) + 6 \text{Tr}(\bar{\psi}_e \phi_f \bar{\phi}^{\{a} \phi_{\{c} \bar{\phi}^{b\}} \phi_{d\}}) - 3 \text{Tr}(\bar{\psi}_e \phi_f) \text{Tr}(\bar{\phi}^{\{a} \phi_{\{c} \bar{\phi}^{b\}} \phi_{d\}}) \nonumber \\
&- 8 \text{Tr}(\bar{\psi}_e \phi_f \bar{\phi}^{\{a} \phi_{\{c}) \text{Tr}(\bar{\phi}^{b\}} \phi_{d\}}) + 3 \text{Tr}(\bar{\psi}_e \phi_f) \text{Tr}(\bar{\phi}^{\{a} \phi_{\{c}) \text{Tr}(\bar{\phi}^{b\}} \phi_{d\}}) \Big ] \nonumber \\
&- \epsilon_{ef} \Big [ 3 \text{Tr}(\psi^e \bar{\phi}^{\{a} \phi_{\{c} \bar{\phi}^{|f|} \phi_{d\}} \bar{\phi}^{b\}}) + 6 \text{Tr}(\psi^e \bar{\phi}^f \phi_{\{c} \bar{\phi}^{\{a} \phi_{d\}} \bar{\phi}^{b\}}) - 3 \text{Tr}(\psi^e \bar{\phi}^f) \text{Tr}(\phi_{\{c} \bar{\phi}^{\{a} \phi_{d\}} \bar{\phi}^{b\}}) \nonumber \\
&- 8 \text{Tr}(\psi^e \bar{\phi}^f \phi_{\{c} \bar{\phi}^{\{a}) \text{Tr}(\phi_{d\}} \bar{\phi}^{b\}}) + 3 \text{Tr}(\psi^e \bar{\phi}^f) \text{Tr}(\phi_{\{c} \bar{\phi}^{\{a}) \text{Tr}(\phi_{d\}} \bar{\phi}^{b\}}) \Big ] \label{abjm-fort2}
\end{align}
will be considered together.

Interestingly, \eqref{zfort-n3} includes enough terms for all possible dressings by an $n = 1$ graviton multiplet. Indeed, by testing linear combinations of the contractions in
\begin{align}
\begin{split}
& \;\; \mathcal{O}, \quad \text{Tr}(\bar{\phi}^a \phi_b) \\
& \mathcal{O}^{ab}_{cd}, \quad \text{Tr}(\bar{\phi}^e \phi_f),
\end{split}
\end{align}
we have found that nothing is in the $Q$-exact and monotone span. As such,
\begin{align}
\begin{split}
& \textcolor{teal}{\epsilon^{df} \epsilon_{be} \mathcal{O}^{\{ab\}}_{\{cd\}} \text{Tr}(\bar{\phi}^e \phi_f) + \alpha \mathcal{O} \text{Tr}(\bar{\phi}^a \phi_c)}, \quad \textcolor{teal}{\mathcal{O}^{\{ab}_{\{cd} \text{Tr}(\bar{\phi}^{e\}} \phi_{f\}})}, \\
& \hspace{1.2cm} \textcolor{teal}{\epsilon^{df} \mathcal{O}^{\{ab}_{\{cd\}} \text{Tr}(\bar{\phi}^{e\}} \phi_f)}, \quad \textcolor{teal}{\epsilon_{be} \mathcal{O}^{\{ab\}}_{\{cd} \text{Tr}(\bar{\phi}^e \phi_{f\}})}
\end{split}
\end{align}
are all fortuitous primaries and dressings by single-graviton operators that have one fermion and one boson can at most result in descendants.

For our last brute-force check, we will anticipate the previous observation that two $n = 1$ gravitons and one $n = 2$ graviton can sometimes produce the same fortuitous operator at $O(x^6)$ when dressing one at $O(x^4)$. As such, the lines of
\begin{align}
\begin{split}\label{dressing-ingredients4}
& \mathcal{O}, \quad \text{Tr}(\bar{\phi}^a \phi_b), \quad \text{Tr}(\bar{\phi}^c \phi_d) \\
& \mathcal{O}, \quad \text{Tr}(\bar{\phi}^{\{a} \phi_{\{b} \bar{\phi}^{c\}} \phi_{d\}}) \\
& \mathcal{O}^{ab}_{cd}, \quad \text{Tr}(\bar{\phi}^e \phi_f), \quad \text{Tr}(\bar{\phi}^g \phi_h) \\
& \mathcal{O}^{ab}_{cd}, \quad \text{Tr}(\bar{\phi}^{\{e} \phi_{\{f} \bar{\phi}^{g\}} \phi_{h\}})
\end{split}
\end{align}
should not be considered separately. When redundant operators of a given representation are discarded, all four lines should be used to express the starting candidates. Of the nine $\mathfrak{su}(2) \oplus \mathfrak{su}(2)$ representations that can be made from \eqref{dressing-ingredients4}, seven contribute to the fortuitous partition function. Among these, it is only $\chi^+_4 \chi^-_4$ which produces fully independent dressings. For all other irreps, there is at least one linear relation between the operators up to monotone and $Q$-exact terms. Finding these relations over the course of several days shows that
\begin{align}
& \hspace{3cm} \textcolor{teal}{\epsilon_{ae} \epsilon_{bg} \mathcal{O}^{ab}_{\{cd} \text{Tr}(\bar{\phi}^e \phi_f) \text{Tr}(\bar{\phi}^g \phi_{h\}})}, \quad \textcolor{teal}{\epsilon^{cf} \epsilon^{dh} \mathcal{O}^{\{ab}_{cd} \text{Tr}(\bar{\phi}^e \phi_f) \text{Tr}(\bar{\phi}^{g\}} \phi_h)}, \nonumber \\
& \hspace{4.2cm} \textcolor{teal}{\epsilon_{ag} \mathcal{O}^{a \{b}_{\{cd} \text{Tr}(\bar{\phi}^{e\}} \phi_f \bar{\phi}^g \phi_{h\}})}, \quad \textcolor{teal}{\epsilon^{ch} \mathcal{O}^{\{ab}_{c \{d} \text{Tr}(\bar{\phi}^e \phi_{f\}} \bar{\phi}^{g\}} \phi_h)}, \nonumber \\
& \hspace{4.4cm} \textcolor{teal}{\mathcal{O}^{\{ab}_{\{cd} [ \text{Tr}(\bar{\phi}^e \phi_f) \text{Tr}(\bar{\phi}^{g\}} \phi_{h\}}) + \alpha \text{Tr}(\bar{\phi}^e \phi_f \bar{\phi}^{g\}} \phi_{h\}}) ]}, \nonumber \\
& \textcolor{teal}{\mathcal{O} \text{Tr}(\bar{\phi}^{\{a} \phi_{\{b} \bar{\phi}^{c\}} \phi_{d\}}) + \alpha \epsilon_{eg} \epsilon^{fh} \mathcal{O}^{e \{a}_{f \{b} [ \text{Tr}(\bar{\phi}^{c\}} \phi_{d\}} \bar{\phi}^g \phi_h) + \text{Tr}(\bar{\phi}^{c\}} \phi_{|h|} \bar{\phi}^g \phi_{d\}}) ] + \beta \epsilon_{eg} \epsilon^{fh} \mathcal{O}^{ab}_{cd} \text{Tr}(\bar{\phi}^e \phi_f \bar{\phi}^g \phi_h)}, \nonumber \\
& \hspace{2.8cm} \textcolor{teal}{\epsilon_{ac} \epsilon^{bd} \mathcal{O} \text{Tr}(\bar{\phi}^a \phi_b) \text{Tr}(\bar{\phi}^c \phi_d) + \alpha \epsilon_{ae} \epsilon_{bg} (\epsilon^{cf} \epsilon^{dh} + \epsilon^{ch} \epsilon^{df}) \mathcal{O}^{ab}_{cd} \text{Tr}(\bar{\phi}^e \phi_f \bar{\phi}^g \phi_h)} \label{dressing-last}
\end{align}
is a valid basis of fortuitous representatives.\footnote{Certain irreps produced an especially large matrix whose rank had to be computed in several steps. Even though a basis for the row space was small enough to fit in one node's memory, the matrix before row reduction could only be stored on the disk.}

Now that we have recursively constructed fortuitous primary operators in ${\rm U}(2)_k \times {\rm U}(2)_{-k}$ and ${\rm U}(3)_k \times {\rm U}(3)_{-k}$ ABJM separately, there is more to be learned by matching operators in the two theories. In particular, we will be able to make a basic prediction about continuous $N$ trajectories which will be studied in the next section. The starting point is the observation that both operators in \eqref{abjm-fort2} are $Q$-closed for $N = 2$ and $N = 3$. For this reason, \cite{2512.23603} referred to them as representatives for fortuitous operators in both theories --- a statement which needs to be amended. For $x^4 y^3$, which appears in the $N = 2$ partition function once, we have already explained this appearance in terms of the first entry in \eqref{dressing1} rather than $\mathcal{O}$. It must therefore be checked whether the two can be used interchangeably. In other words, does the $N = 2$ limit of $\mathcal{O}$ make it equivalent to the bi-singlet obtained from $\mathcal{O}^a_b$ and $\text{Tr}(\bar{\phi}^c \phi_d)$? We have checked that the answer is \textit{yes} up to $Q$-exact and monotone operators. The situation is reversed for $\mathcal{O}^{ab}_{cd}$ --- $N = 2$ does \textit{not} send this to the same fortuitous cohomology class as the bi-triplet from $\mathcal{O}^a_b$ and $\text{Tr}(\bar{\phi}^c \phi_d)$. Instead, what happens is that $\mathcal{O}^{ab}_{cd}$ becomes equivalent to \textit{zero} modulo monotone and $Q$-exact operators while the bi-triplet dressing steps in to fill this void.

\subsection{Comparison to Hochschild cohomology}
Before studying the BPS operators of $O(k^{-1})$ ABJM theory in more detail, it is interesting to explore an alternative method for obtaining their cohomology classes. This comes from changing the nature of the objects which dress fortuitous operators and commute with $Q$. Instead of taking them to be traces, we can also take them to be \textit{vector fields}. The ones that preserve the number of traces have the index contraction pattern
\begin{align}
A \frac{\partial}{\partial B} \equiv A^{ij} \frac{\partial}{\partial B^{ij}}, \quad \bar{A} \frac{\partial}{\partial \bar{B}} \equiv \bar{A}_{ij} \frac{\partial}{\partial \bar{B}_{ij}}.
\end{align}
With this notation, one can easily write
\begin{align}
Q = \sum_{n = 0}^\infty \left ( [Q, D^n \phi_a] \frac{\partial}{\partial D^n \phi_a} + [Q, D^n \bar{\phi}^a] \frac{\partial}{\partial D^n \bar{\phi}^a} + \{ Q, D^n \psi^a \} \frac{\partial}{\partial D^n \psi^a} + \{ Q, D^n \bar{\psi}_a \} \frac{\partial}{\partial D^n \bar{\psi}_a} \right ) \label{q-vf}
\end{align}
which anti-commutes with itself. When another vector field has a vanishing graded commutator with $Q$, this defines what it means for it to be closed in the \textit{Hochschild cohomology}. Similarly, a vector field is exact when it is obtained through the action of \eqref{q-vf} on a witness. This leads to an immediate analogue of the analysis undertaken for the cyclic cohomology. Hochschild cohomology classes can also be used to move from one fortuitous operator to the next or to establish a partial no-hair theorem in the cases where this is not possible.

One difficulty of using the Hochschild cohomology is the presence of infinite sums. The single-graviton operators used previously began with a trace composed entirely of scalars fields. The fermionic fields and derivatives entered one-by-one as a result of descending through the multiplet. With $Q$-closed vector fields, on the other hand, it is unnatural for them to have any sort of bound on the number of covariant derivatives. This is exhibited by two of the simplest examples which are the triplets
\begin{align}
\begin{split}\label{r-sym-vf}
J_{ab} &= \sum_{n = 0}^\infty \epsilon_{c \{a} \left ( D^n \phi_{b\}} \frac{\partial}{\partial D^n \phi_c} + D^n \bar{\psi}_{b\}} \frac{\partial}{\partial D^n \bar{\psi}_c} \right ) \\
J^{ab} &= \sum_{n = 0}^\infty \epsilon^{c \{a} \left ( D^n \bar{\phi}^{b\}} \frac{\partial}{\partial D^n \bar{\phi}^c} + D^n \psi^{b\}} \frac{\partial}{\partial D^n \bar{\psi}^c} \right ).
\end{split}
\end{align}
The $n = 0$ term of this sum ensures that $[Q, J_{ab}]$ and $[Q, J^{ab}]$ have trivial actions on undifferentiated letters. The $n = 1$ term then makes the action trivial on single derivatives and so on. One can then show inductively that these commutators are indeed zero. More physically, the necessity of the full sum follows from the fact that \eqref{r-sym-vf} are nothing but the generators of the R-symmetry in $\mathfrak{osp}(4|2)$. This symmetry interpretation turns out to hold more generally. Elements of the centralizer algebra (at least when they (anti-)commute with $Q$ genuinely, not just up to gauge transformations) are the Hochschild cohomology classes which come from symmetries of the full ABJM theory. The others should be regarded as symmetries of the minimal theory with encodes the $Q$-cohomology. This more symmetric theory is the holomorphic-topological twist of \cite{2005.00083}.

To approach this problem, we have searched for the $n = 0$ terms of vector fields which raise the scaling dimension $E$ by a small number. For each solution, we have iteratively computed several $n > 0$ corrections without finding obstructions. On the contrary, solutions sometimes become more degenerate as $n$ is increased. At $E = 0$, there are two solutions other than \eqref{r-sym-vf}. One of them is the generator $E + J$ from the $\mathfrak{u}(1|1)$ part of the centralizer. The other is the $(C\bar{C}, \bar{C}C) = (1, 0)$ or $(C\bar{C}, \bar{C}C) = (0, 1)$ case of
\begin{align}
\Lambda = \sum_{n = 0}^\infty & \left [ D^n (C\bar{C} \phi_a - \phi_a \bar{C}C) \frac{\partial}{\partial D^n \phi_a} + D^n (C\bar{C} \psi^a - \psi^a \bar{C}C) \frac{\partial}{\partial D^n \psi^a} \right. \nonumber \\
&\left. + D^n (\bar{C}C \bar{\phi}^a - \bar{\phi}^a C\bar{C}) \frac{\partial}{\partial D^n \bar{\phi}^a} + D^n (\bar{C}C \bar{\psi}_a - \bar{\psi}_a C\bar{C}) \frac{\partial}{\partial D^n \bar{\psi}_a} \right ] \label{gauge-vf}
\end{align}
which describes a gauge transformation whenever $C\bar{C}$ and $\bar{C}C$ are closed. This implies that it has a trivial action on traces. At $E = \frac{1}{2}$, we do not find $Q^a_b$ since these symmetries no longer commute with $Q$ when they are lifted to actions on gauge-variant letters. Nevertheless, there appears to be an interesting solution in the singlet representation which looks like $Q$ except with different signs. This transformation, which we will call $\tilde{Q}$, acts as
\begin{align}
& [\tilde{Q}, \phi_a] = 0, \quad \quad \quad \quad \quad [\tilde{Q}, \bar{\phi}^a] = 0 \label{q-action} \\
& \{ \tilde{Q}, \psi^a \} = \epsilon^{bc} \phi_b \bar{\phi}^a \phi_c, \quad \{ \tilde{Q}, \bar{\psi}_a \} = -\epsilon_{bc} \bar{\phi}^b \phi_a \bar{\phi}^c \nonumber \\
& [\tilde{Q}, D \chi \} = (\epsilon_{ab} \psi^a \bar{\phi}^b + \epsilon^{ab} \phi_a \bar{\psi}_b) \chi + (-1)^{|\chi|} \chi (\epsilon^{ab} \bar{\psi}_a \phi_b + \epsilon_{ab} \bar{\phi}^a \psi^b) + D [\tilde{Q}, \chi \} \nonumber \\
& [\tilde{Q}, D \bar{\chi} \} = -(\epsilon^{ab} \bar{\psi}_a \phi_b + \epsilon_{ab} \bar{\phi}^a \psi^b) \bar{\chi} - (-1)^{|\chi|} \bar{\chi} (\epsilon_{ab} \psi^a \bar{\phi}^b + \epsilon^{ab} \phi_a \bar{\psi}_b) + D [\tilde{Q}, \bar{\chi} \}. \nonumber
\end{align}
As with all vector fields that have the
\begin{align}
f(\phi, \bar{\phi}) \frac{\partial}{\partial \psi^a} + \bar{f}(\phi, \bar{\phi}) \frac{\partial}{\partial \bar{\psi}_a} + \dots \label{automatic-vf}
\end{align}
form, $\tilde{Q}$ automatically anti-commutes with $Q$ at leading order. However, it is also automatic that such vector fields will not have an interesting action on the fortuitous cohomologies identified in \cite{2512.04146,2512.23603}. Indeed, all terms of \eqref{abjm-fort1} and \eqref{abjm-fort2} only have one fermion and no derivatives. The result of applying $\tilde{Q}$ must therefore be a monotone operator. It so happens that all other anti-commuting vector fields at $E = \frac{1}{2}$ also look like \eqref{automatic-vf} so these do not produce new fortuitous operators either.

We have gone up to $E = 1$ which starts to show new behaviour. Quotienting by the gauge transformations \eqref{gauge-vf} leaves us with several vector fields including
\begin{align}
\begin{split}\label{mixed-vf}
V^{abc}_d &= \epsilon_{ed} \bar{\phi}^{\{a} \psi^b \bar{\phi}^{c\}} \frac{\partial}{\partial \bar{\psi}_e} - \epsilon^{e\{a} \bar{\phi}^b \phi_d \bar{\phi}^{c\}} \frac{\partial}{\partial \bar{\phi}^e} - \epsilon^{e\{a} (\phi_d \bar{\phi}^b \psi^{c\}} + \psi^b \bar{\phi}^{c\}} \phi_d) \frac{\partial}{\partial \psi^e} + \dots \\
V_{abc}^d &= \epsilon^{ed} \phi_{\{a} \bar{\psi}_b \phi_{c\}} \frac{\partial}{\partial \psi^e} - \epsilon_{e\{a} \phi_b \bar{\phi}^d \phi_{c\}} \frac{\partial}{\partial \phi_e} - \epsilon_{e\{a} (\bar{\phi}^d \phi_b \bar{\psi}_{c\}} + \bar{\psi}_b \phi_{c\}} \bar{\phi}^d ) \frac{\partial}{\partial \bar{\psi}_e} + \dots
\end{split}
\end{align}
and several bi-doublets. Since the main goal is to act on the aforementioned operators without derivatives, the higher-order terms in \eqref{mixed-vf} are only important for establishing $[Q, V^{abc}_d] = [Q, V_{abc}^d] = 0$. We have checked that they are unique at $O(D)$ and doubly degenerate at $O(D^2)$. Looking at the $N = 2$ fortuitous operator \eqref{abjm-fort1}, the effect of \eqref{mixed-vf} is to add two scalars to every trace with a fermion, thereby producing a trace with three bilinears. Trace relations for $N = 2$ are then allowed to kick in and produce triple-trace terms. It is therefore conceivable that
\begin{align}
\epsilon_{ef} V^{abe}_{\{c} \mathcal{O}^f_{d\}} \sim \mathcal{O}^{\{a}_{\{c} \text{Tr}(\bar{\phi}^{b\}} \phi_{d\}}) \sim \epsilon^{ef} V_{cde}^{\{a} \mathcal{O}_f^{b\}}
\end{align}
could hold as an equivalence of fortuitous cohomology classes. The same membership test we have been using to match the partition function \eqref{zfort-n2} confirms that all of these operators are proportional up to $Q$-exact and monotone terms. The Hochschild cohomology can therefore provide an additional proof that the $x^4 y^3 \chi^+_2 \chi^-_2$ contribution is a non-core primary. It would be very interesting to find fortuitous operators which are core primaries from the cyclic cohomology point of view but still constructible from lighter primaries after dressing them with vector fields.

\section{Beyond Cohomology}
\label{sec:hamiltonian}

Up until now we have studied the cohomology classes of fortuitous BPS operators in ABJM theory. As already explained, in each cohomology class there exists one and only one BPS operator, namely one which creates a state annihilated by the Hamiltonian $H = \frac{1}{4} \{Q,Q^\dagger\}$. With this normalization, the Hamiltonian is equal to the dilation operator $E$ when restricted to the classically BPS Hilbert space. In this section, we want to go beyond cohomology and diagonalize this Hamiltonian in several charge sectors. For sectors containing the fortuitous classes previously found in \cite{2512.23603}, this leads to a determination of the actual BPS state. Additionally, it provides access to classically BPS states which acquire a two-loop anomalous dimension for all values of $N$. Just as fortuitous BPS operators are candidate descriptions of extremal black holes, those which are only classically BPS should be associated with \textit{near extremal} black holes \cite{2306.04673}. In both cases, the Hamiltonian allows us to go even further and follow these states in $N$ as was done for $\mathcal{N} = 4$ SYM in \cite{2306.04693}. This type of analytic continuation passes through non-unitary CFTs.

In the language of holographic coverings \cite{2402.10129}, one can consider $Q$ and $H$ to both be linear maps on the covering space $\tilde{\mathcal{H}}$. Given a subspace of trace relations $I_N$, the finite-$N$ Hilbert space and its BPS part are $\tilde{\mathcal{H}}/I_N$ and $\text{ker} \, H / I_N$ respectively. The physically interesting decomposition of the latter comes from a restriction of $\text{ker} \, Q$. Specifically, we should take
\begin{align}
\text{ker} \, H/I_N = \text{ker} \, Q \big |_{\text{ker} \, H / I_N} \oplus \Big (\text{ker} \, Q \big |_{\text{ker} \, H / I_N} \Big )^\perp \label{orthogonal-complement}
\end{align}
so that the first summand is spanned by finite-$N$ BPS states which are killed by the full $Q \in \text{End}(\tilde{\mathcal{H}})$. Since these are monotone by definition, the fortuitous states must span the orthogonal complement which is the second summand.

In practice, we will classify eigenstates by analytically continuing $N$ instead of taking finite-$N$ quotients.
In this approach, the Hamiltonian in a given charge sector becomes a finite-dimensional matrix acting on formal multi-traces whose entries are polynomial in $N$. If $\gamma_i(N)$ are its eigenvalues, the fortuitous operators are those for which $\gamma_i(N)$ has a root at some critical $N = N_* \in \mathbb{N}$. The corresponding eigenvector, which also depends on $N$, defines a one-parameter family of states interpolating between $N_*$ and the large-$N$ limit where it can be studied through integrability methods.

\subsection{Hamiltonian as a function of $N$}

To carry out the procedure above, we first need to describe how to compute the matrix of $H$ in a charge sector.
The only new ingredient that we need for this is the Gram matrix which allows us to build the matrix of $Q^\dagger$. Appendix \ref{app:gram} explains how it may be assembled from basic inner products which read
\begin{align}
\left < 0 \left | (D^m \phi_a)^\dagger D^n \phi_b \right | 0 \right > = \delta^{mn} \delta_{ab} \frac{2}{k} (2n)!, \quad
\left < 0 \left | (D^m \bar{\phi}^a)^\dagger D^n \bar{\phi}^b \right | 0 \right > = \delta^{mn} \delta^{ab} \frac{2}{k} (2n)!
\end{align}
for bosons and
\begin{align}
\left < 0 \left | (D^m \psi^a)^\dagger D^n \psi^b \right | 0 \right > = \delta^{mn} \delta^{ab} \frac{2}{k} (2n + 1)!, \quad
\left < 0 \left | (D^m \bar{\psi}_a)^\dagger D^n \bar{\psi}_b \right | 0 \right > = \delta^{mn} \delta_{ab} \frac{2}{k}(2n + 1)!
\end{align}
for fermions.

Suppose we have a vector space $V$ of multi-traces which has a basis of words $\{w_i\}$. The action of $Q$ on it is given by the matrix $\textbf{Q}$ defined by
\begin{equation}
    Q w_i = \sum_j \textbf{Q}_{ji}w_j.
\end{equation}
Taking the inner product with $w_k$, we see that
\begin{equation}
    \langle w_k,Q w_i\rangle = \sum_j \textbf{Q}_{ji} \textbf{G}_{kj}
\end{equation}
where $\textbf{G}_{ij} \equiv \langle w_i,w_j\rangle$ are components of the Gram matrix on $V$. On the other hand, we can also define a matrix for $Q^\dagger$, which we shall denote by $\tilde{\textbf{Q}}$:
\begin{equation}
    Q^\dagger w_i = \sum_j \tilde{\textbf{Q}}_{ji}w_j.
\end{equation}
Now, when we take the inner product with $w_k$, we can use the fact that
\begin{equation}
    \langle w_k, Q^\dagger w_i\rangle  = \langle Q w_k,w_i\rangle = \langle w_i, Qw_k\rangle^*
\end{equation}
to say that
\begin{equation}
    \sum_j \textbf{Q}_{jk}^*\textbf{G}_{ij}^* = \sum_j \tilde{\textbf{Q}}_{ji} \textbf{G}_{kj}.
\end{equation}
In terms of matrix multiplication this is
\begin{equation}
    (\textbf{G}^* \textbf{Q}^*)_{ik}= (\textbf{G} \tilde{\textbf{Q}})_{ki} \Longrightarrow \textbf{G} \tilde{\textbf{Q}} = (\textbf{G}^* \textbf{Q}^*)^{\intercal}.
\end{equation}
Using the fact that $\textbf{G}$ is Hermitian yields the matrix of $Q^\dagger$ as
\begin{equation}
    \tilde{\textbf{Q}} = \textbf{G}^{-1} \textbf{Q}^\dagger \textbf{G}.
\end{equation}
The coordinate matrix of $H$ is now given by
\begin{equation}
        \textbf{H} = \frac{1}{4} \{\textbf{Q},\tilde{\textbf{Q}}\} = \frac{1}{4} \textbf{Q} \textbf{G}^{-1} \textbf{Q}^\dagger \textbf{G} + \frac{1}{4}  \textbf{G}^{-1} \textbf{Q}^\dagger \textbf{G} \textbf{Q}. \label{h-matrix}
\end{equation}

Strictly speaking, the discussion above requires $V$ to be infinite-dimensional so that $Q$, $Q^\dagger$ and $H$ can all be taken as linear operators in $\text{End}(V)$. It is more useful in practice to restrict the Hamiltonian to a subspace of classically BPS words which have the Cartan charges $(J, H_1, H_2, H_3)$. This makes it necessary to work with
\begin{align}
V_- \equiv V_{J+1/2, H_1-1, H_2, H_3}, \quad V_0 \equiv V_{J, H_1, H_2, H_3}, \quad V_+ \equiv V_{J-1/2, H_1+1, H_2, H_3} \label{weight-spaces}
\end{align}
since $Q$ and $Q^\dagger$ intertwine the weight spaces. Looking at \eqref{h-matrix}, which should now be called $\textbf{H}_0$, it is clear from context that $\textbf{Q}$ is $\text{dim}(V_0) \times \text{dim}(V_-)$ in the first term and $\text{dim}(V_+) \times \text{dim}(V_0)$ in the second. Similarly $\textbf{Q}^\dagger$ is $\text{dim}(V_-) \times \text{dim}(V_0)$ in the first term and $\text{dim}(V_0) \times \text{dim}(V_+)$ in the second. It is also clear that inverting the Gram matrix (which turns out to be a bottleneck) only needs to be done in two of the three spaces from \eqref{weight-spaces}.
For a more symmetric looking expression,
we can consider $H = \frac{1}{4} \{ Q, Q^\dagger \}$ and insert the projectors
\begin{align}
\left | w^-_a \right > (\textbf{G}_-^{-1})_{ab} \left < w^-_b \right |, \quad
\left | w_i \right > (\textbf{G}_0^{-1})_{ij} \left < w_j \right |, \quad
\left | w^+_\alpha \right > (\textbf{G}_+^{-1})_{\alpha\beta} \left < w^+_\beta \right |
\end{align}
which are defined in terms of bases and Gram matrices for $V_-$, $V_0$ and $V_+$. This yields
\begin{equation}\label{eq:middle-pencil-matrix}
    \textbf{M}_{ij}\equiv\langle w_i,Hw_j\rangle
    =\frac{1}{4}\left(\textbf{A}^\dagger \textbf{G}_+^{-1}\textbf{A}+\textbf{B} \textbf{G}_-^{-1} \textbf{B}^\dagger\right)_{ij}
\end{equation}
where the overlap matrices are
\begin{equation}
    \textbf{A}_{\alpha i}=\langle w^+_\alpha,Qw_i\rangle,
    \quad
    \textbf{B}_{ia}=\langle w_i,Qw^-_a\rangle.
\end{equation}
Since $\textbf{A}$ and $\textbf{B}$ are both restrictions of $\textbf{G} \textbf{Q}$, it is easily seen that $\textbf{M}{M} = \textbf{G}_0 \textbf{H}_0$. This formulation leads one to solve the so-called generalized pencil
\begin{equation}\label{eq:middle-pencil}
    \textbf{M} v=\gamma \textbf{G}_0 v
\end{equation}
which is equivalent to the eigenvalue problem for $\textbf{H}_0$ when $\textbf{G}_0$ is invertible. In the calculations that follow, $\textbf{G}_0$ is always invertible and the choice to work with $\textbf{M}$ instead of $\textbf{H}_0$ is simply a matter of taste.\footnote{The way to make $\textbf{G}_0$ singular is to fix an integer value of $N$ which makes the set of multi-traces overcomplete. The space of trace relations, which was referred to as $I_N$ in \eqref{orthogonal-complement}, becomes $\text{ker} \, \textbf{G}_0$ in this case. If one does not mod out by these trace relations right away, it is possible to identify fortuity by looking for jumps in $\dim(\ker\textbf{M})-\dim(\ker \textbf{G}_0)$.}

\subsection{Lowest fortuitous sector for $N = 2$}
\label{sec:lowest}

To start, we should look for the lowest fortuitous operators which have $E = \frac{5}{2}$. They are contained in the $\mathfrak{su}(2) \oplus \mathfrak{su}(2)$ bi-doublet $\mathcal{O}^a_b$. A representative for its cohomology class was written in \eqref{abjm-fort1} and the corresponding eigenvector of the Hamiltonian will be able to complete this to a genuinely BPS state. All four components of $\mathcal{O}^a_b$ come with different Cartans $(J, H_1, H_2, H_3)$. As such, $(\frac{1}{2}, 2, \pm 1, 0)$ and $(\frac{1}{2}, 2, 0, \pm 1)$ should all be equally valid choices for observing the leading fortuitous trajectory. Having tested all four, we will focus on $(\frac{1}{2}, 2, 1, 0)$ which contains the $\mathfrak{su}(2) \oplus \mathfrak{su}(2)$ highest weight. The method above should therefore be applied with
\begin{align}
V_- = V_{1,1,1,0}, \quad V_0 = V_{1/2,2,1,0}, \quad V_+ = V_{0,3,1,0}
\end{align}
which have the dimensions 3, 12 and 10 respectively. These are small enough that there is no reason not to be completely explicit about the multi-trace words. For the bases, we can take
\begin{align}
\{ \text{Tr}(\phi_1 D\bar\phi^1), \text{Tr}(D\phi_1 \bar\phi^1), \text{Tr}(\psi^1 \bar\psi_1) \}
\label{eq:qdagger-basis}
\end{align}
for $V_-$. These have been sorted by the number of traces.
\begin{align}
\begin{gathered}
\{ \text{Tr}(\phi_1 \bar\phi^1 \phi_2 \bar\psi_1), \text{Tr}(\phi_1 \bar\phi^1 \psi^1 \bar\phi^2), \text{Tr}(\phi_1 \bar\phi^1 \psi^2 \bar\phi^1), \text{Tr}(\phi_1 \bar\phi^2 \psi^1 \bar\phi^1), \text{Tr}(\phi_1 \bar\psi_1 \phi_2 \bar\phi^1), \text{Tr}(\phi_1 \bar\phi^1 \phi_1 \bar\psi_2), \\
\text{Tr}(\phi_1 \bar\phi^1)\text{Tr}(\phi_1 \bar\psi_2), \text{Tr}(\phi_1 \bar\phi^1)\text{Tr}(\phi_2 \bar\psi_1), \text{Tr}(\phi_1 \bar\phi^1)\text{Tr}(\psi^1 \bar\phi^2), \\
\text{Tr}(\phi_1 \bar\phi^1)\text{Tr}(\psi^2 \bar\phi^1), \text{Tr}(\phi_1 \bar\phi^2)\text{Tr}(\psi^1 \bar\phi^1), \text{Tr}(\phi_1 \bar\psi_1)\text{Tr}(\phi_2 \bar\phi^1) \}
\end{gathered}
\label{eq:middle-basis}
\end{align}
for $V_0$ and finally
\begin{align}
\begin{gathered}
\{ \text{Tr}(\phi_1 \bar\phi^1 \phi_1 \bar\phi^1 \phi_2 \bar\phi^2), \text{Tr}(\phi_1 \bar\phi^1 \phi_1 \bar\phi^2 \phi_2 \bar\phi^1), \text{Tr}(\phi_1 \bar\phi^1 \phi_2 \bar\phi^1 \phi_1 \bar\phi^2), \text{Tr}(\phi_1 \bar\phi^1)\text{Tr}(\phi_1 \bar\phi^2 \phi_2 \bar\phi^1), \\
\text{Tr}(\phi_1 \bar\phi^1)\text{Tr}(\phi_1 \bar\phi^1 \phi_2 \bar\phi^2), \text{Tr}(\phi_1 \bar\phi^1 \phi_1 \bar\phi^1)\text{Tr}(\phi_2 \bar\phi^2), \text{Tr}(\phi_1 \bar\phi^1 \phi_1 \bar\phi^2)\text{Tr}(\phi_2 \bar\phi^1), \\
\text{Tr}(\phi_1 \bar\phi^1 \phi_2 \bar\phi^1)\text{Tr}(\phi_1 \bar\phi^2), \text{Tr}(\phi_1 \bar\phi^1)\text{Tr}(\phi_1 \bar\phi^1)\text{Tr}(\phi_2 \bar\phi^2), \text{Tr}(\phi_1 \bar\phi^1)\text{Tr}(\phi_1 \bar\phi^2)\text{Tr}(\phi_2 \bar\phi^1) \}
\label{eq:q-basis}
\end{gathered}
\end{align}
for $V_+$.

\begin{figure}[h]
\centering
\includegraphics[scale=0.55]{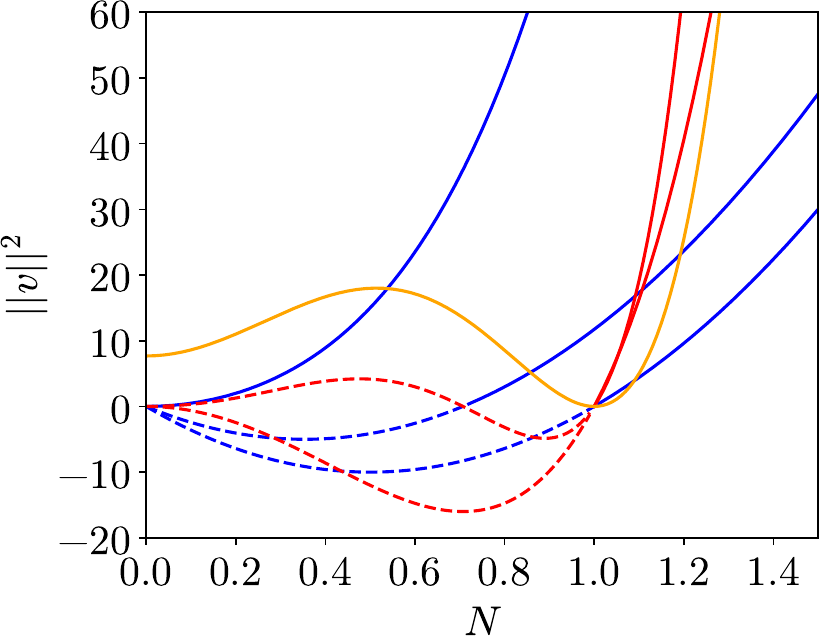} \quad \includegraphics[scale=0.55]{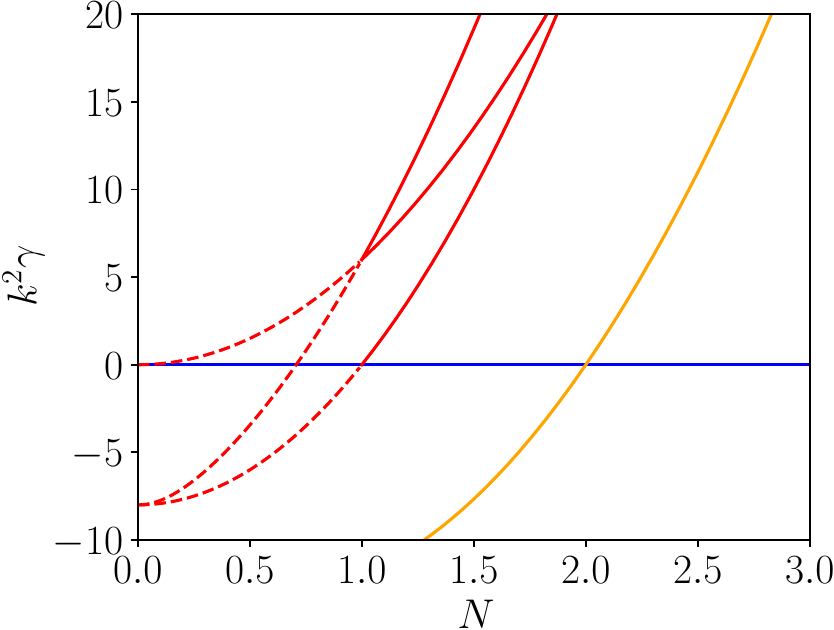}
\caption{Norms and energies of the operators following from $\textbf{G}_0$ in \eqref{eq:G-1420} and $\textbf{H}_0$ in \eqref{eq:H-1420}. Each eigenvalue of $\textbf{H}_0$, plotted in the right panel, is assigned a matrix $\textbf{V}$ whose columns form an eigenbasis. The left panel plots eigenvalues of $\textbf{V}^\dagger \textbf{G}_0 \textbf{V}$. A line is solid (dashed) when $N$ is above (below) the largest value that gives a negative norm. The fortuitous operator is in orange and the 8 monotone operators are in blue. Another 3 operators in red are neither fortuitous nor monotone for $N > 1$ (because they are only classically BPS).}
\label{fig:GH-1420}
\end{figure}

The Gram matrix for $V_{1/2,2,1,0}$ is given by
\begin{equation}
  \resizebox{\linewidth}{!}{%
  \(\displaystyle
\textbf{G}_0(N) = \frac{16}{k^4} \begin{pmatrix}
                   N^4 & 0 & 0 & 0 & N^2 & 0 & 0 & N^3 & 0 & 0 & 0 & N^3 \\
                   0 & N^4 & 0 & N^2 & 0 & 0 & 0 & 0 & N^3 & 0 & N^3 & 0 \\
                   0 & 0 & N^2 (N^2+1) & 0 & 0 & 0 & 0 & 0 & 0 & 2 N^3 & 0 & 0 \\
                   0 & N^2 & 0 & N^4 & 0 & 0 & 0 & 0 & N^3 & 0 & N^3 & 0 \\
                   N^2 & 0 & 0 & 0 & N^4 & 0 & 0 & N^3 & 0 & 0 & 0 & N^3 \\
                   0 & 0 & 0 & 0 & 0 & N^2 (N^2+1) & 2 N^3 & 0 & 0 & 0 & 0 & 0 \\
                   0 & 0 & 0 & 0 & 0 & 2 N^3 & N^2 (N^2+1) & 0 & 0 & 0 & 0 & 0 \\
                   N^3 & 0 & 0 & 0 & N^3 & 0 & 0 & N^4 & 0 & 0 & 0 & N^2 \\
                   0 & N^3 & 0 & N^3 & 0 & 0 & 0 & 0 & N^4 & 0 & N^2 & 0 \\
                   0 & 0 & 2 N^3 & 0 & 0 & 0 & 0 & 0 & 0 & N^2 (N^2+1) & 0 & 0 \\
                   0 & N^3 & 0 & N^3 & 0 & 0 & 0 & 0 & N^2 & 0 & N^4 & 0 \\
                   N^3 & 0 & 0 & 0 & N^3 & 0 & 0 & N^2 & 0 & 0 & 0 & N^4
\end{pmatrix}
  \)
  },
\label{eq:G-1420}
\end{equation}
It is singular only at $N = \pm 1$ which covers the extreme case of fortuity studied in \cite{2512.04146}. Conversely, the rank of $\textbf{G}_+$ drops from 10 to 7 at $N = 2$.
As such we identify three trace relations at $N = 2$ in $V_{0,3,1,0}$. This space is exactly where the image of $Q$ lives and these trace relations are the ones that allow ${\cal O}^1_1 \in V_{1/2,2,1,0}$ to be fortuitous.
The Hamiltonian in the same sector as \eqref{eq:G-1420} is
\begin{equation}
  \resizebox{\linewidth}{!}{%
  \(\displaystyle
  \textbf{H}_0(N)=\frac{1}{k^2}
  \begin{pmatrix}
  4N^2-3 & -1 & N^2-2 & 3-N^2 & 1-3N^2 & 2-N^2 & -N & N & -N & N & 0 & 0 \\
  -1 & 4N^2-3 & 2-N^2 & 1-3N^2 & 3-N^2 & N^2-2 & N & -N & N & -N & 0 & 0 \\
  N^2-2 & 2-N^2 & 2N^2-4 & 2-N^2 & N^2-2 & 4-2N^2 & -2N & 2N & -2N & 2N & 0 & 0 \\
  3-N^2 & 1-3N^2 & 2-N^2 & 4N^2-3 & -1 & N^2-2 & N & -N & N & -N & 0 & 0 \\
  1-3N^2 & 3-N^2 & N^2-2 & -1 & 4N^2-3 & 2-N^2 & -N & N & -N & N & 0 & 0 \\
  2-N^2 & N^2-2 & 4-2N^2 & N^2-2 & 2-N^2 & 2N^2-4 & 2N & -2N & 2N & -2N & 0 & 0 \\
  -N & N & -2N & N & -N & 2N & 2N^2-4 & 4-2N^2 & 2N^2-4 & 4-2N^2 & 0 & 0 \\
  2N & -2N & 4N & -2N & 2N & -4N & -2N^2 & 2N^2 & -2N^2 & 2N^2 & 0 & 0 \\
  -2N & 2N & -4N & 2N & -2N & 4N & 2N^2 & -2N^2 & 2N^2 & -2N^2 & 0 & 0 \\
  N & -N & 2N & -N & N & -2N & 4-2N^2 & 2N^2-4 & 4-2N^2 & 2N^2-4 & 0 & 0 \\
  N & -N & 2N & -N & N & -2N & -4 & 4 & -4 & 4 & 0 & 0 \\
  -N & N & -2N & N & -N & 2N & 4 & -4 & 4 & -4 & 0 & 0
  \end{pmatrix}
  \)
  }
\label{eq:H-1420}
\end{equation}
whose $O(N^2)$ term is block diagonal. This is the usual suppression of mixing between single-trace and double-trace operators at large $N$.
The spectrum of this Hamiltonian quickly reveals the fortuitous operator. We just have to look for an eigenvalue $\gamma(N)$ with the property that $\gamma(2) = 0$ and $\gamma(N) \neq 0$ for $N> 2$. Consider the characteristic polynomial
\begin{equation}
    \gamma^8 \left ( \gamma-8\frac{N^2-1}{k^2} \right ) \left ( \gamma-6\frac{N^2}{k^2} \right ) \left ( \gamma^2+2\frac{10-7N^2}{k^2}\gamma+4\frac{24-54N^2+12N^4}{k^4} \right ). \label{eq:poly-1420}
\end{equation}
The first factor shows 8 operators that are unconditionally BPS (monotone). The next two factors each show an operator which is unconditionally non-BPS. Finally, the last factor has the roots
\begin{equation}
    \gamma_\pm(N) = \dfrac{7N^2-10\pm \sqrt{N^4+76N^2+4}}{k^2}. \label{eq:anom-dim-1420}
\end{equation}
Out of these, $\gamma_-(N)$ has a first-order zero at $N = 2$. This is what makes it fortuitous. It is plotted with an orange line in Figure \ref{fig:GH-1420} alongside the other trajectories.
Even though the actual values of norms are not observable, interesting features become apparent by looking at when they are positive, negative and zero. In particular, they are all positive for $N > 1$. At $N = 1$, the norms of the non-BPS operators have a first-order zero while that of the fortuitous operator has a second-order zero. This is why its negative anomalous dimension at $N = 1$ is not in conflict with unitarity of ${\rm U}(1)_k \times {\rm U}(1)_{-k}$ ABJM.\footnote{Studies which analytically continue spin instead of $N$ have also revealed a conspiracy of zeros which saves unitarity \cite{2211.13754,2312.09283}.} Within the monotone subspace, the metric $\textbf{G}_0$ becomes Lorentzian at $N = 1$ and admits a second ``time'' direction at $N = \frac{1}{\sqrt{2}}$. Although there are two potential non-BPS level crossings, one of them can be discarded because it occurs at $N = 1$ --- this is where the operators decouple anyway. The other one is spurious because we have not distinguished between primaries and descendants at this stage. When we restrict $V_{1/2,2,1,0}$ to centralizer primaries, the trajectory with fortuity at $N = 1$ disappears and there is no level crossing anymore. We will see non-spurious examples in the next subsection.

Before moving on, an important issue to address is the eigenvector itself. An expression which unifies the results of the $(J, H_1, H_2, H_3) = (\frac{1}{2}, 2, \pm 1, 0)$ and $(J, H_1, H_2, H_3) = (\frac{1}{2}, 2, 0, \pm 1)$ calculations is
\begin{align}
\mathcal{O}^a_b(N) &= \frac{10}{7} \left [ \epsilon^{cd} \text{Tr}(\phi_c \bar{\phi}^a) \text{Tr}(\phi_b \bar{\psi}_d) - \epsilon_{cd} \text{Tr}(\phi_b \bar{\phi}^c) \text{Tr}(\psi^a \bar{\phi}^d) \right ] \label{eq:fort-1420} \\
&- \frac{6N^2 + 2 - k^2 \gamma_-(N)}{7} \epsilon_{cd} \text{Tr}(\bar{\phi}^a \phi_b) \text{Tr}(\bar{\phi}^c \psi^d) + \frac{6N^2 - 8 - k^2 \gamma_-(N)}{7} \epsilon^{cd} \text{Tr}(\phi_b \bar{\phi}^a) \text{Tr}(\phi_c \bar{\psi}_d) \nonumber \\
&- \frac{16N^2 - 8 - k^2 \gamma_-(N)}{14N} [\epsilon_{cd} \text{Tr}(\psi^c \bar{\phi}^d \phi_b \bar{\phi}^a) + \epsilon_{cd} \text{Tr}(\psi^c \bar{\phi}^a \phi_b \bar{\phi}^d) - \epsilon^{cd} \text{Tr}(\bar{\psi}_c \phi_d \bar{\phi}^a \phi_b) - \epsilon^{cd} \text{Tr}(\bar{\psi}_c \phi_b \bar{\phi}^a \phi_d)]. \nonumber
\end{align}
This was found by writing the most general ansatz for $\mathcal{O}^a_b(N)$ and demanding that each of its four index specializations be proportional to the fortuitous eigenvector in the corresponding charge sector. The large-$N$ limit of \eqref{eq:fort-1420} is
\begin{align}
\mathcal{O}^a_b(N) &\sim -\frac{5N}{7} [\epsilon_{cd} \text{Tr}(\psi^c \bar{\phi}^d \phi_b \bar{\phi}^a) + \epsilon_{cd} \text{Tr}(\psi^c \bar{\phi}^a \phi_c \bar{\phi}^d) - \epsilon^{cd} \text{Tr}(\bar{\psi}_c \phi_d \bar{\phi}^a \phi_b) - \epsilon^{cd} \text{Tr}(\bar{\psi}_c \phi_c \bar{\phi}^a \phi_d)]
\end{align}
which happens to be single-trace. The other interesting limit is of course $N \to 2$ which turns \eqref{eq:fort-1420} into
\begin{align}
\mathcal{O}^a_b(2) &= \epsilon^{cd} \left [ 4 \text{Tr}(\bar{\psi}_c \phi_d \bar{\phi}^a \phi_b) - 3 \text{Tr}(\bar{\psi}_c \phi_d) \text{Tr}(\bar{\phi}^a \phi_b) \right ] - \epsilon_{cd} \left [ 4 \text{Tr}(\psi^c \bar{\phi}^d \phi_b \bar{\phi}^a) - 3 \text{Tr}(\psi^c \bar{\phi}^d) \text{Tr}(\phi_b \bar{\phi}^a) \right ] \nonumber \\
&- \frac{5}{7} \left [ \epsilon^{cd} \text{Tr}(\phi_c \bar{\psi}_d) - \epsilon_{cd} \text{Tr}(\psi^c \bar{\phi}^d) \right ] \text{Tr}(\phi_b \bar{\phi}^a) + \frac{10}{7} \left [ \epsilon^{cd} \text{Tr}(\phi_c \bar{\phi}^a) \text{Tr}(\phi_b \bar{\psi}_d) - \epsilon_{cd} \text{Tr}(\phi_b \bar{\phi}^c) \text{Tr}(\psi^a \bar{\phi}^d) \right ] \nonumber \\
&+ 2 Q \text{Tr}(\psi^a \bar{\psi}_b).
\end{align}
We have written this in a form which emphasizes our cohomological results. The first line is the representative \eqref{abjm-fort1}. The $Q$-exact correction in the last line, even though it makes the operator BPS, does not make it orthogonal to all of the monotone operators. To accomplish the latter, we need the monotone correction in the middle line which was automatically generated by the analytic continuation down to $N = 2$. Reassuringly, it is manifestly built from single-graviton operators in \eqref{abjm-graviton-n1}.

\subsection{Lowest fortuitous sectors for $N = 3$}
\label{sec:second-lowest}

While $E = \frac{5}{2}$ provided a useful toy model, there should be more richness at higher energies. At $E = \frac{7}{2}$, the $Q$-cohomology has shown us that there are multiple fortuitous operators to consider. First, there is the bi-singlet $\mathcal{O} \in V_{1/2,3,0,0}$ which is a completion of the representative in \eqref{abjm-fort2}. We have previously argued that its $N = 2$ and $N = 3$ versions are both fortuitous. It is therefore conceivable that there is a trajectory which smoothly connects them. The same analysis suggests that bi-triplet fortuitous operators are rather different. Even though the $N = 2$ partition function \eqref{abjm-fort1} and the $N = 3$ partition function \eqref{abjm-fort2} both have this representation appearing once, it should lie on a different trajectory in each case. We will begin with the bi-triplet case where $\mathcal{O}^{ab}_{cd}$ lives in $V_{1/2,3,2,0} \oplus V_{1/2,3,-2,0} \oplus V_{1/2,3,0,2} \oplus V_{1/2,3,0,-2} \oplus V_{1/2,3,1,1} \oplus V_{1.2,3,1,-1} \oplus V_{1.2,3,-1,1} \oplus V_{1/2,3,-1,-1} \oplus V_{1/2,3,0,0}$. Although we will focus on the highest weight as before, there is another filter we can apply to make the calculations faster. Instead of taking $V_0$ to be all of $V_{1/2,3,2,0}$, we set it to the kernel of $S^a_b$ on this space. This amounts to choosing a basis of centralizer primaries --- discarding descendants makes it easier to zoom in on the non-trivial information. In this case, it results in a Hamiltonian which is $10 \times 10$ instead of $28 \times 28$.

\begin{figure}[h]
\centering
\includegraphics[scale=0.55]{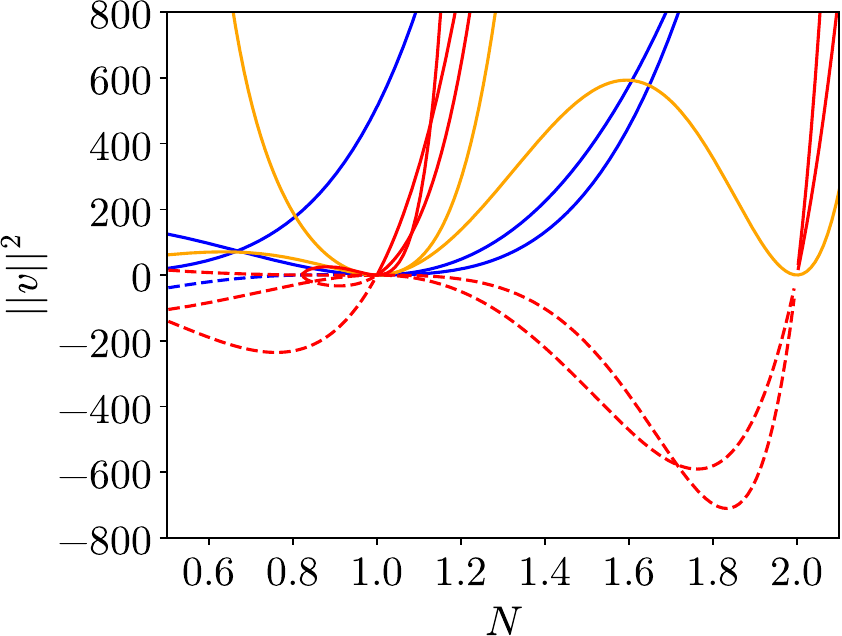} \quad \includegraphics[scale=0.55]{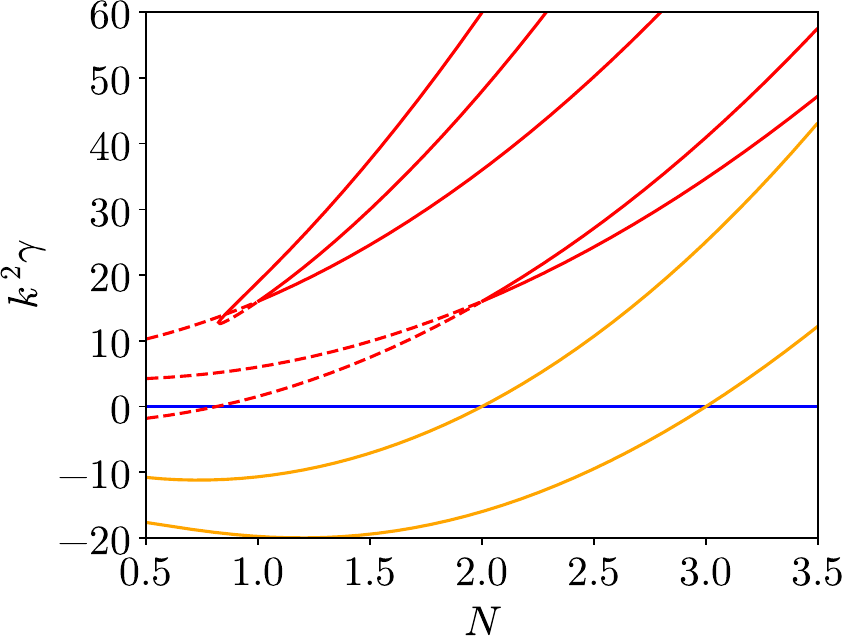}
\caption{Norms and energies of monotone (blue), fortuitous (orange) and non-BPS (red) centralizer primaries in the $(J, H_1, H_2, H_3) = (\frac{1}{2}, 3, 2, 0)$ sector. The right panel plots eigenvalues of $\textbf{H}_0$ while the left panel plots $v^\dagger \textbf{G}_0 v$ where $v$ is an eigenvector. A line is solid (dashed) when $N$ is above (below) the largest value that gives it a negative norm.}
\label{fig:GH-1640}
\end{figure}

Jumping straight to the characteristic polynomial, it is easily found to be
\begin{align}
& \gamma^3 \left ( \gamma^2 - 2 \frac{5N^2 + 6}{k^2} \gamma + 8 \frac{3N^4 + 5N^2 + 4}{k^4} \right ) \left [ \gamma^5 - 2 \frac{17N^2 - 14}{k^2} \gamma^4 \right. \label{eq:poly-1640} \\
&+ 8 \frac{57N^4 - 153N^2 + 28}{k^4} \gamma^3 - 64 \frac{47N^6 - 259N^4 + 178N^2 - 8}{k^6} \gamma^2 \nonumber \\
&\left. + 512 \frac{N^2 (19N^6 - 175N^4 + 316N^2 - 40)}{k^8} \gamma -4096 \frac{N^4 (3N^2-2)(N^2-4)(N^2-9)}{k^{10}} \right ]. \nonumber
\end{align}
The quintic factor indeed becomes proportional to $\gamma$ at $N = 2$ and $N = 3$. For generic values of $N$, we cannot write closed-form expressions for any of its roots but we can still plot them as in Figure \ref{fig:GH-1640}.
Right away, the presence of two orange lines is confirmed. The one associated with $\mathcal{O}^{ab}_{cd}$ in \eqref{abjm-fort2} has $\gamma = 0$ with $||v||^2 > 0$ at $N = 3$ and $\gamma < 0$ with $||v||^2 = 0$ at smaller integers.\footnote{A vanishing norm is equivalent to $Q$-exactness so this reproduces what we previously showed --- the \eqref{abjm-fort2} representative becomes $Q$-exact after its monotone ambiguity is fixed.} There is also $\mathcal{O}^{\{a}_{\{c} \text{Tr}(\bar{\phi}^{b\}} \phi_{d\}})$ which has $\gamma = 0$ with $||v||^2 > 0$ at $N = 2$ and $\gamma < 0$ with $||v||^2 = 0$ at smaller integers. Negative norms start to appear for non-BPS operators at $N = 2$. This is precisely where two of them would otherwise cross. The exact same thing happens for a second non-BPS pair at $N = 1$. Then at $N \approx 0.82$, we see a new phenomenon --- the last non-BPS operator which is still physical annihilates with an unphysical partner. This means that the previously real eigenvalues get an imaginary part and become complex conjugates.
Among monotone operators, the $3 \times 3$ projection of the Gram matrix becomes singular at $N = 1$ and it first obtains an indefinite signature at $N = \sqrt{\frac{2}{3}}$.

There is no formula we can write down for the fortuitous eigenvectors at general $N$. This is again because they are associated with the quintic factor in \eqref{eq:poly-1640}. The most interesting cases, however, are $N \to \infty$, $N = 3$ and $N = 2$. Fortunately, there is no need to diagonalize $\textbf{H}_0$ in all the charge sectors that appear in a bi-triplet. Instead, we can start with the highest weight and act with appropriate combinations of the $\mathfrak{su}(2) \oplus \mathfrak{su}(2)$ lowering operators \eqref{r-sym-vf} to generate all 9 states. The result can then be assembled into a covariant expression by writing an ansatz and fixing coefficients. The $N = 3$ BPS operator that this produces is
\begin{align}
\mathcal{O}^{ab}_{cd}(3) &= \epsilon^{ef} \Big [ 3 \text{Tr}(\bar{\psi}_e \phi_{\{c} \bar{\phi}^{\{a} \phi_{|f|} \bar{\phi}^{b\}} \phi_{d\}}) + 6 \text{Tr}(\bar{\psi}_e \phi_f \bar{\phi}^{\{a} \phi_{\{c} \bar{\phi}^{b\}} \phi_{d\}}) - 3 \text{Tr}(\bar{\psi}_e \phi_f) \text{Tr}(\bar{\phi}^{\{a} \phi_{\{c} \bar{\phi}^{b\}} \phi_{d\}}) \nonumber \\
&- 8 \text{Tr}(\bar{\psi}_e \phi_f \bar{\phi}^{\{a} \phi_{\{c}) \text{Tr}(\bar{\phi}^{b\}} \phi_{d\}}) + 3 \text{Tr}(\bar{\psi}_e \phi_f) \text{Tr}(\bar{\phi}^{\{a} \phi_{\{c}) \text{Tr}(\bar{\phi}^{b\}} \phi_{d\}}) \Big ] \nonumber \\
&- \epsilon_{ef} \Big [ 3 \text{Tr}(\psi^e \bar{\phi}^{\{a} \phi_{\{c} \bar{\phi}^{|f|} \phi_{d\}} \bar{\phi}^{b\}}) + 6 \text{Tr}(\psi^e \bar{\phi}^f \phi_{\{c} \bar{\phi}^{\{a} \phi_{d\}} \bar{\phi}^{b\}}) - 3 \text{Tr}(\psi^e \bar{\phi}^f) \text{Tr}(\phi_{\{c} \bar{\phi}^{\{a} \phi_{d\}} \bar{\phi}^{b\}}) \nonumber \\
&- 8 \text{Tr}(\psi^e \bar{\phi}^f \phi_{\{c} \bar{\phi}^{\{a}) \text{Tr}(\phi_{d\}} \bar{\phi}^{b\}}) + 3 \text{Tr}(\psi^e \bar{\phi}^f) \text{Tr}(\phi_{\{c} \bar{\phi}^{\{a}) \text{Tr}(\phi_{d\}} \bar{\phi}^{b\}}) \Big ] \nonumber \\
&+ \frac{\epsilon^{ef}}{225} \Big [ 198 \text{Tr}(\phi_e \bar{\phi}^{\{a}) \text{Tr}(\phi_{\{c} \bar{\phi}^{b\}} \phi_{d\}} \bar{\psi}_f) + 198 \text{Tr}(\phi_e \bar{\phi}^{\{a} \phi_{\{c} \bar{\phi}^{b\}}) \text{Tr}(\phi_{d\}} \bar{\psi}_f) \nonumber \\
&- 99 \text{Tr}(\phi_{\{c} \bar{\phi}^{\{a} \phi_{d\}} \bar{\phi}^{b\}}) \text{Tr}(\phi_e \bar{\psi}_f) - 132 \text{Tr}(\phi_{\{c} \bar{\phi}^{\{a}) \text{Tr}(\phi_{d\}} \bar{\phi}^{b\}} \phi_e \bar{\psi}_f) \nonumber \\
&- 188 \text{Tr}(\phi_e \bar{\phi}^{\{a}) \text{Tr}(\phi_{\{c} \bar{\phi}^{b\}}) \text{Tr}(\phi_{d\}} \bar{\psi}_f) + 94 \text{Tr}(\phi_{\{c} \bar{\phi}^{\{a}) \text{Tr}(\phi_{d\}} \bar{\phi}^{b\}}) \text{Tr}(\phi_e \bar{\psi}_f) \Big ] \nonumber \\
&- \frac{\epsilon_{ef}}{225} \Big [ 198 \text{Tr}(\bar{\phi}^e \phi_{\{c}) \text{Tr}(\bar{\phi}^{\{a} \phi_{d\}} \bar{\phi}^{b\}} \psi^f) + 198 \text{Tr}(\bar{\phi}^e \phi_{\{c} \bar{\phi}^{\{a} \phi_{d\}}) \text{Tr}(\bar{\phi}^{b\}} \psi^f) \nonumber \\
&- 99 \text{Tr}(\bar{\phi}^{\{a} \phi_{\{c} \bar{\phi}^{b\}} \phi_{d\}}) \text{Tr}(\bar{\phi}^e \psi^f) - 132 \text{Tr}(\bar{\phi}^{\{a} \phi_{\{c}) \text{Tr}(\bar{\phi}^{b\}} \phi_{d\}} \bar{\phi}^e \psi^f) \nonumber \\
&- 188 \text{Tr}(\bar{\phi}^e \phi_{\{c}) \text{Tr}(\bar{\phi}^{\{a} \phi_{d\}}) \text{Tr}(\bar{\phi}^{b\}} \psi^f) + 94 \text{Tr}(\bar{\phi}^{\{a} \phi_{\{c}) \text{Tr}(\bar{\phi}^{b\}} \phi_{d\}}) \text{Tr}(\bar{\phi}^e \psi^f) \Big ] \nonumber \\
&+ Q \Big [ \frac{3}{2} \text{Tr}(\bar{\phi}^{\{a} \phi_{\{c} D \bar{\phi}^{b\}} \phi_{d\}}) - \frac{3}{2} \text{Tr}(\bar{\phi}^{\{a} \phi_{\{c} \bar{\phi}^{b\}} D \phi_{d\}}) \nonumber \\
&- \frac{61}{25} \text{Tr}(\bar{\phi}^{\{a} \phi_{\{c}) \text{Tr}(D \bar{\phi}^{b\}} \phi_{d\}}) + \frac{61}{25} \text{Tr}(\bar{\phi}^{\{a} \phi_{\{c}) \text{Tr}(\bar{\phi}^{b\}} D \phi_{d\}}) \Big ]. \label{eq:long-fort1}
\end{align}
For $N = 2$, we will separate out the representative we know which is a symmetrization between \eqref{abjm-fort1} and the graviton operator $\text{Tr}(\bar{\phi}^a \phi_b)$. Once again the monotone correction in the BPS operator is a lot longer than the $Q$-exact one.
\begin{align}
\mathcal{O}^{ab}_{cd}(2) &= \epsilon^{ef} \text{Tr}(\bar{\phi}^{\{a} \phi_{\{c}) \left [ 4 \text{Tr}(\bar{\psi}_{|e} \phi_{f|} \bar{\phi}^{c\}} \phi_{d\}}) - 3 \text{Tr}(\bar{\psi}_{|e} \phi_{f|}) \text{Tr}(\bar{\phi}^{c\}} \phi_{d\}}) \right ] \nonumber \\
&- \epsilon_{ef} \text{Tr}(\phi_{\{c} \bar{\phi}^{\{a}) \left [ 4 \text{Tr}(\psi^{|e} \bar{\phi}^{|f} \phi_{d\}} \bar{\phi}^{b\}}) - 3 \text{Tr}(\psi^{|e} \bar{\phi}^{f|}) \text{Tr}(\phi_{d\}} \bar{\phi}^{b\}}) \right ] \nonumber \\
&+ \frac{\epsilon^{ef}}{20} \Big [ 24 \text{Tr}(\phi_e \bar{\phi}^{\{a}) \text{Tr}(\phi_{\{c} \bar{\phi}^{b\}} \phi_{d\}} \bar{\psi}_f) + 24 \text{Tr}(\phi_e \bar{\phi}^{\{a} \phi_{\{c} \bar{\phi}^{b\}}) \text{Tr}(\phi_{d\}} \bar{\psi}_f) \nonumber \\
&+ 12 \text{Tr}(\phi_{\{c} \bar{\phi}^{\{a} \phi_{d\}} \bar{\phi}^{b\}}) \text{Tr}(\phi_e \bar{\psi}_f) - 16 \text{Tr}(\phi_{\{c} \bar{\phi}^{\{a}) \text{Tr}(\phi_{d\}} \bar{\phi}^{b\}} \phi_e \bar{\psi}_f) \nonumber \\
&- 6 \text{Tr}(\phi_e \bar{\phi}^{\{a}) \text{Tr}(\phi_{\{c} \bar{\phi}^{b\}}) \text{Tr}(\phi_{d\}} \bar{\psi}_f) - 3 \text{Tr}(\phi_{\{c} \bar{\phi}^{\{a}) \text{Tr}(\phi_{d\}} \bar{\phi}^{b\}}) \text{Tr}(\phi_e \bar{\psi}_f) \Big ] \nonumber \\
&- \frac{\epsilon_{ef}}{20} \Big [ 24 \text{Tr}(\bar{\phi}^e \phi_{\{c}) \text{Tr}(\bar{\phi}^{\{a} \phi_{d\}} \bar{\phi}^{b\}} \psi^f) + 24 \text{Tr}(\bar{\phi}^e \phi_{\{c} \bar{\phi}^{\{a} \phi_{d\}}) \text{Tr}(\bar{\phi}^{b\}} \psi^f) \nonumber \\
&+ 12 \text{Tr}(\bar{\phi}^{\{a} \phi_{\{c} \bar{\phi}^{b\}} \phi_{d\}}) \text{Tr}(\bar{\phi}^e \psi^f) - 16 \text{Tr}(\bar{\phi}^{\{a} \phi_{\{c}) \text{Tr}(\bar{\phi}^{b\}} \phi_{d\}} \bar{\phi}^e \psi^f) \nonumber \\
&- 6 \text{Tr}(\bar{\phi}^e \phi_{\{c}) \text{Tr}(\bar{\phi}^{\{a} \phi_{d\}}) \text{Tr}(\bar{\phi}^{b\}} \psi^f) - 3 \text{Tr}(\bar{\phi}^{\{a} \phi_{\{c}) \text{Tr}(\bar{\phi}^{b\}} \phi_{d\}}) \text{Tr}(\bar{\phi}^e \psi^f) \Big ] \nonumber \\
&+ \frac{8}{5} Q \Big [ \text{Tr}(\bar{\phi}^{\{a} \phi_{\{c}) \text{Tr}(\bar{\phi}^{b\}} \phi_{d\}}) \Big ]. \label{eq:long-fort2}
\end{align}
When taking the large-$N$ limit, there are two things we could mean: the $N \to \infty$ eigenvector on the line passing through $(3, 0)$ or the $N \to \infty$ eigenvector on the line passing through $(2, 0)$. The former is
\begin{align}
\mathcal{O}^{ab}_{cd}(N) \sim \epsilon^{ef} \text{Tr}(\phi_{\{c} \bar{\phi}^{\{a} \phi_{|e|} \bar{\phi}^{b\}} \phi_{d\}} \bar{\psi}_f) - \epsilon_{ef} \text{Tr}(\bar{\phi}^{\{b} \phi_{\{c} \bar{\phi}^{|e|} \phi_{c\}} \bar{\phi}^{a\}} \psi^f)
\end{align}
while the latter is
\begin{align}
\mathcal{O}^{ab}_{cd}(N)' &\sim \epsilon^{ef} \text{Tr}(\phi_{\{c} \bar{\phi}^{\{a}) \left [ \text{Tr}(\phi_{|e|} \bar{\phi}^{b\}} \phi_{c\}} \bar{\psi}_f) + \text{Tr}(\phi_{c\}} \bar{\phi}^{b\}} \phi_e \bar{\psi}_f) \right ] \nonumber \\
&- \epsilon_{ef} \text{Tr}(\phi_{\{c} \bar{\phi}^{\{a}) \left [ \text{Tr}(\bar{\phi}^{|e|} \phi_{d\}} \bar{\phi}^{b\}} \psi^f) + \text{Tr}(\bar{\phi}^{b\}} \phi_{d\}} \bar{\phi}^e \psi^f) \right ].
\end{align}
We have used a prime to denote the one with higher energy. Both are remarkably simple but one of them is double-trace.

\begin{figure}[h]
\centering
\includegraphics[scale=0.53]{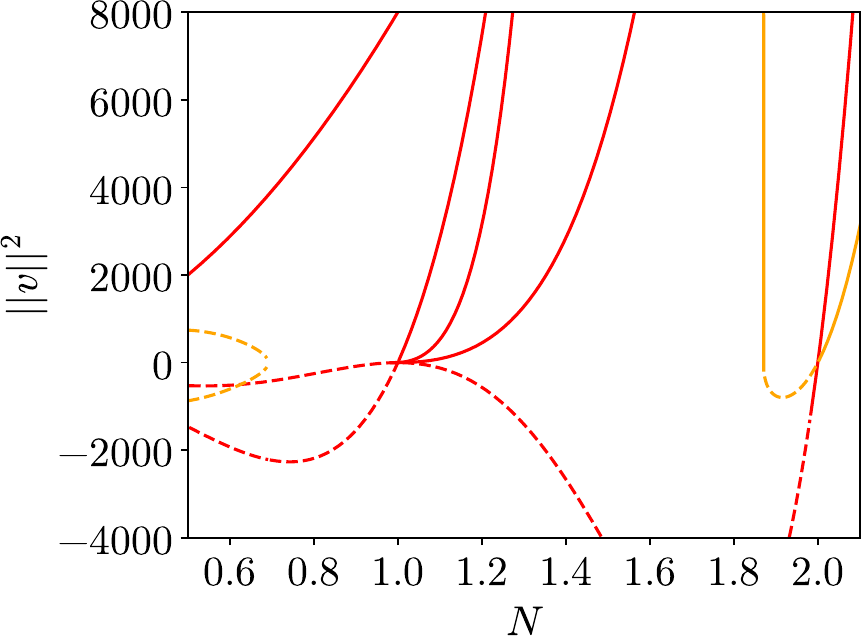} \quad \includegraphics[scale=0.53]{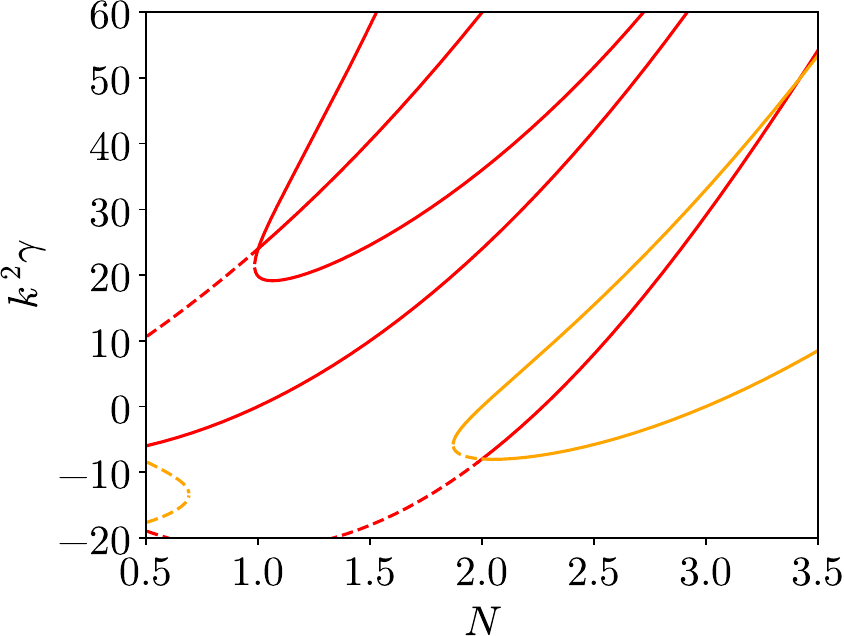}
\caption{Norms and energies of fortuitous (orange) and non-BPS (red) centralizer primaries in the $(J, H_1, H_2, H_3) = (\frac{1}{2}, 3, 0, 0)$ sector that are also R-symmetry highest weights. The right panel plots eigenvalues of $\textbf{H}_0$ while the left panel plots $v^\dagger \textbf{G}_0 v$ where $v$ is an eigenvector. A line is solid (dashed) when $N$ is above (below) the largest value that gives it a negative norm.}
\label{fig:GH-1600}
\end{figure}

Next, let us look for the bi-singlet by formulating a Gram matrix and Hamiltonian for centralizer primaries in $V_{1/2,3,0,0}$. This results in a $28 \times 28$ matrix. Although this is significantly larger than what we had before, some of the information in it is redundant. Bi-triplets also have a component in $V_{1/2,3,0,0}$ so there will be repeated trajectories from Figure \ref{fig:GH-1640} that need to be filtered out.
Doing so leads to the characteristic polynomial
\begin{align}
& \gamma^7 \left ( \gamma - 8 \frac{N^2 - 1}{k^2} \right ) \left [ \gamma^2 - 4 \frac{3N^2 - 1}{k^2} \gamma + 36 \frac{N^2 (N^2 - 2)}{k^4} \right ]^2 \label{eq:poly-1600} \\
& \left ( \gamma - 4 \frac{2N^2 + 1}{k^2} \right )^2 \left ( \gamma - 2 \frac{3N^2 - 8}{k^2} \right )^2 \left ( \gamma - 6 \frac{N^2 + 2}{k^2} \right )^2 \nonumber \\
& \left [ \gamma^3 - 2 \frac{17N^2 - 6}{k^2} \gamma^2 + 120 \frac{N^2 (3N^2 - 7)}{k^4} \gamma - 1152 \frac{N^2 (N^4 - 4N^2 - 6)}{k^6} \right ] \nonumber \\
& \left [ \gamma^3 - 2 \frac{9N^2 - 10}{k^2} \gamma^2 + 8 \frac{N^2 (13N^2 - 67)}{k^4} \gamma - 192 \frac{N^2 (N^2 - 4) (N^2 - 9)}{k^6} \right ] \nonumber
\end{align}
but we are not done yet. The five repeated roots other than $\gamma = 0$ strongly suggest that \eqref{eq:poly-1600} is still including non-singlet representations. An explicit check has confirmed this, prompting us to keep just the linear and two cubic factors and plot seven lines in Figure \ref{fig:GH-1600}. Notably, none of the monotone operators have survived this restriction.
The first surprise is that $\mathcal{O}^{ab}_{cd}$ is not the only $N = 3$ fortuitous operator whose trajectory becomes null at $N = 2$ --- the one for $\mathcal{O}$ does this as well. The difference is that one can still reach the $N = 2$ version of $\mathcal{O} \sim \epsilon_{ac} \epsilon^{bd} \mathcal{O}^a_b \text{Tr}(\bar{\phi}^c \phi_d)$ by moving onto a different sheet. The bifurcation happens at $N \approx 1.87$ where the anomalous dimensions for the two orange lines move into the complex plane. They rejoin the real line at $N \approx 0.69$. There is a similar annihilation of red lines at $N \approx 0.98$, situated just after a decoupling at $N = 1$ which prevents a level crossing. It should be noted that \cite{2306.04693} also observed singular behaviour at values of $N$ which are slightly displaced from integers. Looking at the right side of Figure \ref{fig:GH-1600}, there is a red line crossing an orange line near $N \approx 3.41$ where they belong to physical operators, Level crossing is a finely tuned phenomenon in quantum mechanics and non-perturbative studies including \cite{1512.05362,1709.03967,2207.10118,2312.12576,2601.04310,2607.00079} have shown from a number of perspectives that it is avoided in field theory.\footnote{Level repulsion is also part of the argument for why the CFT operators dual to near extremal black holes are classically BPS \cite{2306.04673}. We will have more to say about them at the end of this section.} We therefore expect this example to be resolved at four loops. This is what happens in $\mathcal{N} = 4$ SYM --- \cite{2306.04693} showed that one-loop level crossings are an artifact of perturbation theory, at least in the $\mathfrak{su}(2)$ sector where the Hamiltonian is known at two loops and finite $N$ \cite{hep-th/0303060,hep-th/0306054}.

The eigenvectors we are interested in can be expressed at $N \to \infty$, $N = 3$ and $N = 2$ as before. Since the important factors of \eqref{eq:poly-1600} are cubic, it should be possible to write a general expression for the $N$ dependence but we do not expect it to be enlightening. The $N = 3$ specialization is
\begin{align}
\mathcal{O}(3) &= \epsilon^{ab} \epsilon_{cd} \epsilon^{ef} \left [ 12 \text{Tr}(\bar{\psi}_a \phi_b \bar{\phi}^c \phi_e \bar{\phi}^d \phi_f) - 8 \text{Tr}(\bar{\psi}_a \phi_b \bar{\phi}^c \phi_e) \text{Tr}(\bar{\phi}^d \phi_f) + 3 \text{Tr}(\bar{\psi}_a \phi_b) \text{Tr}(\bar{\phi}^c \phi_e) \text{Tr}(\bar{\phi}^d \phi_f) \right ] \nonumber \\
&- \epsilon_{ab} \epsilon^{cd} \epsilon_{ef} \left [ 12 \text{Tr}(\psi^a \bar{\phi}^b \phi_c \bar{\phi}^e \phi_d \bar{\phi}^f) - 8 \text{Tr}(\psi^a \bar{\phi}^b \phi_c \bar{\phi}^e) \text{Tr}(\phi_d \bar{\phi}^f) + 3 \text{Tr}(\psi^a \bar{\phi}^b) \text{Tr}(\phi_c \bar{\phi}^e) \text{Tr}(\phi_d \bar{\phi}^f) \right ] \nonumber \\
&+ \epsilon_{ab} \epsilon^{cd} Q \Big [ 3 \text{Tr}(\phi_c \bar{\phi}^a \phi_d D \bar{\phi}^b) - 3 \text{Tr}(\phi_c \bar{\phi}^a D \phi_d \bar{\phi}^b) + 12 \text{Tr}(\phi_c \bar{\phi}^a \psi^b \bar{\psi}_d) \nonumber \\
&+ 2 \text{Tr}(\bar{\phi}^a \phi_c) \text{Tr}(D \bar{\phi}^b \phi_d) - 2 \text{Tr}(\bar{\phi}^a \phi_c) \text{Tr}(\bar{\phi}^b D \phi_d) \Big ] \label{fort2-full1}
\end{align}
which shows that \eqref{abjm-fort2} does not need any (inexact) monotone correction. Next, the $N = 2$ solution
\begin{align}
\mathcal{O}(2)' &= \epsilon^{ab} \epsilon_{cd} \epsilon^{ef} \left [ 12 \text{Tr}(\bar{\psi}_a \phi_b \bar{\phi}^c \phi_e \bar{\phi}^d \phi_f) - 8 \text{Tr}(\bar{\psi}_a \phi_b \bar{\phi}^c \phi_e) \text{Tr}(\bar{\phi}^d \phi_f) + 3 \text{Tr}(\bar{\psi}_a \phi_b) \text{Tr}(\bar{\phi}^c \phi_e) \text{Tr}(\bar{\phi}^d \phi_f) \right ] \nonumber \\
&- \epsilon_{ab} \epsilon^{cd} \epsilon_{ef} \left [ 12 \text{Tr}(\psi^a \bar{\phi}^b \phi_c \bar{\phi}^e \phi_d \bar{\phi}^f) - 8 \text{Tr}(\psi^a \bar{\phi}^b \phi_c \bar{\phi}^e) \text{Tr}(\phi_d \bar{\phi}^f) + 3 \text{Tr}(\psi^a \bar{\phi}^b) \text{Tr}(\phi_c \bar{\phi}^e) \text{Tr}(\phi_d \bar{\phi}^f) \right ] \nonumber \\
&+ \epsilon_{ab} \epsilon^{cd} Q \left [ \text{Tr}(\bar{\phi}^a \phi_c) \text{Tr}(D \bar{\phi}^b \phi_d) - \text{Tr}(\bar{\phi}^a \phi_c) \text{Tr}(\bar{\phi}^b D \phi_d) + 6 \text{Tr}(\psi^a \bar{\phi}^b) \text{Tr}(\phi_c \bar{\psi}_d) \right ] \label{fort2-full2}
\end{align}
shows the same representative and the same vanishing monotone correction but a different $Q$-exact term.
The $N \to \infty$ operator connected to \eqref{fort2-full1} is the single-trace
\begin{align}
\mathcal{O}(N) \sim \epsilon^{ab} \epsilon_{cd} \epsilon^{ef} \text{Tr}(\phi_a \bar{\phi}^c \phi_e \bar{\phi}^d \phi_b \bar{\psi}_f) + \epsilon_{ab} \epsilon^{cd} \epsilon_{ef} \text{Tr}(\bar{\phi}^d \phi_c \bar{\phi}^e \phi_d \bar{\phi}^c \psi^f)
\end{align}
while the $N \to \infty$ operator connected to \eqref{fort2-full2} is the double-trace
\begin{align}
\mathcal{O}(N)' \sim (\epsilon^{ab} \epsilon_{cd} \epsilon^{ef} + \epsilon^{ae} \epsilon_{cd} \epsilon^{bf}) \text{Tr}(\phi_a \bar{\phi}^c) \text{Tr}(\phi_b \bar{\phi}^d \phi_e \bar{\psi}_f) - (\epsilon_{ab} \epsilon^{cd} \epsilon_{ef} + \epsilon_{ae} \epsilon^{cd} \epsilon_{bf}) \text{Tr}(\bar{\phi}^a \phi_c) \text{Tr}(\bar{\phi}^b \phi_d \bar{\phi}^e \psi^f).
\end{align}

\subsection{More fortuitous operators}
\label{sec:more}

The last two subsections deal with sectors that have $E + J = 3$ and $E + J = 4$ respectively. We will now repeat much of this analysis for all of the charge sectors containing fortuitous operators with $E + J = 5$. This will cover the terms of $Z_2^{\text{fort}}$, $Z_3^{\text{fort}}$ and $Z_4^{\text{fort}}$ which appeared in the partition function calculation of \cite{2512.23603}. Beyond restricting to $\mathfrak{osp}(4|2)\oplus \mathfrak{u}(1|1)$ centralizer primaries (which are the same as $\mathfrak{osp}(4|2)$ primaries when they are BPS), we will further restrict to $\mathfrak{so}(4)$ highest weights, In doing so, we describe 18 fortuitous trajectories and identify the planar limit of their continuations in $N$. The charge sectors and associated multiplicities are shown in Table~\ref{tab:fortuitous-sectors}. Acting with lowering generators reconstructs all 252 of the $E + J \leq 5$ states in the partition function of \cite{2512.23603} and one can of course descend further.
\begin{table}[ht!]
    \centering
    \begin{tabular}{ccc}
        \hline
        \textbf{Charge sector} & \textbf{Centralizer primaries} & \textbf{Planar eigenvalue} \\
        \hline
        $\left(\frac{1}{2},2,1,0\right)$ & 1 & 6 \\
        $\left(\frac{1}{2},3,0,0\right)$ & 2 & 4, 6 \\
        $\left(\frac{1}{2},3,2,0\right)$ & 2 & 4, 6 \\
        $\left(\frac{1}{2},4,1,0\right)$ & 3 & $8 - 2\sqrt{5}$, 4, 4 \\
        $\left(\frac{1}{2},4,2,-1\right)$ & 1 & 4 \\
        $\left(\frac{1}{2},4,2,1\right)$ & 1 & 4 \\
        $\left(\frac{1}{2},4,3,0\right)$ & 3 & $5 - \sqrt{5}$, 4, 6 \\
        $\left(1,3,1,0\right)$ & 2 & 6, 6 \\
        $\left(1,3,2,-1\right)$ & 1 & $6-\tfrac{\sqrt{21}}{2}$ \\
        $\left(1,3,2,1\right)$ & 1 & $6-\tfrac{\sqrt{21}}{2}$ \\
        $\left(\frac{3}{2},2,1,0\right)$ & 1 & 4\\
        \hline
    \end{tabular}
    \caption{Charge sectors containing fortuitous primaries of the $\mathfrak{osp}(4|2)$ centralizer, together with their multiplicities and $\Delta_2(\infty)$ values.}
    \label{tab:fortuitous-sectors}
\end{table}

\begin{figure}[p]
    \centering
    \includegraphics[width=0.98\textwidth]{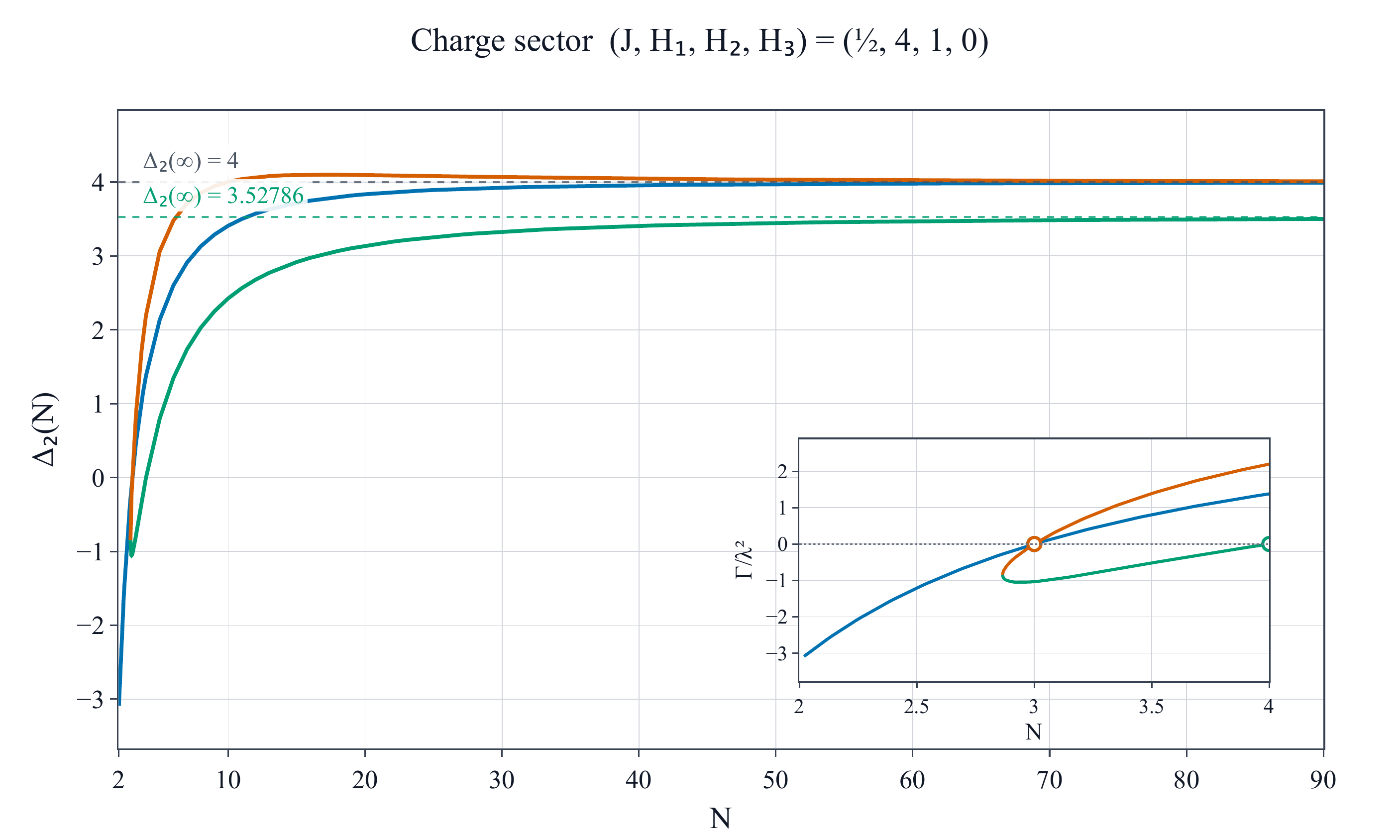}
    \includegraphics[width=0.98\textwidth]{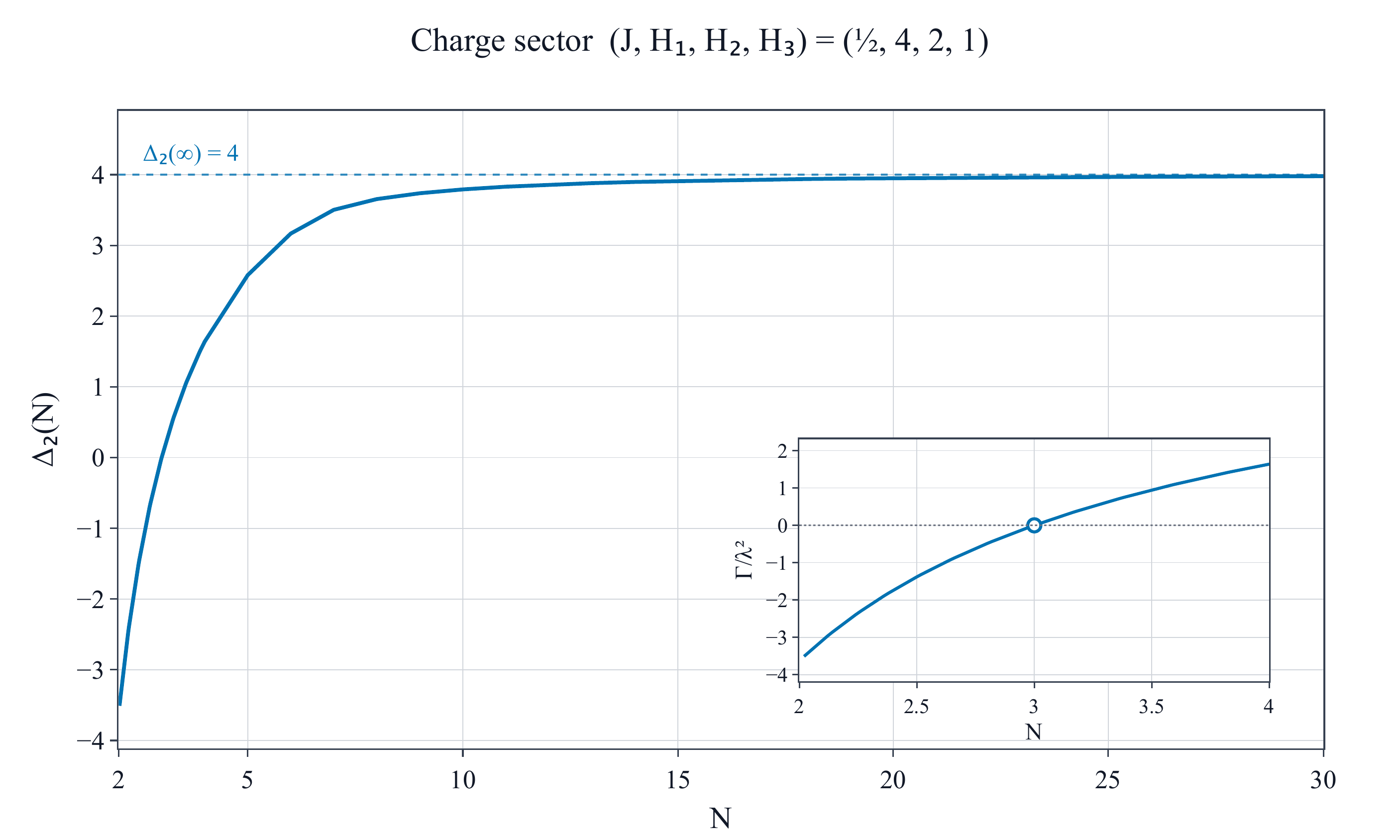}
    \caption{Centralizer-primary trajectories in $(J,H_1,H_2,H_3)=(\tfrac{1}{2},4,1,0)$ and $(\tfrac{1}{2},4,2,1)$.}
    \label{fig:numerical-trajectories-1}
\end{figure}

\begin{figure}[p]
    \centering
    \includegraphics[width=0.98\textwidth]{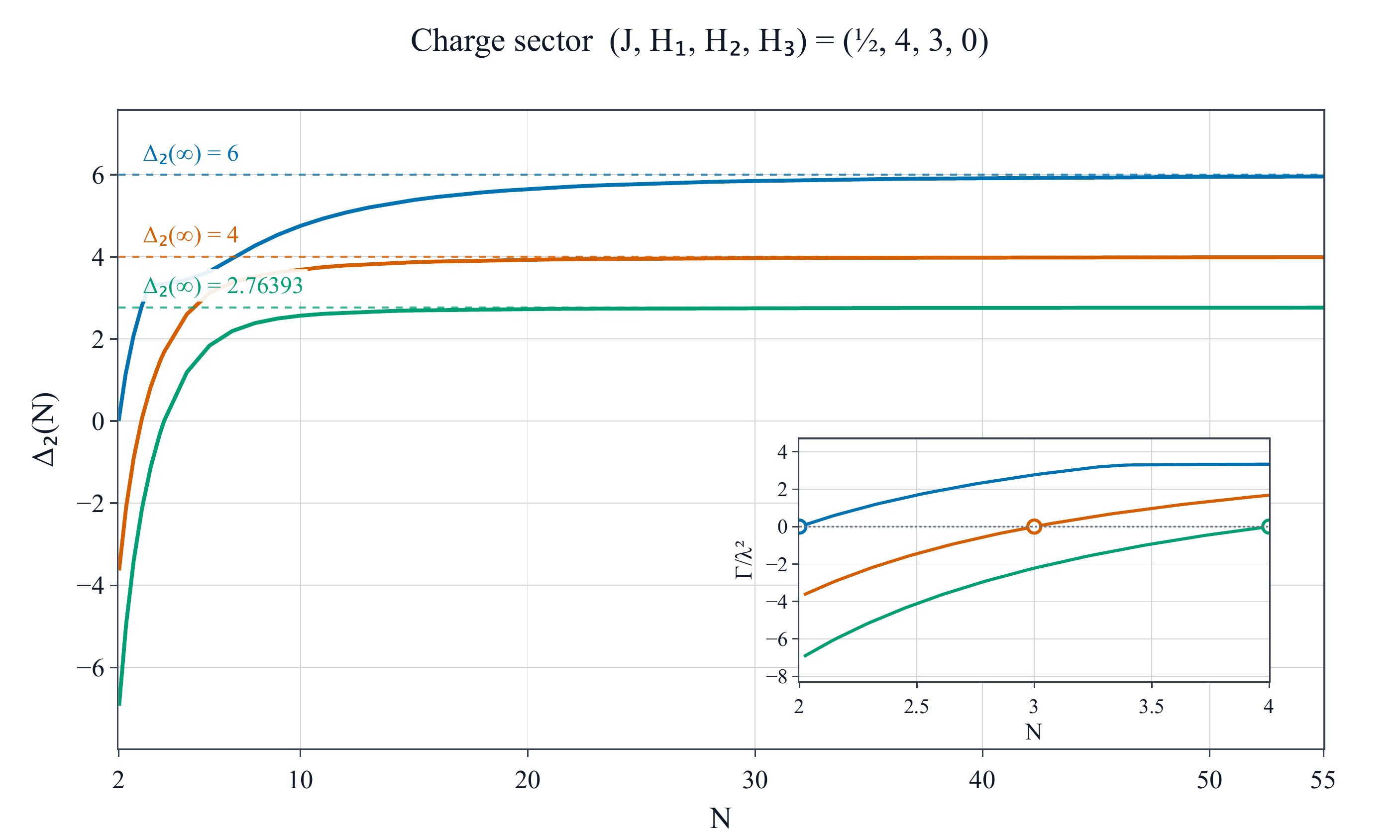}
    \includegraphics[width=0.98\textwidth]{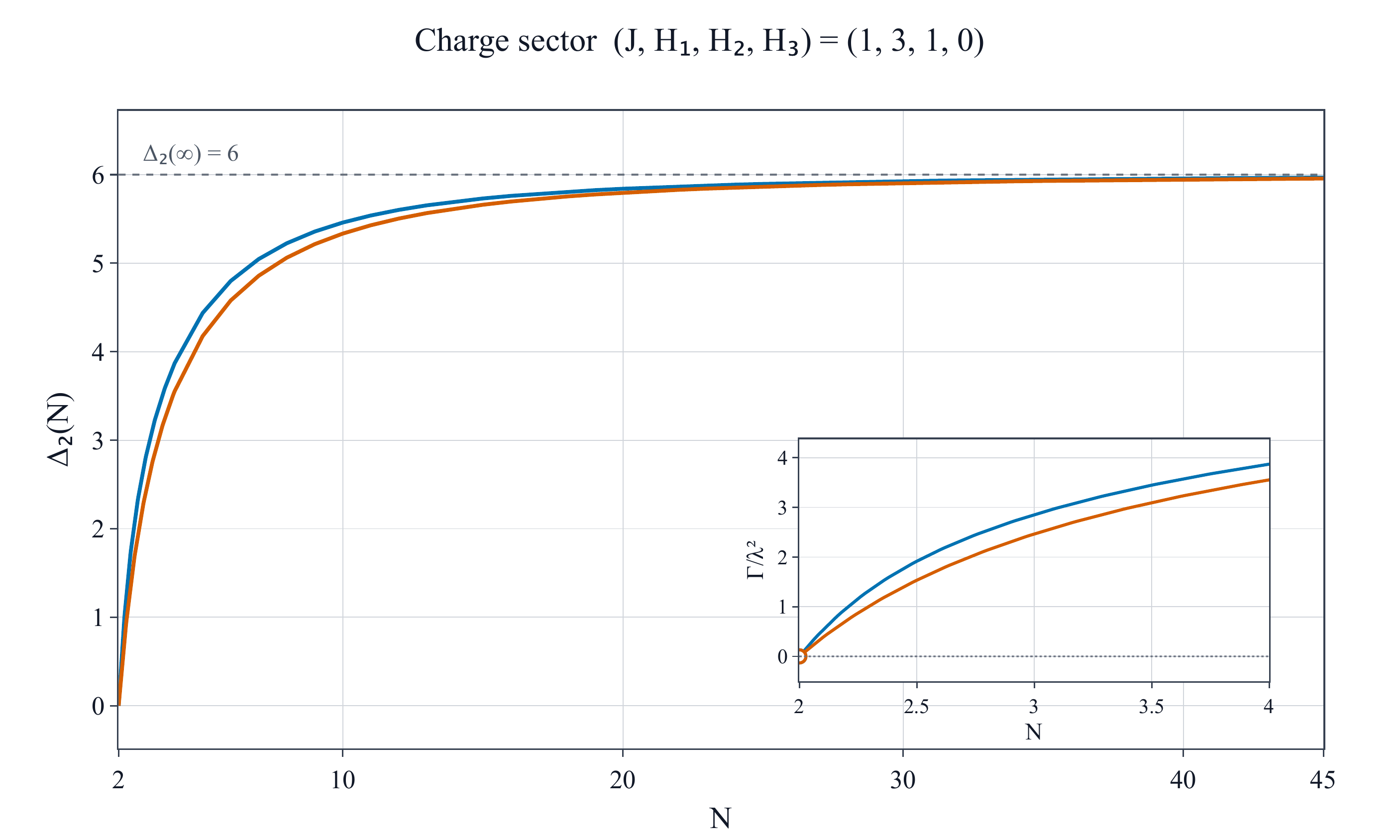}
    \caption{Centralizer-primary trajectories in $(J,H_1,H_2,H_3)=(\tfrac{1}{2},4,3,0)$ and $(1,3,1,0)$.}
    \label{fig:numerical-trajectories-2}
\end{figure}

\begin{figure}[p]
    \centering
    \includegraphics[width=0.98\textwidth]{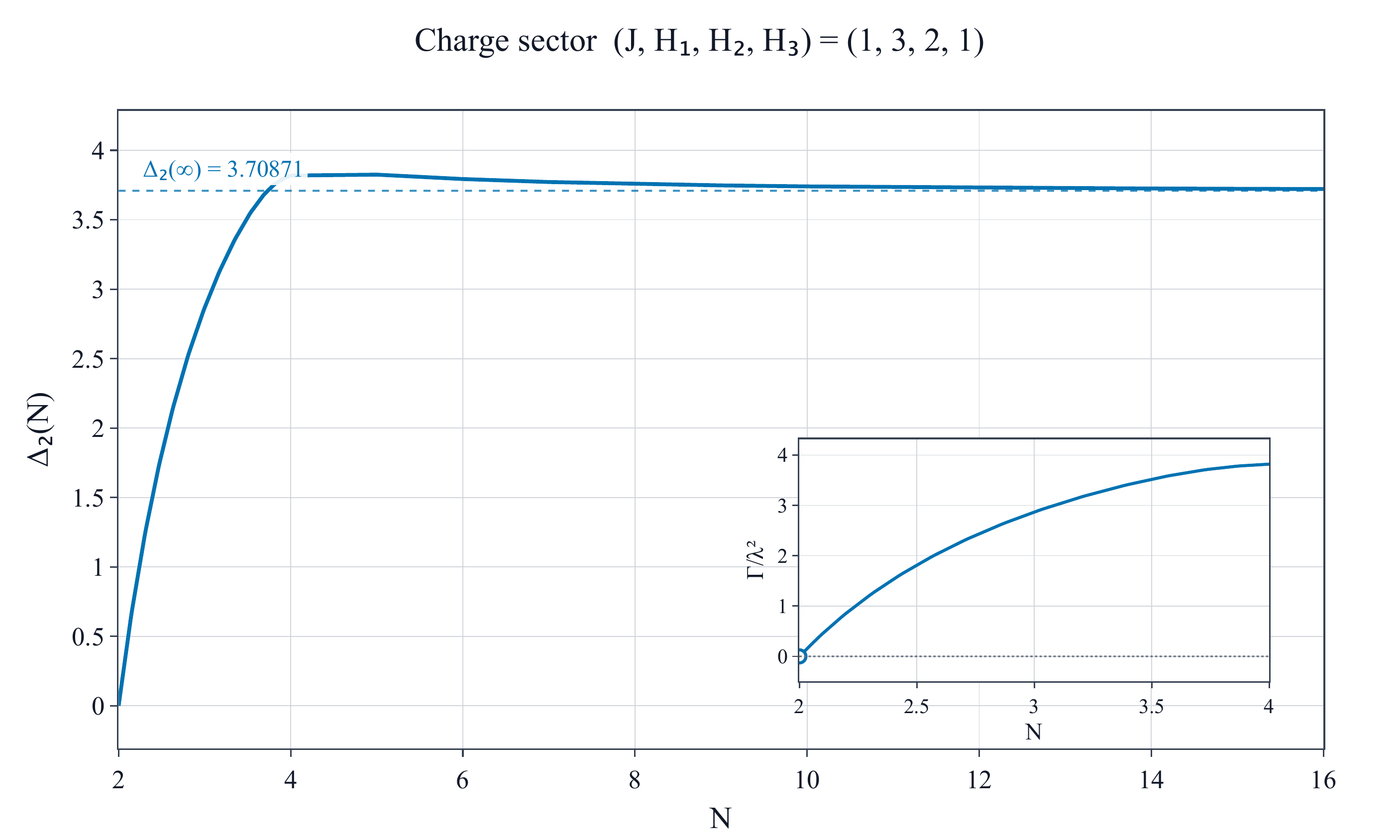}
    \includegraphics[width=0.98\textwidth]{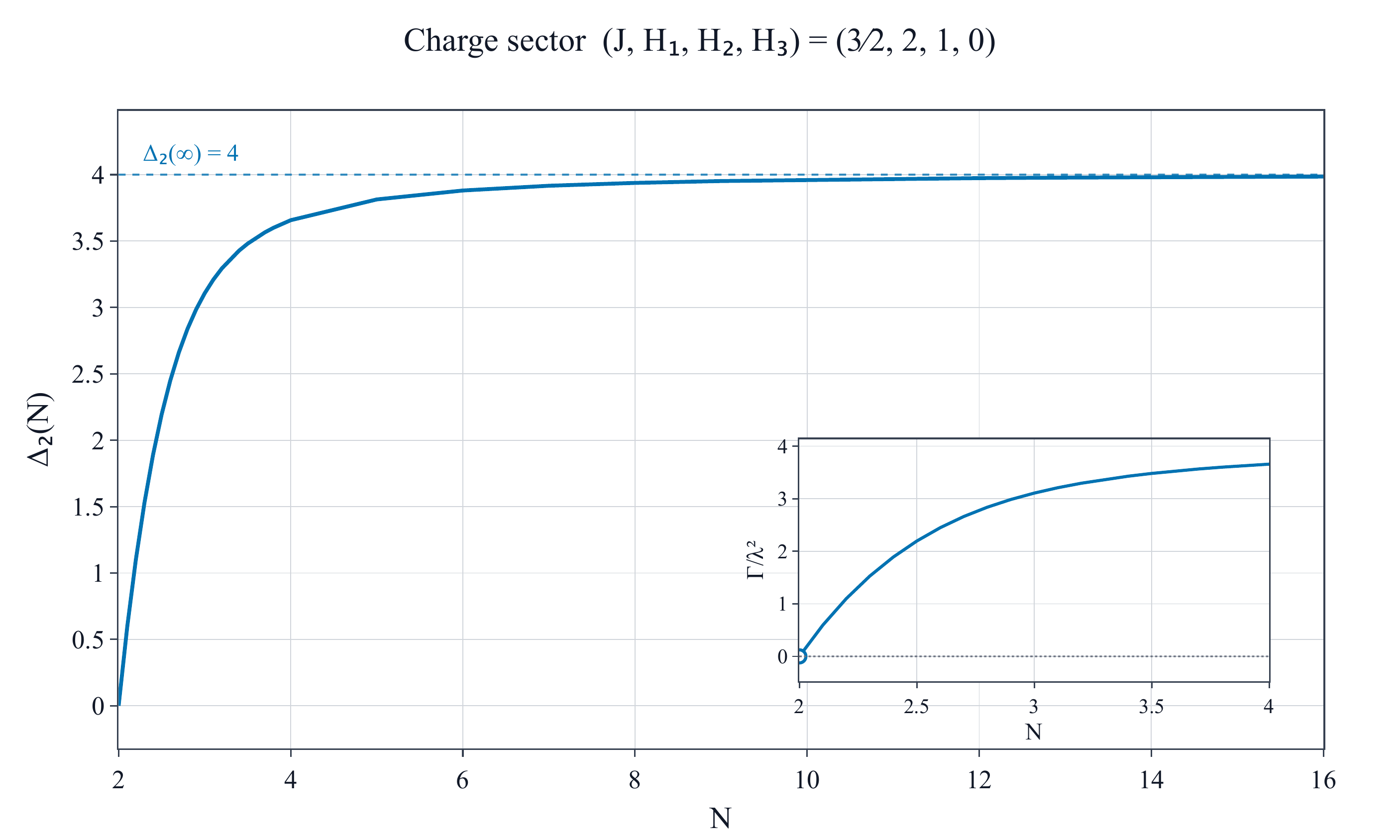}
    \caption{Centralizer-primary trajectories in $(J,H_1,H_2,H_3)=(1,3,2,1)$ and $(\tfrac{3}{2},2,1,0)$.}
    \label{fig:numerical-trajectories-3}
\end{figure}

The analysis is similar to what we have done above, except that due to the size of the matrices involved we turned to a numerical analysis in which the eigenvalues and eigenvectors were computed at many values of $N$. The numerical trajectories and their associated planar limits are shown in Figure \ref{fig:numerical-trajectories-1} through Figure \ref{fig:numerical-trajectories-3}. For each charge sector, the main panel displays the two-loop anomalous dimension in coupling units
\begin{equation}
\Delta_2(N) \equiv \gamma(N) / \lambda^2, \quad \lambda \equiv N/k \label{eq:coupling-units}
\end{equation}
and its approach to the large-$N$ limit. Dashed horizontal lines indicate the corresponding values of $\Delta_2(\infty)$. The lower-right inset magnifies the finite-rank window $2\leq N\leq4$, and open circles mark integer ranks at which the operator becomes BPS. The highest-weight condition already limits us to
\begin{equation}
H_2 \geq |H_3| \label{eq:h-swap1}
\end{equation}
and we have further chosen to only show plots with $H_3 \geq 0$. This is because the discrete automorphism $H_3 \leftrightarrow -H_3$ simply acts as
\begin{equation}
\phi_a \leftrightarrow \bar{\phi}^a, \quad \psi^a \leftrightarrow \bar{\psi}_a \label{eq:h-swap2}
\end{equation}
and leaves the shape of a trajectory unchanged.

As a first observation, most of the planar eigenvalues in Table \ref{tab:fortuitous-sectors} are integers, while the non-integer values are $8-2\sqrt{5}$, $5-\sqrt{5}$, and $6-\sqrt{21}/2$, all quadratic irrationals. 
Another fact about the eigenvalues is that in the charge sectors we have studied we never found an example of a trajectory that vanishes at multiple ranks. Further trends can be seen in the complete eigenvectors ${\cal O}(N)$ which were computed in the same fixed multi-trace bases and then covariantized. The length of \eqref{eq:long-fort1} and \eqref{eq:long-fort2} suggest that their finite-$N$ expressions will be cumbersome. We are mostly interested in patterns that hold away from integer $N$ so we have given selected large-$N$ expressions below. Each one is projective and hence defined only up to an overall normalization.

Skipping the eigenvectors that have already been shown, $(J, H_1, H_2, H_3) = (\tfrac{1}{2}, 4, 1, 0)$ has the single-trace
\begin{align}
&\mathcal{O}^a_b(N)={}
\left(3\sqrt{5}-5\right)\,\epsilon_{cd} \epsilon^{ef} \epsilon^{gh} \operatorname{Tr}\!\left(\phi_{e} \bar\phi^{c} \phi_{b} \bar\phi^{a} \phi_{h} \bar\phi^{d} \phi_{f} \bar\psi_{g}\right)
{}-\left(3\sqrt{5}-5\right)\,\epsilon_{cd} \epsilon_{ij} \epsilon^{ef} \operatorname{Tr}\!\left(\phi_{e} \bar\phi^{c} \phi_{b} \bar\phi^{a} \phi_{f} \bar\phi^{i} \psi^{d} \bar\phi^{j}\right)
\notag\\[-1mm]&\quad{}-2\left(2\sqrt{5}-4\right)\,\epsilon_{cd} \epsilon^{ef} \epsilon^{gh} \operatorname{Tr}\!\left(\phi_{e} \bar\phi^{c} \phi_{g} \bar\phi^{d} \phi_{f} \bar\phi^{a} \phi_{h} \bar\psi_{b}\right)
{}+2\left(2\sqrt{5}-4\right)\,\epsilon_{cd} \epsilon_{ij} \epsilon^{ef} \operatorname{Tr}\!\left(\phi_{b} \bar\phi^{c} \phi_{e} \bar\phi^{i} \phi_{f} \bar\phi^{d} \psi^{a} \bar\phi^{j}\right)
\notag\\[-1mm]&\quad{}-2\left(7-3\sqrt{5}\right)\,\epsilon_{cd} \epsilon^{ef} \epsilon^{gh} \operatorname{Tr}\!\left(\phi_{e} \bar\phi^{c} \phi_{g} \bar\phi^{a} \phi_{h} \bar\phi^{d} \phi_{b} \bar\psi_{f}\right)
{}+2\left(7-3\sqrt{5}\right)\,\epsilon_{cd} \epsilon_{ij} \epsilon^{ef} \operatorname{Tr}\!\left(\phi_{e} \bar\phi^{c} \phi_{b} \bar\phi^{d} \phi_{f} \bar\phi^{i} \psi^{j} \bar\phi^{a}\right)
\notag\\[-1mm]&\quad{}+2\left(\sqrt{5}-2\right)\,\epsilon_{cd} \epsilon^{ef} \epsilon^{gh} \operatorname{Tr}\!\left(\phi_{e} \bar\phi^{c} \phi_{f} \bar\phi^{a} \phi_{g} \bar\phi^{d} \phi_{h} \bar\psi_{b}\right)
{}-2\left(\sqrt{5}-2\right)\,\epsilon_{cd} \epsilon_{ij} \epsilon^{ef} \operatorname{Tr}\!\left(\phi_{e} \bar\phi^{c} \phi_{b} \bar\phi^{i} \phi_{f} \bar\phi^{j} \psi^{a} \bar\phi^{d}\right)
\notag\\[-1mm]&\quad{}+\left(7-3\sqrt{5}\right)\,\epsilon_{cd} \epsilon^{ef} \epsilon^{gh} \operatorname{Tr}\!\left(\phi_{e} \bar\phi^{a} \phi_{b} \bar\phi^{c} \phi_{h} \bar\phi^{d} \phi_{g} \bar\psi_{f}\right)
{}-\left(7-3\sqrt{5}\right)\,\epsilon_{cd} \epsilon_{ij} \epsilon^{ef} \operatorname{Tr}\!\left(\phi_{b} \bar\phi^{a} \phi_{e} \bar\phi^{c} \phi_{f} \bar\phi^{d} \psi^{i} \bar\phi^{j}\right)
\notag\\[-1mm]&\quad{}-\left(7-3\sqrt{5}\right)\,\epsilon_{cd} \epsilon^{ef} \epsilon^{gh} \operatorname{Tr}\!\left(\phi_{e} \bar\phi^{c} \phi_{f} \bar\phi^{d} \phi_{b} \bar\phi^{a} \phi_{g} \bar\psi_{h}\right)
{}+\left(7-3\sqrt{5}\right)\,\epsilon_{cd} \epsilon_{ij} \epsilon^{ef} \operatorname{Tr}\!\left(\phi_{e} \bar\phi^{d} \phi_{f} \bar\phi^{a} \phi_{b} \bar\phi^{i} \psi^{j} \bar\phi^{c}\right)
\notag\\[-1mm]&\quad{}-\left(5\sqrt{5}-11\right)\,\epsilon_{cd} \epsilon^{ef} \epsilon^{gh} \operatorname{Tr}\!\left(\phi_{e} \bar\phi^{a} \phi_{g} \bar\phi^{c} \phi_{h} \bar\phi^{d} \phi_{f} \bar\psi_{b}\right)
{}+\left(5\sqrt{5}-11\right)\,\epsilon_{cd} \epsilon_{ij} \epsilon^{ef} \operatorname{Tr}\!\left(\phi_{e} \bar\phi^{c} \phi_{f} \bar\phi^{d} \phi_{b} \bar\phi^{i} \psi^{a} \bar\phi^{j}\right)
\end{align}
as its lowest-lying fortuitous trajectory. The other two
become double-trace at large $N$
and are
\begin{align}
\mathcal{O}^a_b(N)' &= \epsilon^{cd}\epsilon_{ef}\epsilon_{gh} \text{Tr}(\bar\phi^a\phi_b) \text{Tr}(\bar\phi^e\phi_c\bar\phi^g\phi_d\bar\phi^f\psi^h) - \epsilon_{cd}\epsilon^{ef}\epsilon^{gh} \text{Tr}(\phi_b\bar\phi^a) \text{Tr}(\phi_e\bar\phi^c\phi_g\bar\phi^d\phi_f\bar\psi_h) \nonumber \\
\mathcal{O}^a_b(N)'' &= \epsilon^{cd}\epsilon_{ef}\epsilon_{gh} \text{Tr}(\bar\phi^e\phi_c) \Bigl[ \text{Tr}(\bar\phi^a\phi_b\bar\phi^g\phi_d\bar\phi^f\psi^h) + \text{Tr}(\bar\phi^f\phi_d\bar\phi^g\phi_b\bar\phi^a\psi^h) \Bigr] \nonumber \\
&- \epsilon_{cd}\epsilon^{ef}\epsilon^{gh} \text{Tr}(\phi_e\bar\phi^c) \Bigl[ \text{Tr}(\phi_b\bar\phi^a\phi_g\bar\phi^d\phi_f\bar\psi_h) + \text{Tr}(\phi_f\bar\phi^d\phi_g\bar\phi^a\phi_b\bar\psi_h) \Bigr].
\end{align}
Next, $(J, H_1, H_2, H_3) = (\tfrac{1}{2}, 4, 3, 0)$ starts with the single-trace
\begin{align}
&\mathcal{O}^{abc}_{def}(N)={}
\left(1+\sqrt{5}\right)\,\epsilon^{gh} \operatorname{Tr}\!\left(\phi_{\{d} \bar\phi^{\{a} \phi_{e} \bar\phi^{b} \phi_{|g|} \bar\phi^{c\}} \phi_{f\}} \bar\psi_{h}\right)
{}-\left(1+\sqrt{5}\right)\,\epsilon_{ij} \operatorname{Tr}\!\left(\phi_{\{d} \bar\phi^{|i|} \phi_{e} \bar\phi^{\{a} \phi_{f\}} \bar\phi^{b} \psi^{|j|} \bar\phi^{c\}}\right)
\notag\\[-1mm]&\quad{}-\left(1+\sqrt{5}\right)\,\epsilon^{gh} \operatorname{Tr}\!\left(\phi_{g} \bar\phi^{\{a} \phi_{|h|} \bar\phi^{b} \phi_{\{d} \bar\phi^{c\}} \phi_{e} \bar\psi_{f\}}\right)
{}+\left(1+\sqrt{5}\right)\,\epsilon_{ij} \operatorname{Tr}\!\left(\phi_{\{d} \bar\phi^{\{a} \phi_{e} \bar\phi^{|j|} \phi_{f\}} \bar\phi^{|i|} \psi^{b} \bar\phi^{c\}}\right)
\notag\\[-1mm]&\quad{}-\left(3-\sqrt{5}\right)\,\epsilon^{gh} \operatorname{Tr}\!\left(\phi_{\{d} \bar\phi^{\{a} \phi_{e} \bar\phi^{b} \phi_{f\}} \bar\phi^{c\}} \phi_{h} \bar\psi_{g}\right)
{}+\left(3-\sqrt{5}\right)\,\epsilon_{ij} \operatorname{Tr}\!\left(\phi_{\{d} \bar\phi^{\{a} \phi_{e} \bar\phi^{b} \phi_{f\}} \bar\phi^{c\}} \psi^{i} \bar\phi^{j}\right)
\notag\\[-1mm]&\quad{}+2\,\epsilon^{gh} \operatorname{Tr}\!\left(\phi_{g} \bar\phi^{\{a} \phi_{\{d} \bar\phi^{b} \phi_{e} \bar\phi^{c\}} \phi_{|h|} \bar\psi_{f\}}\right)
{}-2\,\epsilon_{ij} \operatorname{Tr}\!\left(\phi_{\{d} \bar\phi^{\{a} \phi_{e} \bar\phi^{b} \phi_{f\}} \bar\phi^{|i|} \psi^{c\}} \bar\phi^{j}\right)
\end{align}
and then has the double-trace
\begin{align}
\mathcal{O}^{abc}_{def}(N)' ={}&
\epsilon^{gh} \operatorname{Tr}\!\left(\phi_{\{d} \bar\phi^{\{a}\right) \operatorname{Tr}\!\left(\phi_{e} \bar\phi^{b} \phi_{|g|} \bar\phi^{c\}} \phi_{f\}} \bar\psi_{h}\right)-\epsilon_{ij} \operatorname{Tr}\!\left(\phi_{\{d} \bar\phi^{\{a}\right) \operatorname{Tr}\!\left(\phi_{e} \bar\phi^{|i|} \phi_{f\}} \bar\phi^{b} \psi^{|j|} \bar\phi^{c\}}\right).
\end{align}
The remaining trajectory, which is BPS at $N=2$, has another
double-trace endpoint, with $\Delta_2(\infty)=6$:
\begin{align}\label{eq:planar-1860-N2}
\mathcal{O}^{abc}_{def}(N)'' ={}&
\epsilon^{gh}
\operatorname{Tr}\!\left(\phi_{\{d} \bar\phi^{\{a} \phi_e \bar\phi^b\right)
\operatorname{Tr}\!\left(\phi_{f\}} \bar\phi^{c\}} \phi_g \bar\psi_h\right)
{}+\epsilon^{gh}
\operatorname{Tr}\!\left(\phi_{\{d} \bar\phi^{\{a} \phi_e \bar\phi^b\right)
\operatorname{Tr}\!\left(\phi_{|g|} \bar\phi^{c\}} \phi_{f\}} \bar\psi_h\right) \\
&-\epsilon_{ij}
\operatorname{Tr}\!\left(\phi_{\{d} \bar\phi^{\{a} \phi_e \bar\phi^b\right)
\operatorname{Tr}\!\left(\bar{\phi}^{|i|} \phi_{f\}} \bar\phi^{c\}} \psi^j \right )
{}-\epsilon_{ij}
\operatorname{Tr}\!\left(\phi_{\{d} \bar\phi^{\{a} \phi_e \bar\phi^b\right)
\operatorname{Tr}\!\left(\bar{\phi}^{c\}} \phi_{f\}} \bar\phi^{i} \psi^{j} \right). \nonumber
\end{align} 
The five sectors where we have looked at explicit eigenvectors so far all have one thing in common: the lightest trajectory is single-trace. The action of \eqref{eq:h-swap2} also reveals a pattern. All 11 large-$N$ eigenvectors discussed up to this point are odd under this $\mathbb{Z}_2$ parity.\footnote{It seems that this parity does not have to be definite at the operator level for finite $N$. In \eqref{eq:fort-1420}, there is a parity-even part which becomes $Q$-exact at $N = 2$.} Both of these properties change when we look at $(J, H_1, H_2, H_3) = (1, 3, 1, 0)$ where we find two trajectories. Despite the appearance of floating-point numbers, we can write the planar endpoint of the first trajectory as
\begin{align}\label{eq:parity-even}
\mathcal{O}^a_b(N) &= 2\eta\,\epsilon_{cd} \epsilon^{ef} \operatorname{Tr}\!\left(\phi_e \bar\phi^a\right) \operatorname{Tr}\!\left(\phi_f \bar\phi^c \phi_b D\bar\phi^d\right)
+\eta\,\epsilon_{cd} \epsilon^{ef} \operatorname{Tr}\!\left(\phi_b \bar\phi^a\right) \operatorname{Tr}\!\left(\phi_e \bar\phi^c \phi_f D\bar\phi^d\right)
\notag\\[-1mm]&\quad{}+2(1-\eta)\,\epsilon^{ef} \epsilon^{gh} \operatorname{Tr}\!\left(\phi_e \bar\phi^a\right) \operatorname{Tr}\!\left(\phi_b \bar\psi_g \phi_f \bar\psi_h\right)
-2\epsilon^{ef} \epsilon^{gh} \operatorname{Tr}\!\left(\phi_e \bar\phi^a \phi_g \bar\psi_h\right) \operatorname{Tr}\!\left(\phi_f \bar\psi_b\right)
\notag\\[-1mm]&\quad{}-2\epsilon_{cd} \epsilon^{ef} \operatorname{Tr}\!\left(\phi_b \bar\phi^c\right) \operatorname{Tr}\!\left(\phi_e \bar\phi^a \psi^d \bar\psi_f\right) +2\epsilon_{cd} \epsilon^{ef} \operatorname{Tr}\!\left(\phi_e \bar\phi^a\right) \operatorname{Tr}\!\left(\phi_b \bar\phi^c \psi^d \bar\psi_f\right)
\notag\\[-1mm]&\quad{}-2\epsilon^{ef} \epsilon^{gh} \operatorname{Tr}\!\left(\phi_e \bar\phi^a \phi_g \bar\psi_f\right) \operatorname{Tr}\!\left(\phi_b \bar\psi_h\right)
+\epsilon^{ef} \epsilon^{gh} \operatorname{Tr}\!\left(\phi_e \bar\phi^a \phi_f \bar\psi_b\right) \operatorname{Tr}\!\left(\phi_g \bar\psi_h\right)
\notag\\[-1mm]&\quad{}+2\epsilon_{cd} \epsilon^{ef} \operatorname{Tr}\!\left(\phi_e \bar\phi^c \phi_b \bar\psi_f\right) \operatorname{Tr}\!\left(\psi^a \bar\phi^d\right)
+2\epsilon_{cd} \epsilon^{ef} \operatorname{Tr}\!\left(\phi_b \bar\phi^c \phi_e \bar\psi_f\right) \operatorname{Tr}\!\left(\psi^d \bar\phi^a\right)
\notag\\[-1mm]&\quad{}-\epsilon_{cd} \epsilon^{ef} \operatorname{Tr}\!\left(\phi_e \bar\phi^a \phi_f \bar\psi_b\right) \operatorname{Tr}\!\left(\psi^c \bar\phi^d\right)
+\dots
\end{align}
where $\eta=1.2911781864$, and that of the second trajectory as
\begin{align}\label{eq:parity-odd}
\mathcal{O}^a_b(N)' &= 2(2\eta_1 - \eta_2 + 1)\,\epsilon_{cd} \epsilon^{ef} \operatorname{Tr}\!\left(\phi_e \bar\phi^c\right) \operatorname{Tr}\!\left(\phi_b \bar\phi^a \phi_f D\bar\phi^d\right)
+(\eta_2 - 1)\,\epsilon_{cd} \epsilon^{ef} \operatorname{Tr}\!\left(\phi_e \bar\phi^c\right) \operatorname{Tr}\!\left(\phi_b \bar\phi^d \phi_f D\bar\phi^a\right)
\notag\\[-1mm]&\quad{}+(\eta_2 - 1)\,\epsilon_{cd} \epsilon^{ef} \operatorname{Tr}\!\left(\phi_e \bar\phi^c\right) \operatorname{Tr}\!\left(\phi_f \bar\phi^d \phi_b D\bar\phi^a\right)
-(2\eta_1 - \eta_2 + 1)\,\epsilon_{cd} \epsilon^{ef} \operatorname{Tr}\!\left(\phi_b \bar\phi^c\right) \operatorname{Tr}\!\left(\phi_e \bar\phi^a \phi_f D\bar\phi^d\right)
\notag\\[-1mm]&\quad{}+4\eta_1\,\epsilon_{cd} \epsilon^{ef} \operatorname{Tr}\!\left(\phi_e \bar\phi^c\right) \operatorname{Tr}\!\left(\phi_b \bar\phi^a \psi^d \bar\psi_f\right)
-2\eta_2\,\epsilon_{cd} \epsilon^{ef} \operatorname{Tr}\!\left(\phi_b \bar\phi^c\right) \operatorname{Tr}\!\left(\phi_e \bar\phi^a \psi^d \bar\psi_f\right)
\notag\\[-1mm]&\quad{}-2\eta_2\,\epsilon_{cd} \epsilon^{ef} \operatorname{Tr}\!\left(\phi_e \bar\phi^a\right) \operatorname{Tr}\!\left(\phi_b \bar\phi^c \psi^d \bar\psi_f\right)
-2\epsilon^{ef} \epsilon^{gh} \operatorname{Tr}\!\left(\phi_e \bar\phi^a\right) \operatorname{Tr}\!\left(\phi_b \bar\psi_g \phi_f \bar\psi_h\right)
\notag\\[-1mm]&\quad{}-(\eta_1 + 2\eta_2)\,\epsilon^{ef} \epsilon^{gh} \operatorname{Tr}\!\left(\phi_e \bar\phi^a \phi_g \bar\psi_f\right) \operatorname{Tr}\!\left(\phi_h \bar\psi_b\right)
+(3\eta_1 - 2\eta_2)\,\epsilon^{ef} \epsilon^{gh} \operatorname{Tr}\!\left(\phi_b \bar\phi^a \phi_e \bar\psi_f\right) \operatorname{Tr}\!\left(\phi_g \bar\psi_h\right)
\notag\\[-1mm]&\quad{}+\eta_1\,\epsilon^{ef} \epsilon^{gh} \operatorname{Tr}\!\left(\phi_e \bar\phi^a \phi_f \bar\psi_g\right) \operatorname{Tr}\!\left(\phi_b \bar\psi_h\right)
+(\eta_1 - 2\eta_2)\,\epsilon^{ef} \epsilon^{gh} \operatorname{Tr}\!\left(\phi_e \bar\phi^a \phi_g \bar\psi_b\right) \operatorname{Tr}\!\left(\phi_f \bar\psi_h\right)
\notag\\[-1mm]&\quad{}+(2\eta_2 - \eta_1)\,\epsilon^{ef} \epsilon^{gh} \operatorname{Tr}\!\left(\phi_e \bar\phi^a \phi_b \bar\psi_g\right) \operatorname{Tr}\!\left(\phi_h \bar\psi_f\right)
-4\eta_2\,\epsilon_{cd} \epsilon^{ef} \operatorname{Tr}\!\left(\phi_e \bar\phi^c \phi_b \bar\psi_f\right) \operatorname{Tr}\!\left(\psi^a \bar\phi^d\right)
\notag\\[-1mm]&\quad{}-2\eta_1\,\epsilon_{cd} \epsilon^{ef} \operatorname{Tr}\!\left(\phi_b \bar\phi^a \phi_e \bar\psi_f\right) \operatorname{Tr}\!\left(\psi^c \bar\phi^d\right)
+2\eta_2\,\epsilon_{cd} \epsilon^{ef} \operatorname{Tr}\!\left(\phi_e \bar\phi^c \phi_f \bar\psi_b\right) \operatorname{Tr}\!\left(\psi^d \bar\phi^a\right)
\notag\\[-1mm]&\quad{}+2(2\eta_2-\eta_1)\,\epsilon_{cd} \epsilon^{ef} \operatorname{Tr}\!\left(\phi_e \bar\phi^a \phi_b \bar\psi_f\right) \operatorname{Tr}\!\left(\psi^c \bar\phi^d\right)
-\dots
\end{align}
where $\eta_1=0.7602113506$, $\eta_2=0.5393124178$, and $\eta_3=(1-\eta_2)/2$. In both expressions, the ellipsis denotes the image of the
displayed terms under \eqref{eq:h-swap2}, with the sign explicitly indicated. Upper and lower indices are exchanged in this image, and the free indices are relabelled $a\leftrightarrow b$ to restore the index placement of $\mathcal{O}^a_b$. Both trajectories become double-trace at large $N$, with $\Delta_2(\infty)=6$, but have opposite parities under \eqref{eq:h-swap2}.

Let us now look at the $H_3 \neq 0$ sectors where operators form doublets under \eqref{eq:h-swap2}. The first is $(J, H_1, H_2, H_3) = (\tfrac{1}{2}, 4, 2, 1)$ where
\begin{align}
\mathcal{O}^{abc}_d(N) &= 2\epsilon^{ef} \epsilon^{gh} \operatorname{Tr}\!\left(\phi_{e} \bar\phi^{\{a}\right) \operatorname{Tr}\!\left(\phi_{|d|} \bar\phi^{b} \phi_{|g|} \bar\phi^{c\}} \phi_{f} \bar\psi_{h}\right)-\epsilon^{ef} \epsilon^{gh} \operatorname{Tr}\!\left(\phi_{d} \bar\phi^{\{a}\right) \operatorname{Tr}\!\left(\phi_{|e|} \bar\phi^{b} \phi_{|g|} \bar\phi^{c\}} \phi_{f} \bar\psi_{h}\right) \nonumber
\\[-1mm]&\quad{}-\epsilon_{ij} \epsilon^{ef} \operatorname{Tr}\!\left(\phi_{e} \bar\phi^{\{a}\right) \operatorname{Tr}\!\left(\phi_{|d|} \bar\phi^{|i|} \phi_{|f|} \bar\phi^{b} \psi^{|j|} \bar\phi^{c\}}\right)-\epsilon_{ij} \epsilon^{ef} \operatorname{Tr}\!\left(\phi_{e} \bar\phi^{\{a}\right) \operatorname{Tr}\!\left(\phi_{|f|} \bar\phi^{|i|} \phi_{|d|} \bar\phi^{b} \psi^{|j|} \bar\phi^{c\}}\right).
\end{align}
This is another example where the lightest (only) fortuitous trajectory becomes double-trace. Next, the sector $(J, H_1, H_2, H_3) = (1,3,2,1)$ contains a very important operator. Defining $\xi=(\sqrt{21}-3)/6$, its planar endpoint is
\begin{align}\label{eq:planar-2642}
\mathcal{O}^{abc}_d(N) &= \xi\epsilon^{ef} \operatorname{Tr}\!\left(\phi_d \bar\phi^{\{a} \phi_{|e|} \bar\phi^b \phi_{|f|} D\bar\phi^{c\}}\right)
-\xi\epsilon^{ef} \operatorname{Tr}\!\left(\phi_e \bar\phi^{\{a} \phi_{|f|} \bar\phi^b \phi_{|d|} D\bar\phi^{c\}}\right)
+\epsilon^{ef} \operatorname{Tr}\!\left(\phi_e \bar\phi^{\{a} \phi_{|d|} \bar\phi^b D\phi_{|f|} \bar\phi^{c\}}\right)
\notag\\[-1mm]&\quad{}+\epsilon^{ef} \operatorname{Tr}\!\left(\phi_d \bar\phi^{\{a} \phi_{|e|} \bar\phi^b D\phi_{|f|} \bar\phi^{c\}}\right)
+2\xi\,\epsilon^{ef} \operatorname{Tr}\!\left(\phi_d \bar\phi^{\{a} \phi_{|e|} \bar\phi^b \psi^{c\}} \bar\psi_f\right)
-2\xi\,\epsilon^{ef} \operatorname{Tr}\!\left(\phi_e \bar\phi^{\{a} \phi_{|d|} \bar\psi_{|f|} \psi^b \bar\phi^{c\}}\right)
\notag\\[-1mm]&\quad{}-2\xi\,\epsilon^{ef} \operatorname{Tr}\!\left(\phi_d \bar\phi^{\{a} \psi^b \bar\phi^{c\}} \phi_e \bar\psi_f\right)
-2\xi\,\epsilon^{ef} \operatorname{Tr}\!\left(\phi_e \bar\phi^{\{a} \psi^b \bar\phi^{c\}} \phi_d \bar\psi_f\right)
\notag\\[-1mm]&\quad{}-2\xi\,\epsilon_{gh} \operatorname{Tr}\!\left(\phi_d \bar\phi^g \psi^{\{a} \bar\phi^b \psi^{c\}} \bar\phi^h\right)
-2\epsilon_{gh} \operatorname{Tr}\!\left(\phi_d \bar\phi^{\{a} \psi^{|g|} \bar\phi^b \psi^{|h|} \bar\phi^{c\}}\right).
\end{align}
When continued down to $N = 2$, this is a core primary because we have shown in section \ref{sec:dressing} that it cannot be made by dressing another fortuitous operator. We expect that the monotone and $Q$-exact terms can be adjusted at this value of $N$ to yield a cohomology representative with rational coefficients. Finally, the sector $(J,H_1,H_2,H_3)=(\tfrac32,2,1,0)$ also contains a fortuitous primary at $N=2$. Its single-trace planar endpoint, with $\Delta_2(\infty)=4$, is 
\begin{align}\label{eq:planar-3420}
\mathcal{O}^a_b(N) &= 2\,\epsilon^{ef} \operatorname{Tr}\!\left(\phi_b \bar\phi^a \phi_e D\bar\psi_f\right)
+\epsilon^{ef} \operatorname{Tr}\!\left(\phi_e \bar\phi^a \phi_f D\bar\psi_b\right)
-2\,\epsilon^{ef} \operatorname{Tr}\!\left(\phi_b \bar\phi^a D\phi_e \bar\psi_f\right)
+\epsilon^{ef} \operatorname{Tr}\!\left(\phi_e \bar\phi^a D\phi_f \bar\psi_b\right)
\notag\\[-1mm]&\quad{}+2\,\epsilon_{cd} \operatorname{Tr}\!\left(\phi_b \bar\phi^a \psi^c D\bar\phi^d\right)
+\epsilon_{cd} \operatorname{Tr}\!\left(\phi_b \bar\phi^c \psi^a D\bar\phi^d\right)
-2\,\epsilon_{cd} \operatorname{Tr}\!\left(\phi_b \bar\phi^a D\psi^c \bar\phi^d\right)
+\epsilon_{cd} \operatorname{Tr}\!\left(\phi_b \bar\phi^c D\psi^a \bar\phi^d\right)
\notag\\[-1mm]&\quad{}-2\,\epsilon^{ef} \operatorname{Tr}\!\left(\phi_b D\bar\phi^a \phi_e \bar\psi_f\right)
-\epsilon^{ef} \operatorname{Tr}\!\left(\phi_e D\bar\phi^a \phi_f \bar\psi_b\right)
-2\,\epsilon_{cd} \operatorname{Tr}\!\left(\phi_b D\bar\phi^c \psi^d \bar\phi^a\right)
-\epsilon_{cd} \operatorname{Tr}\!\left(\phi_b D\bar\phi^c \psi^a \bar\phi^d\right)
\notag\\[-1mm]&\quad{}+2\,\epsilon^{ef} \operatorname{Tr}\!\left(\phi_b \bar\psi_e D\phi_f \bar\phi^a\right)
+\epsilon^{ef} \operatorname{Tr}\!\left(\phi_e \bar\psi_b D\phi_f \bar\phi^a\right)
+2\,\epsilon^{ef} \operatorname{Tr}\!\left(\phi_b \bar\psi_e \psi^a \bar\psi_f\right)
+2\,\epsilon_{cd} \operatorname{Tr}\!\left(D\phi_b \bar\phi^a \psi^c \bar\phi^d\right)
\notag\\[-1mm]&\quad{}-\epsilon_{cd} \operatorname{Tr}\!\left(D\phi_b \bar\phi^c \psi^a \bar\phi^d\right)
-2\,\epsilon_{cd} \operatorname{Tr}\!\left(\psi^c \bar\phi^a \psi^d \bar\psi_b\right).
\end{align}
It is nice that all core primaries observed so far are single-trace but the dataset is small enough that this might be a coincidence.

\subsection{OPE coefficients and decoupling}
We have previously mentioned a check of our results: the ${\rm U}(N)_k \times {\rm U}(N)_{-k}$ theory for integer $N$ should not have any classically BPS operators with a negative anomalous dimension. Fortuitous operators are the main examples where this needs to be checked.--- the only way for them to avoid negative anomalous dimensions is through a finely tuned second-order zero which does not appear in this work. Subsections \ref{sec:lowest} and \ref{sec:second-lowest} argue that fortuitous operators decouple at the dangerous values of $N$ by showing that their norms approach zero. A more convincing approach is to show this at the level of operator product expansion (OPE) coefficients.

If one wishes to test an OPE coefficient where all three operators are BPS, there is an additional constraint --- the $J$ and $H_1$ Cartans must sum to zero for the three-point function to survive. When combined with the BPS condition, this implies that the three-point function is extremal. A fortuitous-monotone-monotone case like $\left < F^\dagger(x_1) M_1(x_2) M_2(x_3) \right >$ is then seen to be uninteresting. Regularity as $x_2 \to x_3$ shows that this is also equal to a mixed two-point function. These have been chosen to vanish through the prescription \eqref{orthogonal-complement}.
Instead, the three-point functions that we will follow in $N$ couple
two fortuitous operators $F_1$ and $F_2$ with a monotone operator $G$. In the extremal correlator $\langle F_2^\dagger(\infty)G(1)F_1(0)\rangle$,
it is natural to choose $F_1$ and $F_2$ to be such that $F_1$ belongs to a cohomology class that can be dressed with gravitons to produce the class of $F_2$ as discussed in section \ref{sec:dressing}. For concreteness we will take $F_1$ to be the lowest fortuitous trajectory in the $(J, H_1, H_2, H_3) = (\tfrac{1}{2},2,1,0)$ sector and consider four possibilities for $F_2$, namely, the two fortuitous trajectories in the $(J, H_1, H_2, H_3) = (\tfrac{1}{2},3,0,0)$ and $(J, H_1, H_2, H_3) = (\tfrac{1}{2},3,2,0)$, always focusing on centralizer primaries. Once $F_1$ and $F_2$ are fixed, we can scan through the possible choices of $G$ that produce an extremal correlator and we find out that when $F_2$ belongs to the $(\tfrac{1}{2},3,0,0)$ sector, then $G = \operatorname{Tr}\left[\bar\phi^2\phi_2\right]$ is uniquely selected, whereas when $F_2$ belongs to the $(\tfrac{1}{2},3,2,0)$ sector, then $G = \operatorname{Tr}\left[\bar\phi^1\phi_1\right]$ is uniquely selected.
\begin{figure}[h]
    \centering
    \includegraphics[width=\textwidth]{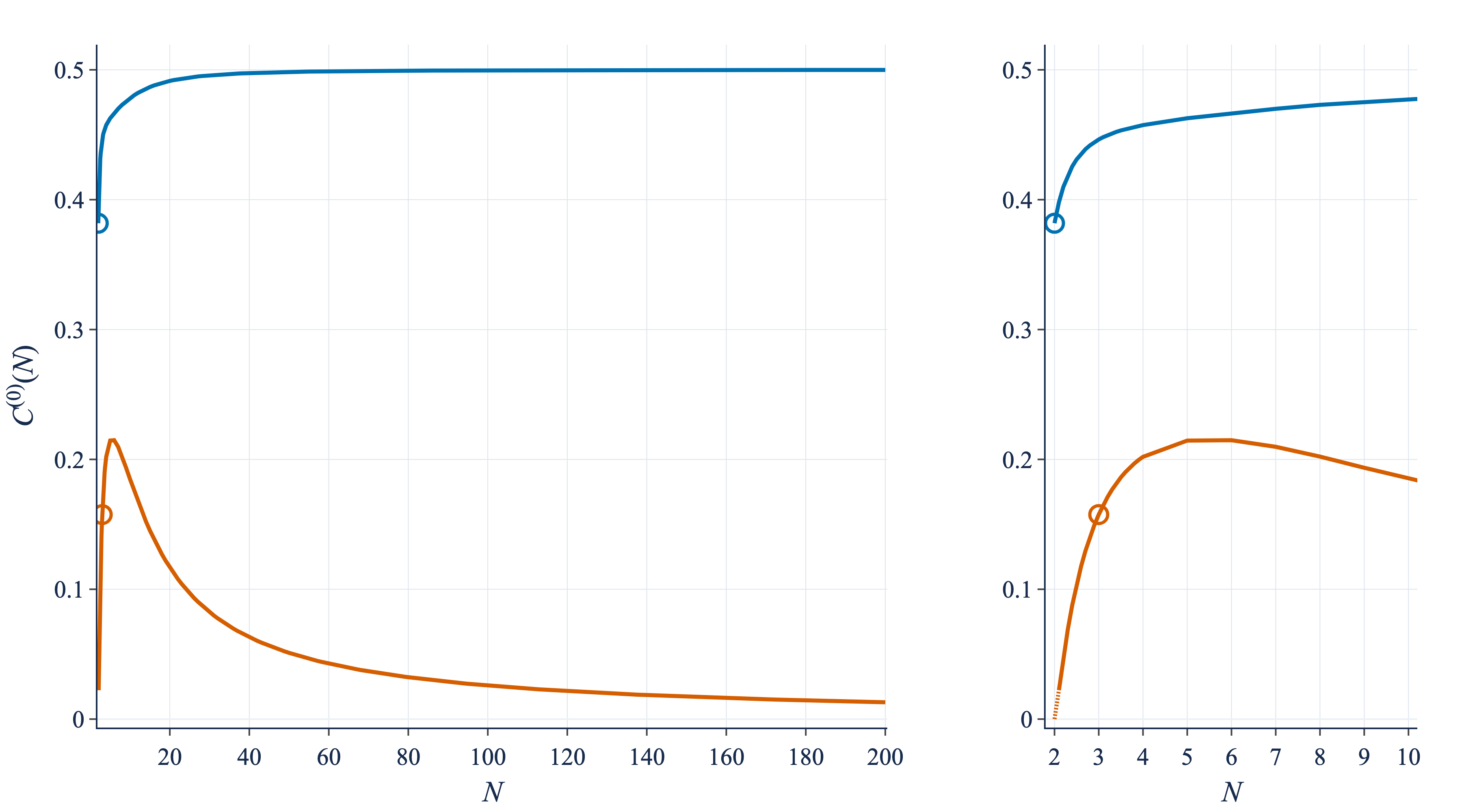}
    \caption{OPE coefficient $C^{(0)}(N)$ extracted from the extremal correlator $\langle F_2(\infty)G(1)F_1(0)\rangle$ with $F_1$ the lowest fortuitous trajectory in the $(J, H_1, H_2, H_3) = (\tfrac{1}{2},2,1,0)$ sector, $F_2$ taken to be one of the two centralizer primary fortuitous trajectories in the $(J, H_1, H_2, H_3) = (\tfrac{1}{2},3,0,0)$ sector, and $G = \operatorname{Tr}\left[\bar\phi^2\phi_2\right]$. The left panel show the approach to the planar limit whereas the right panel shows the behavior in the near-BPS region. The blue line shows the dressed trajectory that is BPS at $N=2$ whereas the orange line shows the core trajectory that is BPS at $N=3$.}
    \label{fig:ope-coefficients-F2-half-3-0-0-comparison}
\end{figure}

We can check some other expeted features in the OPE coefficients: firstly, since $F_1$ becomes a single trace in the planar limit and $G$ is also single trace, it turns out that we only expect a non-trivial OPE coefficient in the large $N$ limit for the fortuitous trajectories which become double trace at large $N$. In the $(J, H_1, H_2, H_3) = (\tfrac{1}{2},3,0,0)$ sector, whose OPE coefficient plots are depicted in Figure \ref{fig:ope-coefficients-F2-half-3-0-0-comparison}, we have two bi-singlets: the core fortuitous trajectory that is BPS at $N = 3$ and becomes single-trace at large $N$, and the dressed fortuitous trajectory that is BPS at $N=2$ and becomes double-trace at large $N$. As expected, the OPE coefficient for the core goes to zero in the planar limit, whereas the OPE coefficient for the dressed trajectory approaches a non-trivial value. 

\begin{figure}[h]
    \centering
    \includegraphics[width=\textwidth]{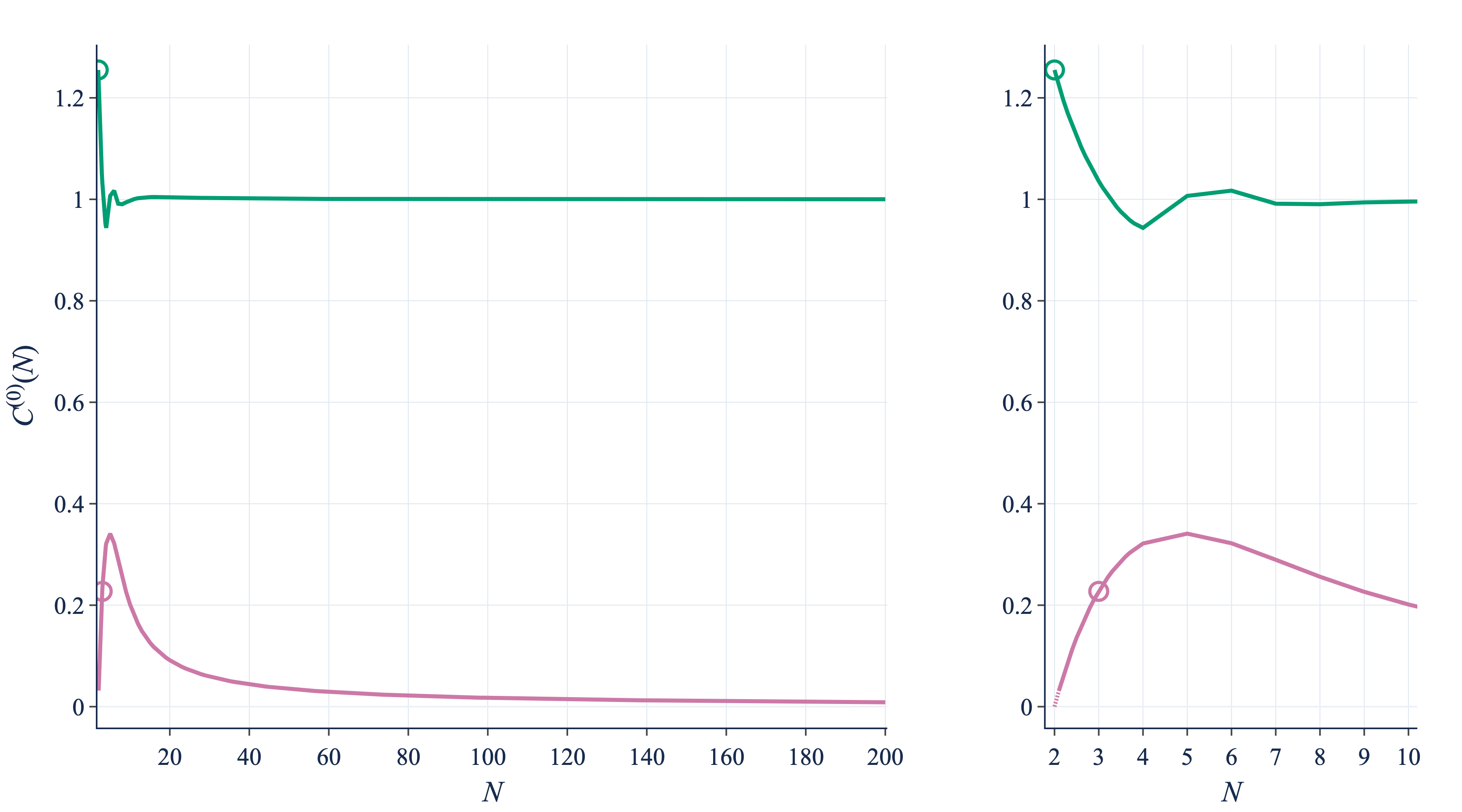}
    \caption{OPE coefficient $C^{(0)}(N)$ extracted from the extremal correlator $\langle F_2(\infty)G(1)F_1(0)\rangle$ with $F_1$ the lowest fortuitous trajectory in the $(J, H_1, H_2, H_3) = (\tfrac{1}{2},2,1,0)$ sector, $F_2$ taken to be one of the two centralizer primary fortuitous trajectories in the $(J, H_1, H_2, H_3) = (\tfrac{1}{2},3,2,0)$ sector, and $G = \operatorname{Tr}\left[\bar\phi^1\phi_1\right]$. The left panel show the approach to the planar limit whereas the right panel shows the behavior in the near-BPS region.}
    \label{fig:ope-coefficients-F2-half-3-2-0-comparison}
\end{figure}

Likewise, in the $(J, H_1, H_2, H_3) = (\tfrac{1}{2},3,2,0)$ sector, shown in Figure \ref{fig:ope-coefficients-F2-half-3-2-0-comparison}, there are two bi-triplets. One of them is a core bi-triplet that becomes BPS at $N=3$ and turns single-trace at large $N$, while the other is a dressed bi-triplet that becomes BPS at $N=2$ and turns double-trace at large $N$. Once again, the OPE coefficient for the core goes to zero in the planar limit, whereas the OPE coefficient for the dressed trajectory approaches a non-trivial value.


Finally, we come to the feature which motivated this subsection --- the decoupling of operators once the anomalous dimension becomes unphysical. Indeed both the core bi-singlet in $(\tfrac{1}{2},3,0,0)$ and the core bi-triplet in $(\tfrac{1}{2},3,2,0)$ become BPS at $N=3$ and then their anomalous dimension becomes negative for $2 < N < 3$. The corresponding OPE coeffcients extracted from the extremal correlator can be seem to approach zero at the physical $N=2$ value, signaling the decoupling of these operators as expected.
The OPE analysis provides a more physical alternative to the norms which were used for this type of consistency check in $\mathcal{N}=4$ SYM
\cite{2306.04693}.\footnote{Various three-point functions coupling two single-trace operators to a double-trace operator (and sometimes four-point functions coupling three single-traces to a triple-trace in particular kinematics) were computed in \cite{2306.04693} as a heuristic probe of black hole disassociation.} It is also reminiscent of a decoupling pattern for Regge trajectories explained in \cite{2211.13754,2312.09283}, although the analytic continuations discussed in these works required nonlocal operators. As such, there was no nice analogue of the overcomplete multi-trace basis and different methods had to be used.

\subsection{A survey of classically BPS operators}

The fact that $Q$-cohomology discards non-BPS states is both a blessing and a curse. Although there is no longer a known criterion like fortuity which can tell them apart, small-$N$ operators above the BPS bound can still be seen as highly quantum versions of black holes and multi-gravitons. The expectation that the former should outnumber the latter as we move toward large $N$ (while scaling their charges appropriately) is universal and independent of supersymmetry. This is one of the motivations for collecting eigenvalue statistics about lifted states from the Hamiltonian. Carrying this out for $\mathcal{N} = 4$ SYM, the authors of \cite{2306.04673} found evidence that the minimal anomalous dimension $\gamma_{\text{gap}}$ grows more slowly than $N$ so that $g^{-1}_{\text{YM}} \gamma_{\text{gap}} / N$ shrinks. It is then possible to make speculative comparisons to the spectral gap predicted by the gravitational path integral \cite{2003.02860,2011.01953,2203.01331}.

In ABJM theory, one can test whether the analogous quantity $k^2 \gamma_{\text{gap}} / N$ shrinks as well. To do so, we have defined
\begin{equation}
\delta = \frac{k^2 \gamma}{N}
\end{equation}
and examined the distribution of eigenvalues at $N = 2$ and $N = 10$, covering many more charge sectors than the ones containing fortuitous states. We have chosen the truncation level to match the partition function \eqref{zfort-n2}. The $O(x^7)$ there means that the last included sectors have $E + J = 6$. Since we are working with non-BPS operators where $E$ receives corrections, it is better to say that we have imposed
\begin{equation}
2J + H_1 \leq 6 \label{eq:hist-cutoff}
\end{equation}
here. Additionally, it makes sense to ensure that our eigenvalue distributions are dynamical rather than kinematical. Since a  centralizer multiplet is seeded by an $\mathfrak{so}(4)$ multiplet worth of $\mathfrak{osp}(4|2) \oplus \mathfrak{u}(1|1)$ primaries, we should keep only the $\mathfrak{so}(4)$ highest weight as in the previous subsection. This once again limits us to sectors with \eqref{eq:h-swap1} and allows us to use \eqref{eq:h-swap2} to nearly double the speed of the computation. One more point is that $J = 0$ can be completely ignored. When a classically BPS operator is a scalar, it is built from $\phi_a$ and $\bar{\phi}^a$ letters which means it is $Q$-closed at one loop. The only way for it to receive an anomalous dimension is for it to lie outside the $Q^\dagger$ kernel and therefore outside the set of $\mathfrak{osp}(4|2) \oplus \mathfrak{u}(1|1)$ primaries.\footnote{Some non-scalar sectors like $(J, H_1, H_2, H_3) = (\tfrac{1}{2}, 5, 0, 0)$ also show a reduction in the number of highest weights when we include the $\mathfrak{u}(1|1)$ part but these are harder to predict.}

\begin{figure}[h]
\centering
\includegraphics[scale=0.6]{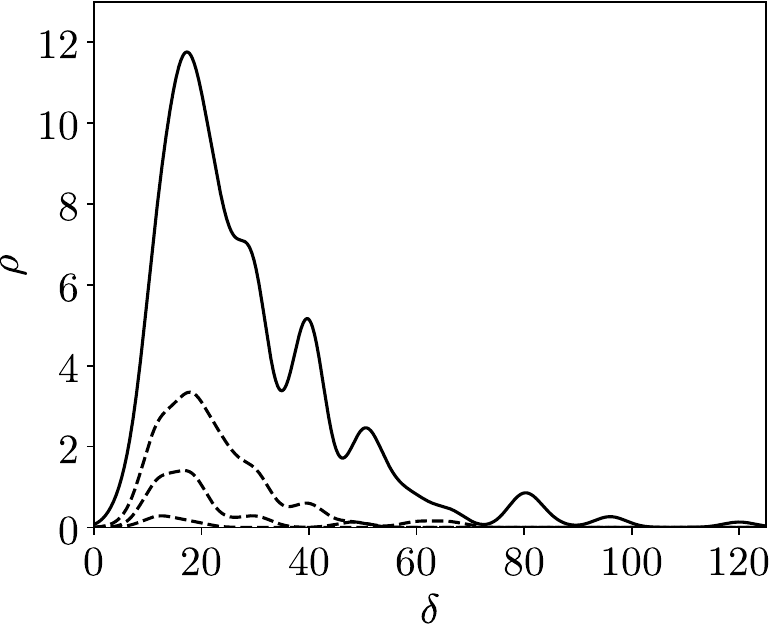} \quad \includegraphics[scale=0.6]{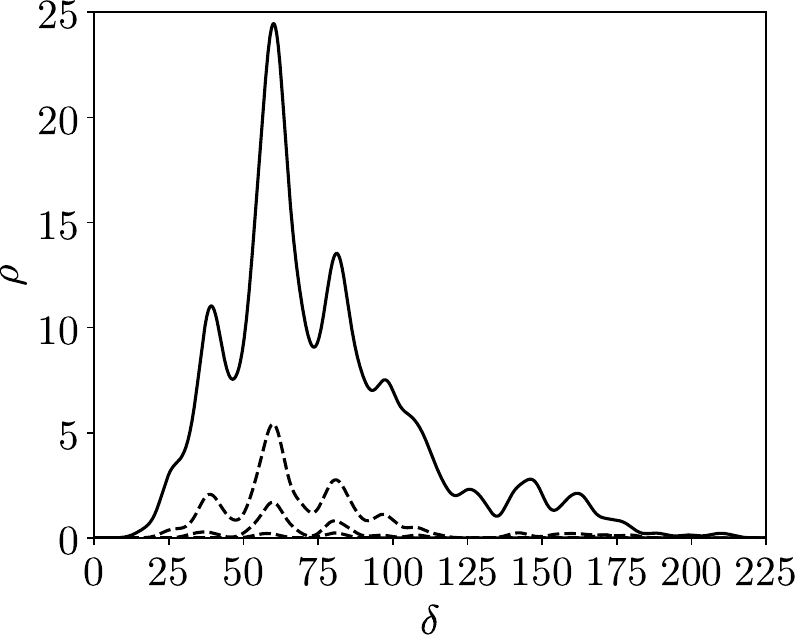}
\caption{Anomalous dimensions for lifted highest-weight operators at two-loops in ${\rm U}(2)_k \times {\rm U}(2)_{-k}$ (left) and ${\rm U}(10)_k \times {\rm U}(10)_{-k}$ (right) ABJM theory. The solid line represents data taken over all charge sectors with $2J + H_1 \leq 6$. The successively smaller regions bounded by a dotted line show this for $2J + H_1 \leq 5,4,3$. All of these use $\sigma = 3$.}
\label{fig:hist1}
\end{figure}

\begin{figure}[h]
\centering
\includegraphics[scale=0.6]{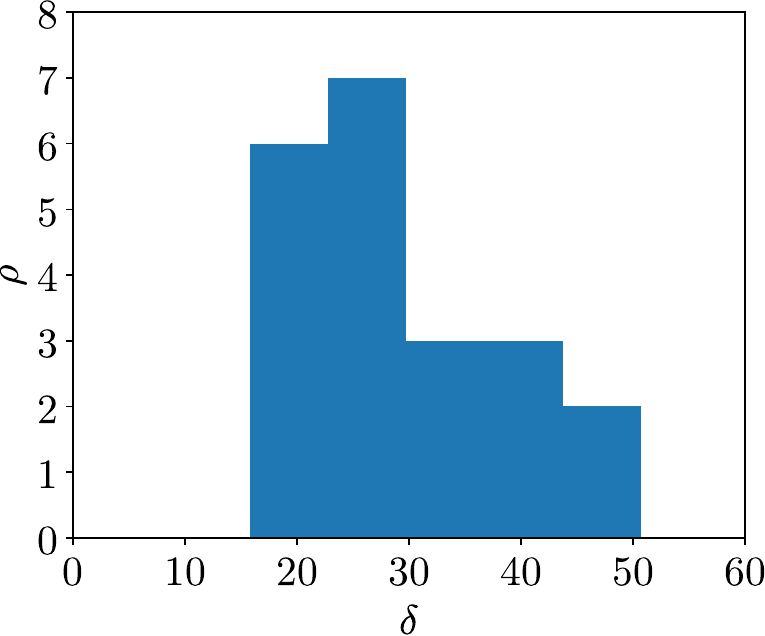} \quad \includegraphics[scale=0.6]{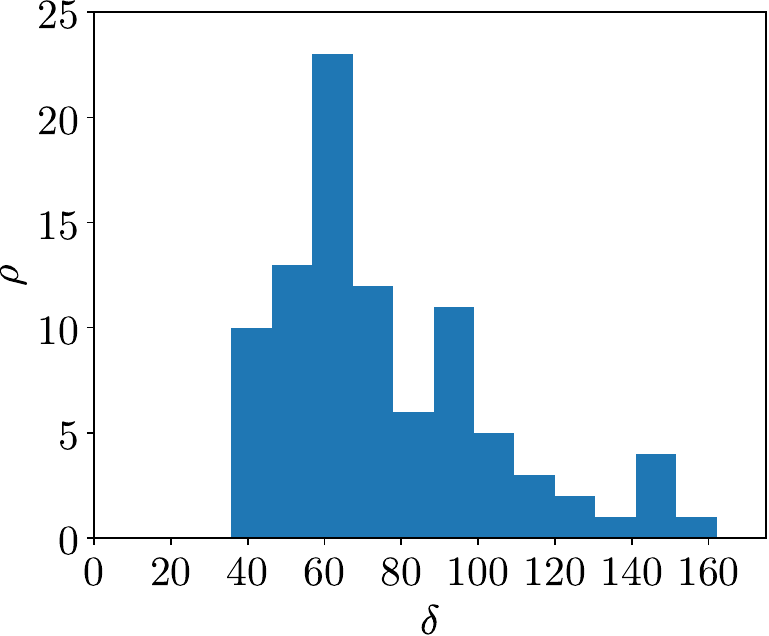}
\caption{Two-loop energy spectra for classically BPS operators of highest-weight type with $(J, H_1, H_2, H_3) = (1, 4, 2, 0)$ charges. The plot with $10$ eigenvalues on the left is for ${\rm U}(2)_k \times {\rm U}(2)_{-k}$ ABJM. The plot with $82$ eigenvalues on the right is for ${\rm U}(10)_k \times {\rm U}(10)_{-k}$ ABJM.}
\label{fig:hist2}
\end{figure}

With these choices, there are $303$ non-zero eigenvalues for $N = 2$ and $1032$ for $N = 10$. Following \cite{2306.04673}, Figure \ref{fig:hist1} plots histograms for $\delta$ which have been smoothed according to
\begin{align}
\rho(\delta) = \sum_i \frac{1}{\sqrt{2 \pi \sigma^2}} \exp \left [ -\frac{(\delta - \delta_i)^2}{2\sigma^2} \right ].
\end{align}
In the $N = 2$ plot, there is no gap on the left side for sufficiently large values of the cutoff \eqref{eq:hist-cutoff}. The gap in the $N = 10$ plot is much more robust.
In this respect, the weak coupling data for ABJM and $\mathcal{N} = 4$ SYM are qualitatively different. For the most populous sector being included, which is $(J, H_1, H_2, H_3) = (1, 4, 2, 0)$, we have also compared the true histograms in Figure \ref{fig:hist2}. Moving from $N = 2$ to $N = 10$, the spectrum becomes more sharply peaked around $\gamma \approx 6 \lambda^2$.

Although $N = 10$ and $k \gg 1$ are far from the values which recover classical gravity, it should be noted that BPS black holes should \textit{never} be treated semi-classically. More generally, near-extremal black holes have temperatures which are small compared to the fluctuations in temperature. This breakdown of the saddle-point approximation was noticed already in \cite{psstw91}. It was revisited in \cite{2003.02860} which approached the problem using an effective field theory for the near-horizon dynamics known as the Schwarzian theory. As a result, it is now possible to count states using a precise low-temperature expansion of the gravitational path integral which includes quantum corrections.

In this context, the extremal to near-extremal gap is an especially interesting observable. For the non-supersymmetric backgrounds studied in \cite{2003.02860}, there is no such gap and the density of states approaches zero continuously at the minimum energy. The situation is very different in supergravity \cite{2011.01953}. Looking at 5d gauged supergravity in particular, \cite{2203.01331} showed that for fixed angular momenta and R-symmetry charges, there is a degeneracy of black holes at the BPS energy separated from a continuum by $E_{\text{gap}} = O(N^{-2})$.\footnote{One could object that the spectrum of a CFT is always discrete and its bulk dual only appears to exhibit a continuum because gravity is a coarse-grained description. The precise statement is that gaps above $E_{\text{gap}}$ are exponentially small in $N$ while those below $E_{\text{gap}}$ are parametrically larger.} The conjecture of \cite{2306.04673} is that the 't Hooft coupling dependence turns this into
\begin{equation}
E_{\text{gap}} = \frac{E_{\text{SYM}}(\lambda)}{N^2}, \quad E_{\text{SYM}}(\lambda) = O(\lambda) \label{eq:sym-gap}
\end{equation}
which implies an $O(N^{-1})$ gap for $g_{\text{YM}} \ll 1$. An ABJM version of this conjecture can be formulated based on \cite{2412.03697}, which found $E_{\text{gap}} = O(N^{-3/2})$ from a similar analysis of 4d gauged supergravity. Even though the main focus was on $k = 1$, it is highly plausible that the gap for general $k$ scales as
\begin{align}
E_{\text{gap}} \sim \frac{G_N}{L^2} \sim k^{-1/2} N^{-3/2} = \lambda^{1/2} N^{-2}.
\end{align}
If this $\lambda^{1/2}$ is part of an interpolating function which starts at two loops when expanded around $\lambda = 0$, then \eqref{eq:sym-gap} becomes
\begin{equation}
E_{\text{gap}} = \frac{E_{\text{ABJM}}(\lambda)}{N^2}, \quad E_{\text{ABJM}}(\lambda) = O(\lambda^2) \label{eq:abjm-gap}
\end{equation}
which implies an $O(1)$ gap for $k \gg 1$. While we do not think either \eqref{eq:sym-gap} or \eqref{eq:abjm-gap} has been eestablished, it is at least encouraging that our data shows the $\mathcal{N} = 4$ SYM gap closing faster than the ABJM gap as we take $N \to \infty$.

A final point is that extremality, even in supergravity, need not imply a preserved supercharge. Looking at the AdS$_5$ black holes of \cite{hep-th/0401042,hep-th/0401129} for instance, it is easy to fix generic Cartans and see that zero temperature is achieved at an energy strictly above the BPS bound. The two only coincide when the charges satisfy an additional nonlinear relation. This relation is what diagnoses whether or not the near-extremal black hole spectra of \cite{2011.01953,2203.01331,2412.03697} have gaps. On the ohter hand, there has never been an independent derivation of it in the fortuity literature.\footnote{Some prescriptions \cite{1810.12067,2405.17648} for enumerating states with the charge relation imposed have turned out to be surprisingly simple but they are still \textit{ad hoc}.} Is such a derivation even possible in principle? Some of the more recently found black hole solutions \cite{1806.01849,1809.04084} violate the charge relation but they do so in a rather innocuous way. As discussed in \cite{2305.08922}, there are many dressings of black hole solutions which move one away from the special charge locus. The simplest way to demonstrate this is to take conformal descendants which affect the spin of an operator but not its R-symmetry representation. If all violations of the charge relation arise through this mechanism, it is reasonable to expect that they will become negligible at high energies in much the same way that descendants become rare compared to primaries in a CFT. Given this state of affairs, it would be interesting to look for evidence of a gap at weak coupling by restricting to charges where the standard condition for extremality is (approximately) compatible with the BPS bound. The difficulty of reducing type IIA supergravity on $\mathbb{CP}^3$ \cite{1001.4089} means that ABJM theory is probably not the best starting point for this type of analysis.

\section{Comparison to Known CFT Data}
\label{sec:integrability}

A widely studied aspect of ABJM theory, noticed shortly after the original paper \cite{0806.1218}, is that it can be solved at large $N$ through a correspondence with integrable spin chains \cite{0806.3951}. This has been used to study operators with a fixed number of fields as well as those where the number of fields scales with $N$ \cite{0806.3391}.\footnote{The integrability literature sometimes calls these short and long operators respectively. We will avoid this terminology because it has nothing to do with whether operators belong to long or short superconformal multiplets.} Whether fortuitous operators leave enough of an imprint at large $N$ to be detected by integrability is an open question. However, it is still instructive to see that the anomalous dimensions found above can be reproduced from the spin chain point of view. Following a brief review, we will carry out this check for some low-lying single traces.

In the large-$N$ (planar) limit, it is clear that matrix elements of the Hamiltonian vanish when they relate operators with different numbers of traces. If one also restricts to two loops, the ABJM Lagrangian can be used to show that the Hamiltonian acts within the space of single traces that only involve scalars. Clearly, two such traces can only mix if they have the same number of fields $2L$. This can be interpreted as the length of a periodic and alternating spin chain which has one field on each site. Its ground state is represented, not by the field theory vacuum, but by a trace where all $L$ barred fields are $\bar{\phi}^1$ and all $L$ unbarred fields are $\phi_1$.
Excited states have ``particles'' at locations where the scalar fields differ from $\phi_1$ and $\bar{\phi}^1$. Schematically, we can consider the $K$-particle state
\begin{align}
\left | \Psi(p_1, \dots, p_K) \right > = \sum_{1 \leq n_1 < \dots < n_K \leq L} \sum_{\sigma \in S_K} A_\sigma(p_1, \dots, p_K) e^{i [ p_1 n_{\sigma(1)} + \dots + p_K n_{\sigma(K)}]} \left | n_1, \dots, n_K \right > \label{eq:bethe-state}
\end{align}
which is labelled by a set of momenta \cite{1501.06805}. If this is to be an energy eigenstate at all, we can work out its eigenvalue using the terms where the $n_j$ separation exceeds the range of the Hamiltonian.\footnote{The range of the Hamiltonian typically increases as one goes to higher loop order. Consequently, traces of a fixed length will eventually experience wrapping effects which need to be accounted for separately.} For a large class of spin chains, it is given by
\begin{align}
\gamma \propto 2K - 2 \sum_{j = 1}^K \cos(p_j) = \sum_{j = 1}^K \frac{1}{u_j^2 + \tfrac{1}{4}} \label{eq:bethe-energy1}
\end{align}
where we have traded the momentum $p_j$ for the rapidity $u_j = \frac{1}{2} \cot (p_j/2)$. Nearby site indices then lead to constraints on the coefficients of \eqref{eq:bethe-state} which may be viewed as S-matrix elements for the $K$ particles. For ABJM theory, the biggest change is that we need separate degrees of freedom for the even sites and odd sites:
\begin{align}
\gamma = \lambda^2 \left [ \sum_{j = 1}^K \frac{1}{u_j^2 + \tfrac{1}{4}} + \sum_{j = 1}^{\bar{K}} \frac{1}{\bar{u}_j^2 + \tfrac{1}{4}} \right ]. \label{eq:bethe-energy2}
\end{align}
Along with this energy, there is a momentum given by
\begin{align}
e^{2\pi i P} = \prod_{j = 1}^K \frac{u_j + i/2}{u_j - i/2} \prod_{j = 1}^{\bar{K}} \frac{\bar{u}_j + i/2}{\bar{u}_j - i/2} \label{eq:bethe-momentum}
\end{align}
which must vanish for physical operators.

Focusing on the scalar traces above, \cite{0806.3951} computed the Hamiltonian using Feynman diagrams and showed that it takes an integrable next-nearest-neighbour form. This implies the key result that $(K, \bar{K})$-particle S-matrices factor into those which only permute two particles at a time. Subsequent work \cite{0901.0411,0901.1142} found that this structure persists for arbitrary field content, indicating (perhaps surprisingly) that notions of length and particle number remain well defined --- any mixing with traces of length $2L \pm 2$ must be suppressed by powers of $N^{-1}$ or $\lambda^2$.\footnote{If one only uses BPS letters then $L = H_1$. If one only uses
non-BPS letters then $L = -H_1$. Both of these are clearly preserved by $\textbf{H}$. The non-trivial statement about length concerns traces where some, but not all, of the letters are BPS.}

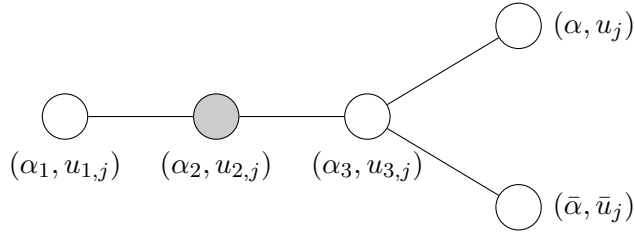
\begin{figure}[h]
\centering
\begin{tikzpicture}[
    every node/.style={font=\small},
    even/.style={circle, draw, minimum size=6mm, inner sep=0pt},
    odd/.style={circle, draw, fill=black!20, minimum size=6mm, inner sep=0pt}
]

\node[even,label=below:{$(\alpha_1,u_{1,j})$}] (a1) at (0,0) {};
\node[odd,label=below:{$(\alpha_2,u_{2,j})$}]  (a2) at (2,0) {};
\node[even,label=below:{$(\alpha_3,u_{3,j})$}] (a3) at (4,0) {};

\node[even,label=right:{$(\alpha,u_j)$}] (atop) at (6,1.2) {};
\node[even,label=right:{$(\bar{\alpha},\bar{u}_j)$}] (abot) at (6,-1.2) {};

\draw (a1) -- (a2) -- (a3);
\draw (a3) -- (atop);
\draw (a3) -- (abot);

\end{tikzpicture}
\caption{The distinguished $\mathfrak{osp}(6|4)$ Dynkin diagram with one odd root. Its structure dictates how rapidities in five families are coupled to one another in the Bethe ansatz equations. The ones associated with an even node also couple to themselves.}
\label{fig:dynkin}
\end{figure}
The way to solve an integrable spin chain is to combine periodicity with factorized scattering. This results in the \textit{Bethe ansatz equations} of which the valid rapidities are solutions. In supersymmetric examples, one distinguishes between the momentum-carrying
rapidities
from \eqref{eq:bethe-energy2} and additional internal
rapidities
which are related to R-symmetry. ABJM has three of these which we will call $u_{1,j}$, $u_{2,j}$ and $u_{3,j}$. To study a (not necessarily BPS) state with the $\mathfrak{osp}(6|4)$ Cartans $(-E+J, -E-J, H_1. H_2, H_3)$, the number of rapidity variables we need is equal to the number of raising operators which connect it to the ground state $(-L, -L, L, L, 0)$. Lie superalgebras differ from ordinary Lie algebras in that they can have different types of root systems \cite{hep-th/9607161}.
For a two-loop analysis, there is no reason not to choose the distinguished system which has only one odd root.
Its Dynkin diagram, shown in Figure \ref{fig:dynkin}, comes from taking inner products of the simple roots
\begin{align}
\begin{gathered}
\alpha_1 = (1,-1,0,0,0), \quad \alpha_2 = (0,1,-1,0,0), \quad \alpha_3 = (0,0,1,-1,0), \\
\alpha = (0,0,0,1,-1), \quad \bar{\alpha} = (0,0,0,1,1)
\end{gathered}
\end{align}
using the bilinear form $\text{diag}(-1,-1,1,1,1)$. The difference between a given highest weight state and the spin-chain ground state is now easily seen to be $K \alpha + \bar{K} \bar{\alpha} + \sum_{i = 1}^3 K_i \alpha_i$ with
\begin{align}
\begin{gathered}
K_1 = E - J - L, \quad K_2 = 2(E - L), \quad K_3 = 2E - L - H_1, \\
K = E - \frac{H_1 + H_2 + H_3}{2}, \quad \bar{K} = E - \frac{H_1 + H_2 - H_3}{2}.
\end{gathered}
\label{eq:exc-numbers}
\end{align}

We can now quote the two-loop Bethe ansatz equations as they were first conjectured in \cite{0806.3951} from an uplift of the $(\alpha, \alpha_3, \bar{\alpha})$ results. In terms of the parameters \eqref{eq:exc-numbers}, they are
\begin{align}
\left ( \frac{u_j + i/2}{u_j - i/2} \right )^L &= \prod_{k \neq j}^K \frac{u_j - u_k + i}{u_j - u_k - i} \prod_{k = 1}^{K_3} \frac{u_j - u_{3,k} - i/2}{u_j - u_{3,k} + i/2} \nonumber \\
\left ( \frac{\bar{u}_j + i/2}{\bar{u}_j - i/2} \right )^L &= \prod_{k \neq j}^{\bar{K}} \frac{\bar{u}_j - \bar{u}_k + i}{\bar{u}_j - \bar{u}_k - i} \prod_{k = 1}^{K_3} \frac{\bar{u}_j - u_{3,k} - i/2}{\bar{u}_j - u_{3,k} + i/2} \nonumber \\
1 &= \prod_{k \neq j}^{K_3} \frac{u_{3,j} - u_{3,k} + i}{u_{3,j} - u_{3,k} - i} \prod_{k = 1}^K \frac{u_{3,j} - u_k - i/2}{u_{3,j} - u_k + i/2} \prod_{k = 1}^{\bar{K}} \frac{u_{3,j} - \bar{u}_k - i/2}{u_{3,j} - \bar{u}_k + i/2} \prod_{k = 1}^{K_2} \frac{u_{3,j} - u_{2,k} - i/2}{u_{3,j} - u_{2,k} + i/2} \nonumber \\
1 &= \prod_{k = 1}^{K_3} \frac{u_{2,j} - u_{3,k} - i/2}{u_{2,j} - u_{3,k} + i/2} \prod_{k = 1}^{K_1} \frac{u_{2,j} - u_{1,k} + i/2}{u_{2,j} - u_{1,k} - i/2} \nonumber \\
1 &= \prod_{k \neq j}^{K_1} \frac{u_{1,j} - u_{1,k} - i}{u_{1,j} - u_{1,k} + i} \prod_{k = 1}^{K_2} \frac{u_{1,j} - u_{2,k} + i/2}{u_{1,j} - u_{2,k} - i/2}. \label{eq:bae}
\end{align}
To appreciate the role of these equations, let us recall that the anomalous dimensions in this paper were previously obtained as roots of a single polynomial --- the characteristic polynomial for a block of $\textbf{H}$. Writing this down required a case-by-case computation of supercharge actions and Wick contractions which become expensive for large values of the Cartans. Although \eqref{eq:bae} has the added complexity of being a \textit{coupled system} of nonlinear equations, it can be written down right away and yields the same values for anomalous dimensions in the planar limit. Moreover, it can be generalized to arbitrary loop order as was done in \cite{0807.0777,0807.1924}.

\subsection{Integrability at large $N$}

As a demonstration of what \eqref{eq:bae} can do, consider the four non-graviton parts of Figure \ref{fig:GH-1420}. One is a centralizer descendant, as previously discussed, but the other three have anomalous dimensions given by one of
\begin{align}
    \gamma &= \gamma_-(N) = \lambda^2 \left [ 6-\frac{48}{N^2}+\frac{720}{N^4} + O(N^{-6}) \right ] + O(\lambda^4) \nonumber \\
    \gamma &= 6 \lambda^2 + O(\lambda^4) \label{eq:dim-expansions} \\
    \gamma &= \gamma_+(N) = \lambda^2 \left [ 8+\frac{28}{N^2}-\frac{720}{N^4} + O(N^{-6}) \right ] + O(\lambda^4). \nonumber
\end{align}
The last one belongs to an eigenvector which is double-trace at large $N$ so the prediction is that the Bethe ansatz equations should find $\gamma = 6\lambda^2$ for all $(J,H_1,H_2,H_3) = (\frac{1}{2},2,1,0)$ states. To start, it is important to recognize that these Cartans need to be shifted by those of certain simple roots. Plugging $(J,H_1,H_2,H_3) = (\frac{1}{2},2,1,0)$, along with $E = \frac{5}{2}$ and $L = 2$, into \eqref{eq:exc-numbers} would not be correct. The problem is that such a state only has a highest weight with respect to the $\mathfrak{osp}(4|2) \oplus \mathfrak{u}(1|1)$ centralizer, not the full $\mathfrak{osp}(6|4)$.
The required shift is determined by taking $A_1[2J]_{J + H_1 + 1}^{(H_1, H_2, H_3)}$, the 3d $\mathcal{N} = 6$ multiplet to which classically BPS states must belong, and decomposing it into centralizer multiplets. This tells us to act with the unique supercharge which restores the BPS condition and keeps $H_2$ and $H_3$ maximal. Translated into the language of simple roots, the first two lines of \eqref{eq:dim-expansions} are associated with
\begin{align}
(-E+J, -E-J, H_1, H_2, H_3) + \alpha_2 = (-2, -2, 1, 1, 0)
\end{align}
which implies
\begin{align}
K_1 = 0, \quad K_2 = 0, \quad K_3 = 1, \quad K = 1, \quad \bar{K} = 1. \label{eq:ks1}
\end{align}
These numbers are small enough that \eqref{eq:bae} can be solved by hand. The solutions are
\begin{align}
\begin{gathered}
(u_{3,1}, u_1, \bar{u}_1) = \frac{1}{2\sqrt{3}} (0, \pm 1, \mp 1) \\
(u_{3,1}, u_1, \bar{u}_1) = \pm \frac{1}{2} (1, 1, 1)
\end{gathered}
\end{align}
but only the first pair survives the \eqref{eq:bethe-momentum} condition. The energy it yields via \eqref{eq:bethe-energy2} is precisely $\gamma = 6\lambda^2$.

For added complexity, we can consider Figures \ref{fig:GH-1640} with seven $\mathfrak{osp}(4|2)$ primaries above the BPS bound and Figure \ref{fig:GH-1600} with six. Writing the sets of anomalous dimensions, they are
\begin{align}
\gamma = 4\lambda^2 [1 + O(N^{-2})] + O(\lambda^4), \quad \gamma = 8\lambda^2 [1 + O(N^{-2})] + O(\lambda^4)
\end{align}
and
\begin{align}
\gamma = 4\lambda^2 [1 + O(N^{-2})] + O(\lambda^4), \quad \gamma = 12\lambda^2 [1 + O(N^{-2})] + O(\lambda^4)
\end{align}
respectively once we go to large $N$ and throw out the multi-traces.\footnote{We have also had to take appropriate linear combinations of eigenvectors that become degenerate in the planar limit so that single-traces and multi-traces are separated.}
In the first case, the $\alpha_2$-shifted weight vector corresponding to $(J, H_1, H_2, H_3) = (\frac{1}{2}, 3, 2, 0)$ gives \eqref{eq:ks1} again. The Bethe ansatz equations are different, however, because we now have $L = 3$. Their nine solutions are
\begin{align}
& (u_{3,1}, u_1, \bar{u}_1) = \frac{1}{2} (0, \pm 1, \mp 1), \quad (u_{3,1}, u_1, \bar{u}_1) = \pm \frac{\sqrt{3}}{2} (1, 1, 1), \quad (u_{3,1}, u_1, \bar{u}_1) = (0,0,0), \nonumber \\
& (u_{3,1}, u_1, \bar{u}_1) = \Big ( \pm \frac{\sqrt{3}}{5}, \pm \frac{\sqrt{19 \mp 4 \sqrt{21}}}{10}, \frac{\sqrt{7} \pm 2\sqrt{3}}{10} \Big ), \\
& (u_{3,1}, u_1, \bar{u}_1) = \Big ( \pm \frac{\sqrt{3}}{5}, \frac{\sqrt{7} \pm 2 \sqrt{3}}{10}, \pm \frac{\sqrt{19 \mp 4 \sqrt{21}}}{10} \Big ). \nonumber
\end{align}
The first pair yields $\gamma = 4\lambda^2$, the zero solution yields $\gamma = 8\lambda^2$ and the rest are all fail to satisfy \eqref{eq:bethe-momentum}. Moving onto $(J, H_1, H_2, H_3) = (\frac{1}{2}, 3, 0, 0)$, the $\alpha_2$-shifted weight vector produces
\begin{align}
K_1 = 0, \quad K_2 = 0, \quad K_3 = 1, \quad K = 2, \quad \bar{K} = 2. \label{eq:ks2}
\end{align}
The resulting system should now be solved numerically, which we have done with \texttt{bertini} \cite{bertini}. There are initially 84 real solutions, of which 30 satisfy the momentum constraint, but there is one more condition that should be imposed. It is a version of the Pauli exclusion principle for the Bethe ansatz which states that
rapidities
of the same type must have different values \cite{0911.2220}. Forbidding $u_1 = u_2$ and $\bar{u}_1 = \bar{u}_2$ gets us down to just four solutions. Guessing a closed form based on several digits, they are given by orbits of
\begin{align}
(u_{3,1}, u_1, u_2, \bar{u}_1, \bar{u}_2) = \frac{1}{2\sqrt{3}} (0, 1, -1, 1, -1)
\end{align}
under $u_1 \leftrightarrow u_2$ and $\bar{u}_1 \leftrightarrow \bar{u}_2$. These all have $\gamma = 12\lambda^2$ so how do we get $\gamma = 4\lambda^2$? The resolution to this well known issue \cite{1409.7382} is that \texttt{bertini} solves systems of polynomial equations which require us to clear denominators in \eqref{eq:bae}. This can miss solutions where rapidities sit at the locations of the poles. Four especially simple ones have
\begin{align}
(u_1, u_2) = (\bar{u}_1, \bar{u}_2) = \frac{i}{2} (\pm 1, \mp 1), \quad (u_1, u_2) = -(\bar{u}_1, \bar{u}_2) = \frac{i}{2} (\pm 1, \mp 1) \label{eq:singular-sol}
\end{align}
which obey the two conditions we have been imposing. The appropriate value for $u_{3,1}$ may be obtained by adding a regularization parameter to \eqref{eq:singular-sol}. Then, we may safely take the $x \to -\frac{i}{2}$ limit in
\begin{align}
\gamma = 2\lambda^2 \left [ \frac{1}{x^2 + \tfrac{1}{4}} + \frac{1}{(x + i)^2 + \tfrac{1}{4}} \right ] = \frac{16 \lambda^2}{(2x + 3i)(2x - i)}.
\end{align}

This method for studying single-trace operators forms the starting point for studies of multi-trace operators as well. This is because the planar limit makes anomalous dimensions additive --- gravitationally, there is no binding energy when Newton's constant is taken to zero. In a double-trace, for instance, each appearance of $\text{Tr}(X_1 \bar{X}_1 \dots X_L \bar{X}_L)$ may be expanded in a basis of energy eigenstates. It then becomes well defined to compute the desired anomalous dimension $\gamma = \gamma_1 + \gamma_2$ from a single term in the resulting double sum. The original operator being a primary guarantees that $\gamma$ will be the same for each term even though $(\gamma_1, \gamma_2)$ individually are allowed to differ. What we have not understood is why certain sets of operators should prefer for the lightest one to be single-trace. This is an enticing feature of the large-$N$ trajectories connected to fortuitous operators.

\subsection{Effective vertices at finite $N$}

While integrability is lost at finite $N$, there is sometimes enough structure left at fixed loop order for particular blocks of the Hamiltonian to still be written in a compact way. One way of seeing this is to use the method of effective vertices from \cite{hep-th/0208178}. It has been applied to the spinless classically BPS Hilbert space of ABJM theory in \cite{0811.2150}. These operators, which are necessarily built from only the four scalar letters $\phi_a$ and $\bar{\phi}^a$, form what is called the $\mathfrak{su}(2) \oplus \mathfrak{su}(2)$ sector of ABJM because this is what results from taking the centralizer of $Q$, $Q^\dagger$ and $J$.\footnote{The operators built from all eight scalars, which were the focus of \cite{0806.3951}, are often referred to as an $\mathfrak{su}(4)$ sector in the literature. This is imprecise because they only form a closed subsector due to a two-loop accident. At higher orders, they will mix with other operators in the centralizer of $K_\mu$ and $M_{\mu\nu}$ which is the true $\mathfrak{su}(4)$ sector. A good mnemonic is that loop corrections should make a sector smaller, not bigger.}

For the operators studied in \cite{0811.2150}, which we will not repeat here, it is always possible to make them vanish by reordering the scalars in each trace. This is what explains the anomalous dimensions they receive --- they are $Q$-exact and therefore do not correspond to BPS operators.\footnote{Recall from the previous subsection that non-BPS operators built only from $\phi_a$ and $\bar{\phi}^a$ can at most be primary with respect to $\mathfrak{osp}(4|2)$, not the full centralizer $\mathfrak{osp}(4|2) \oplus \mathfrak{u}(1|1)$.} Going through the various charges, $(J, H_1, H_2, H_3) = (0, 4, 2, 0)$ leads to the exact
\begin{align}
\gamma = \lambda^2 \left [ 5 + \frac{6}{N^2} \pm \sqrt{1 + \frac{20}{N^2} + \frac{4}{N^4}} \right ] + O(\lambda^4)
\end{align}
and quintic roots expanding to
\begin{align}
\begin{gathered}
\gamma = \lambda^2 \left [ 8 \pm \frac{16}{N} + O(N^{-3}) \right ] + O(\lambda^4) ,\quad \gamma = \lambda^2 \left [ 4 - \frac{28}{N^2} + O(N^{-3}) \right ] + O(\lambda^4) \\
\gamma = \lambda^2 \left [ 6 - \frac{64}{N^2} + O(N^{-3}) \right ] + O(\lambda^4), \quad \gamma = \lambda^2 \left [ 8 + \frac{64}{N^2} + O(N^{-3}) \right ] + O(\lambda^4)
\end{gathered}
\end{align}
Next, in $(J, H_1, H_2, H_3) = (0, 4, 1, 1)$, we have found
\begin{align}
\begin{gathered}
\gamma = \lambda^2 \left [ 6 - \frac{2}{N^2} \pm 2 \sqrt{\frac{12}{N^2} + \frac{1}{N^4}} \right ] + O(\lambda^4) \\
\gamma = \lambda^2 \left [ 6 - \frac{16}{N^2} \right ] + O(\lambda^4), \quad \gamma = \lambda^2 \left [ 6 + \frac{12}{N^2} \right ] + O(\lambda^4)
\end{gathered}
\end{align}
exactly where the last appears to be an $\mathfrak{osp}(4|2)$ primary but not an $\mathfrak{so}(4)$ highest weight. Finally, in $(J, H_1, H_2, H_3) = (0, 4, 0, 0)$, we can expand the roots of a cubic to find the approximate eigenvalues
\begin{align}
\begin{gathered}
\gamma = \lambda^2 \left [ 12 - \frac{156}{N^2} + O(N^{-4}) \right ] + O(\lambda^4), \quad \gamma = \lambda^2 \left [ 8 + \frac{64}{N^2} + O(N^{-4}) \right ] + O(\lambda^4) \\
\gamma = \lambda^2 \left [ 6 - \frac{576}{5 N^4} + O(N^{-5}) \right ] + O(\lambda^4), \quad \gamma = \lambda^2 \left [ 6 - \frac{1728}{N^5} + O(N^{-5}) \right ] + O(\lambda^4) \\
\gamma = \lambda^2 \left [ 16 + \frac{144}{N^2} + O(N^{-4}) \right ] + O(\lambda^4), \quad \gamma = \lambda^2 \left [ 4 - \frac{84}{N^2} + O(N^{-4}) \right ] + O(\lambda^4).
\end{gathered}
\end{align}
All of these show a perfect match with the results of \cite{0811.2150}.

The last sector where we can make a comparison is $(J, H_1, H_2, H_3) = (0, 7, 4, 3)$ which was used to test similarities with $\mathcal{N} = 4$ SYM. Since this leads to a much bigger matrix, we have opted to sample it at fixed values of $N$ as in subsection \ref{sec:more}.
This shows that, within the set of eigenvalues, two are consistent with
\begin{align}
\gamma = \lambda^2 \left [ 5 + \frac{15}{N^2} \right ] + O(\lambda^4), \quad \gamma = \lambda^2 \left [ 6 + \frac{24}{N^2} \right ] + O(\lambda^4)
\end{align}
as they should be. A final comment is that all anomalous dimensions in this section hold with the standard $\lambda = N/k$. Since \cite{0811.2150} uses a normalization for the ABJM action which is the same as what we have used in Appendix \ref{app:gram}, we conclude that their non-standard definition of the 't Hooft coupling is a typo.

\section{Conclusion}
\label{sec:conc}

In this paper we studied fortuitous BPS states in ABJM theory. At the level of cohomology classes, we studied the mechanism by which some fortuitous classes at higher levels can be obtained from those at lower levels by adding graviton dressings, distinguishing between core and dressed fortuitous classes and establishing a partial no-hair theorem analogous to that previously found in ${\cal N}=4$ SYM \cite{2304.10155}. Going beyond cohomology, we studied the actual BPS operators by embedding fortuitous states into one-parameter families defined by a continuation of the gauge group rank $N$. This analysis allowed us to extract a large-$N$ limit in which the operators simplify and their anomalous dimensions match known integrability results. By following OPE coefficients as functions of $N$, we could also observe the decoupling of operators as they become unphysical and quantify, at the level of the actual operators, the dressing structure first identified in cohomology.

Graviton dressings of fortuitous states can be thought of as a quantum version of hair \cite{2304.10155}. In studying the dressing structure,
we found expressions for many of the cohomology classes which contribute to the partition functions \eqref{zfort-n2} and \eqref{zfort-n3}. They are simply given by the green expressions starting with \eqref{dressing1} and ending with \eqref{dressing-last}. Several disallwoed dressings, written as red entries, were observed along the way.
This already precludes the possibility that the Hilbert space of black holes has a simple Fock-space structure, built from core fortuitous states by adding arbitrary graviton dressings. But is there a pattern governing which dressings are allowed? Such a pattern would not be in conflict with the chaotic nature of black holes. In discussing \eqref{eq:sym-gap} and \eqref{eq:abjm-gap}, it was important to note that conformal primaries exponentially outnumber descendants when some sufficiently large energy cutoff is imposed. The same would be true for core primaries outnumbering dressed ones even in the case of a Fock space. The partial no-hair theorems only make this more pronounced. An appealing picture is that maximal chaos applies to core black holes while simple patterns are allowed to govern hair at the level of the index or, more ambitiously, the partition function.

In connection with this question, it was shown in ${\cal N}=4$ SYM that some fortuitous cohomologies at leading non-trivial order are lifted above the BPS bound by higher-loop effects \cite{2510.24008,2511.09519}. More recently, it was found that some of the lightest fortuitous states and their dressings survive quantum corrections, while many heavier core primaries are lifted \cite{2606.27955}. Is there a connection between the dressing structure and quantum robustness?

This work has gathered several kinds of data on fortuitous states: explicit operator representatives, trajectories in $N$ of both the states and their anomalous dimensions, and a first set of OPE coefficients. What can we learn from these data? Can we identify new characterizations of fortuitous states? Can we extract further black-hole properties from these states?

Along these lines, our results show that the 18 $\mathfrak{osp}(4|2) \oplus \mathfrak{u}(1|1)$ highest-weight trajectories assume a considerably simpler form at large $N$. Any single-trace primary must be a core primary but the converse also holds in the cases we have seen. Another simplification is that for the trajectories with $H_3=0$, the planar endpoints have definite parity under the $\mathbb{Z}_2$ automorphism exchanging the two
$\mathfrak{su}(2)$ factors of the centralizer R-symmetry, with both even and odd endpoints occurring. It would be interesting to further pursue the idea put forward in ${\cal N}=4$ SYM \cite{2306.04693} of characterizing fortuitous states by what they become when continued to the planar limit. While in general we do not expect a simple pattern for fortuitous states at the BPS value of $N$, since black holes are expected to be maximally chaotic objects, at large $N$ integrability sets in, raising the possibility that these states may leave an identifiable signature in the integrability data.

The OPE coefficient analysis is another possible source of interesting features. Two of the coefficients we have followed would lead to an obvious violation of unitarity if they did not vanish at $N = 2$.
The check we have done resonates with a more general formalism in \cite{2509.05834} which accounts for evanescent operators (the ones that vanish when trace relations are imposed). On the other hand, for the particular three-point functions considered here, the OPE coefficients also effectively probe the dressing structure studied at the level of cohomology. This might also be true for the more general dressing protocols used in \cite{2412.08695}. It would be interesting to study OPE coefficients involving fortuitous states more systematically, especially beyond the free-theory setting employed here. Understanding the bulk counterpart of these continuations in $N$ is another natural next step.

Even though infinitely many fortuitous representatives are known in $\mathcal{N} = 4$ SYM, they appear at high enough levels that only the original class from \cite{2209.06728,2209.12696} has been completed into a BPS state and continued away from $N = 2$. Regarding one one of the main motivations for this study, fortuitous ABJM states are more plentiful at low levels and it would be nice to realize them as part of infinite families. Recently, \cite{2609.19297} gave a recipie to construct families of fortuitous operators which are dual to branes instead of black holes. The simplest example,
\begin{align}
\mathcal{O}^{a_1 \dots a_{N - 1}}_{b_1 \dots b_{N - 1}} = \epsilon \epsilon (\phi_{b_1} \dots \phi_{b_{N - 1}} \phi_c) \epsilon \epsilon (\bar{\psi}_d \bar{\phi}^{a_1} \dots \bar{\phi}^{a_{N - 1}}) \epsilon^{cd} - \epsilon \epsilon (\bar{\phi}^{a_1} \dots \bar{\phi}^{a_{N - 1}} \bar{\phi}^c) \epsilon \epsilon (\psi^d \phi_{b_1} \dots \phi_{b_{N - 1}}) \epsilon_{cd}, \label{gen-monotone}
\end{align}
interpolates between the lowest bi-doublet at $N = 2$, the lowest bi-triplet at $N = 3$ and so on. Even though $U(N)_k \times U(N)_{-k}$ ABJM theory does not have a baryonic $U(1)$ symmetry, one can interpret each term as the product of one baryonic and one anti-baryonic operator. Each such operator becomes $Q$-closed for arbitrary $N$ once a baryonic ground state (a determinant) is factored out. According to the new criteria introduced in \cite{2609.19297}, products of this type are \textit{generalized monotones} while the ones with more complicated $N$ dependence are \textit{strongly fortuitous}. Many of the BPS operators we have studied are generalized monotones for the simple reason that they can be obtained as graviton dressings of \eqref{gen-monotone}. Conversely, the core operator \eqref{fort2-full1} is likely to be strongly fortuitous. More precisely, there is no way to write it in terms of ordinary monotones dressing determinants if we use an ansatz where all terms factor into baryon number $+B$ and $-B$ components as in \eqref{gen-monotone}. The last state to account for is the $N = 2$ version of \eqref{eq:planar-2642}. After a brute force search, it can be written as
\begin{align}
\mathcal{O}^{abc}_d(2) = 3 \epsilon \epsilon (\phi_d \phi_f) \epsilon \epsilon (\bar{\phi}^a D \bar{\phi}^b) \text{Tr}(\phi_e \bar{\phi}^c) \epsilon^{ef} + 2 \epsilon \epsilon (\phi_d \phi_e) \epsilon \epsilon (\bar{\psi}_f \bar{\phi}^a) \text{Tr}(\psi^b \bar{\phi}^c) \epsilon^{ef}
\end{align}
and both terms of this expression can be shown to have determinants as their only fortuitous part. As a result, this core primary is similar to the one in \eqref{gen-monotone}. Interestingly, Figure \ref{fig:GH-1600} shows that generalized monotone and strongly fortuitous trajectories are allowed to annihilate with each other.

There is still much to be learned about fortuitous states and their identification as black-hole microstates. Here we have exhibited concrete ways in which fortuity can be studied beyond state counting and identified a particularly favorable setting for doing so. We hope that the data and results presented here will provide a useful starting point for a deeper understanding of quantum black holes in holographic theories, a direction we intend to pursue in future work.


\section*{Acknowledgements}
We are grateful to Marco Bianchi, Davide Gaiotto, Kristan Jensen, Justin Kulp, Siyul Lee, Harish Murali, Sabrina Pasterski, Gordon Rogelberg, Chiara Toldo and Pedro Vieira for stimulating discussions. C.B. thanks the organizers of IX Quantum Gravity in the Southern Cone where some of these discussions took place. This project received support from the FAPESP Foundation through the grants 2023/04415-2 and 2023/03825-2. Research at Perimeter Institute is supported in part by the Government of Canada through the Department of Innovation, Science and Economic Development and by the Province of Ontario through the Ministry of Colleges and Universities.

\appendix
\section{Constructing the Gram matrix}
\label{app:gram}
Identifying fortuitous cohomology classes requires a matrix representation of the nilpotent supercharge $Q$. For ABJM theory, blocks of this matrix corresponding to several charge sectors were constructed in \cite{2512.04146,2512.23603}. To also solve for the action of $Q^\dagger$, as used in section \ref{sec:hamiltonian}, the new ingredient is the Gram matrix which records all of the possible BPZ inner products. The following describes a simple way to compute it.

\subsection{Generalities}

Suppose we have some collection of linearly independent multi-traces $\{w_i\}$ spanning a space $V$, which we will take to be one ABJM charge sector later. The words are constructed from the BPS letters \eqref{abjm-letters}. Expanding their colour indices, we can write
\begin{equation}
    w_i = \sum_{\mu}A_{\mu i}t_\mu,
\end{equation}
where $\{t_\mu\}$ is a collection of component monomials. The Gram matrix is the BPZ inner product
\begin{equation}\label{eq:gram-matrix}
    G_{ij}\equiv\langle w_i,w_j\rangle
    =\langle 0|w_i^\dagger w_j|0\rangle
    =\sum_{\mu,\nu}A_{\mu i}^*A_{\nu j}\langle 0|t_\mu^\dagger t_\nu|0\rangle .
\end{equation}
Crucially, Hermitian conjugation must not be confused with the bar on a BPS letter: $\phi_a^\dagger$ is proportional to $\bar\Phi_L^a$, whereas the BPS letter $\bar\phi^a$ is proportional to $\bar\Phi_R^a$. The same distinction applies to the fermions. Thus the adjoint of a BPS word is generally built from non-BPS fundamental components.\footnote{In $\mathfrak{osp}(\mathcal{N}|4)$ notation, we are using BPS letters with respect to $Q = Q_{-1} + iQ_{-2}$. Their Hermitian conjugates are BPS with respect to $Q_{-1} - iQ_{-2}$.}

For two component monomials $s=X_1 \bar{X}_1 \cdots X_L \bar{X}_L$ and $t=Y_1 \bar{Y}_1 \cdots Y_L \bar{Y}_L$, Wick's theorem gives
\begin{equation}\label{eq:monomials-inner-products}
    \langle s,t\rangle
    =\sum_{\sigma, \tau \in S_L}(-1)^{N_{\sigma + \tau}}
    \prod_{r=1}^{L}
    \left\langle 0\left|X_r^\dagger Y_{\sigma(r)}\right|0\right\rangle
    \left\langle 0\left|\bar{X}_r^\dagger \bar{Y}_{\tau(r)}\right|0\right\rangle,
\end{equation}
where $N_{\sigma + \tau}$ is the parity of the fermionic permutation. The colour-index expansion and the coefficients $A_{\mu i}$ can already be produced by the code in \cite{2512.23603}.

\subsection{Derivation of letter inner products}

The transformation laws in section \ref{sec:conventions} are obtained from an action which has a $\frac{2\pi}{k}$ multiplying the canonical kinetic terms. In this normalization, the fundamental field two-point functions are
\begin{equation}
    \left\langle (\Phi_A)^{ij}(x)(\bar\Phi^B)_{kl}(0)\right\rangle = \delta_A^B \delta^i_k \delta^j_l \dfrac{1}{2k |x|},\quad
    \left\langle (\Psi^A_\alpha)^{ij}(x)(\bar\Psi_{B\beta})_{kl}(0)\right\rangle = \delta^A_B \delta^i_k \delta^j_l \dfrac{i x_\mu \gamma^\mu_{\alpha\beta}}{2k |x|^3}. \label{lagrangian-2pt}
\end{equation}
To fix the dependence on derivatives $D = iD_{++}$, recall the definitions
\begin{equation}
H = D - M^-_{\phantom{-}-},\quad P = \frac{1}{2}P_{++},\quad K = -\frac{1}{2}K_{++}
\end{equation}
and the associated $\mathfrak{sl}(2)$ algebra
\begin{equation}
[H,P]=2P,\quad [H,K]=-2K,\quad [P,K]=H,
\end{equation}
encountered when studying the $\mathfrak{osp}(4|2)\oplus \mathfrak{u}(1|1)$ centralizer algebra of the BPS Hamiltonian. We note that in the free theory, $D = 2P$. Moreover, $P^\dagger = -K$ is the BPZ adjoint. This means that for a given BPS letter $X$, the equalities
\begin{equation}
    G_X(m,n) \equiv \langle 0|(D^m X)^\dagger D^n X|0\rangle = 2^{m+n}\langle P^m X, P^n X\rangle = 2^{m+n}\langle X, (-1)^m K^m P^n X\rangle
\end{equation}
hold. Next, we have the commutation relation
\begin{equation}
    [K,P^n] = -n(H-n+1) P^{n-1},
\end{equation}
which implies
\begin{equation}
    G_X(m,n) = 2^{m+n}\langle  P^{m-1}X, n (H-n+1) P^{n-1}X\rangle
\end{equation}
since $K$ annihilates $X$. But since $P$ is a raising operator for the dimension we have that
\begin{equation}
    H P^{n-1} X = (h_X+2n-2)P^{n-1}X,\quad HX = h_X X
\end{equation}
which leads to the recursion relation
\begin{equation}
    G_X(m,n) = 2^{m+n}n(h_X+n-1)\langle  P^{m-1}X, P^{n-1}X\rangle = 4n(h_X+n-1)G_X(m-1,n-1).
\end{equation}
Now suppose that $m > n$. Iterating this relation $n$ times reduces it to 
\begin{equation}
G_X(m-n,0) = 2^{m-n}\langle P^{m-n}X,X\rangle = -2^{m-n}\langle P^{m-n-1}X,KX\rangle = 0
\end{equation}
where we have used the fact that $K$ annihilates $X$ again. The same is true for $m<n$. As a result, this can only yield a non-zero result when $m = n$. Denoting henceforth $G_{X}(n,n)=G_X(n)$, its recursion relation
can be solved exactly to give
\begin{equation}
    G_X(n) = 4^nn! (h_X)_n G_X(0) = 4^n\dfrac{2}{k} n! (h_X)_n.
\end{equation}
The initial condition is the same for both bosons and fermions because we have fixed the phase in the fermionic BPZ adjoint to ensure that all norms are positive.\footnote{Also, $G_X(0)$ is not the $\frac{1}{2k}$ present in \eqref{lagrangian-2pt}. It is larger by a factor of 4 because \cite{2512.23603} applied a convenient rescaling to all BPS components of fundamental fields to define the BPS letters.} Using the fact that the bosons have $h = E + J = \frac{1}{2}$, while the fermions have $h = E + J = \frac{3}{2}$, the final inner product result for unbarred letteres (suppressing colour indices) is
\begin{align}
    \left\langle 0\left|\big(D^m \phi_a \big)^\dagger D^n\phi_b\right|0\right\rangle &= \delta^{mn} \delta_{ab} \frac{2}{k}(2n)!, \quad
    \left\langle 0\left|\big(D^m \psi^a \big)^\dagger D^n\psi^b\right|0\right\rangle &= \delta^{mn} \delta^{ab} \frac{2}{k} (2n + 1)!.
\end{align}
For barred fields, the result is exactly the same with the flavour indices in opposite positions.

 \bibliographystyle{utphys}
 \bibliography{references}

\end{document}